\documentclass[11pt]{article}
\usepackage{jheppub}
\pdfoutput=1
\usepackage{amsmath,bbm,array,amsfonts,graphicx,wrapfig,lscape,float,mathtools,multirow,longtable,amsthm}
\usepackage[dvipsnames]{xcolor}
\usepackage{stackrel}
\usepackage[all]{xy}
\usepackage{caption}
\usepackage{subcaption}
\usepackage[bottom]{footmisc}
\usepackage{mathrsfs}
\usepackage{tikz}
\usepackage{bm}
\usepackage{color}
\usepackage{enumerate}
\usetikzlibrary{decorations.pathreplacing}
\usepackage{diagbox}
\numberwithin{equation}{section}
\numberwithin{figure}{section}
\numberwithin{table}{section}
\usepackage{pgfplots}
\pgfplotsset{compat=1.14}
\usepackage{makecell}
\usepackage{lscape}
\usepackage[utf8]{inputenc}
\usepackage{mathtools}
\usepackage{upgreek}
\allowdisplaybreaks
\newcommand{\rvline}{\hspace*{-\arraycolsep}\vline\hspace*{-\arraycolsep}}
\usepackage{mathrsfs}

\usepackage[autostyle=true]{csquotes}

\newtheorem{remark}{Remark}

\usepackage[toc,page]{appendix}

\usetikzlibrary{external}
\tikzset{external/system call={pdflatex \tikzexternalcheckshellescape 
		-halt-on-error
		-interaction=batchmode 
		-jobname "\image" "\texsource"
		&& pdftops -eps "\image.pdf"}}
\usepackage{tocloft}

	\title{More on Affine Dynkin Quiver Yangians}
	
	\author[a,b]{Jiakang Bao}
	
	\affiliation[a]{
		Department of Mathematics, City, University of London, EC1V 0HB, UK}
	\affiliation[b]{
		London Institute for Mathematical Sciences, Royal Institution, London W1S 4BS, UK}
	
	\emailAdd{jiakang.bao@city.ac.uk}
	
	\preprint{
		\begin{flushright}
			
		\end{flushright}
	}

	\abstract{We consider the quiver Yangians associated to general affine Dynkin diagrams. Although the quivers are generically not toric, the algebras have some similar structures. The odd reflections of the affine Dynkin diagrams should correspond to Seiberg duality of the quivers, and we investigate the relations of the dual quiver Yangians. We also mention the construction of the twisted quiver Yangians. It is conjectured that the truncations of the (twisted) quiver Yangians can give rise to certain $\mathcal{W}$-algebras. Incidentally, we give the screening currents of the $\mathcal{W}$-algebras in terms of the free field realization in the case of generalized conifolds. Moreover, we discuss the toroidal and elliptic algebras for any general quivers.
	}

\begin{document}
	\maketitle

\section{Introduction and Summary}\label{intro}
In the setting of Type IIA string theory on toric Calabi-Yau (CY) threefolds, the quiver Yangians \cite{Li:2020rij,Galakhov:2020vyb} were constructed as realizations of the BPS algebras \cite{Harvey:1995fq,Harvey:1996gc}. In particular, they admit the crystal melting \cite{Okounkov:2003sp,Iqbal:2003ds,Ooguri:2009ijd}, whose configurations count the BPS degeneracies, as representations. A concise summary of the quiver Yangians can be found in \cite{Yamazaki:2022cdg}.

One may consider certain extensions of the quiver Yangians. For instance, the shifted quiver Yangians \cite{Galakhov:2021xum} and different framings can naturally incorporate the wall crossing phenomena with various configurations of the crystal models (see also \cite{Aganagic:2010qr,Bao:2022oyn}). There are also trigonometric and elliptic counterparts of the quiver Yangians as introduced in \cite{Noshita:2021ldl,Galakhov:2021vbo}. These algebras, dubbed rational/toroidal/elliptic quiver BPS algebras, form a hierarchy that also appears in the context of integrability. More generally, it could be possible to further extend the algebras associated to the generalized cohomology theories \cite{Galakhov:2021vbo}. See also \cite{Galakhov:2023aev,Li:2023zub} for some recent progress.

Mathematically, the counting of BPS states can be translated into the study of (generalized) Donaldson-Thomas (DT) invariants and hence many other invariants in enumerative geometry. It is also expected that the quiver BPS algebras are closely related to various quantum algebras such as the cohomological Hall algebras (CoHAs) \cite{Joyce:2008pc,Kontsevich:2008fj,Kontsevich:2010px,Rapcak:2018nsl}.

Therefore, we may also consider the BPS algebras associated to the Coulomb branch besides the quiver Yangians of the Higgs branch. In particular, the BPS algebras constructed from scattering $S$-matrices were identified with the BPS Hall algebras in \cite{Galakhov:2018lta}. Such algebras have a shuffle algebra structure regarding the wave functions for the BPS Hilbert spaces via the localization techniques. Under the Higgs-Coulomb duality \cite{Denef:2002ru}, the quiver Yangians and the quiver shuffle algebras are expected to give isomorphic BPS Hilbert spaces.

Another interesting perspective would be the connections of the quiver Yangians to the vertex operator algebras (VOAs)/$\mathcal{W}$-algebras. These VOAs have been extensively studied in various literature in physics and mathematics. It is hard to give an exhaustive list of the references here, and we shall instead mention some when we discuss the relevant aspects in more detail. More specifically, the truncations of the quiver Yangians are expected to give rise to (the universal enveloping algebras) of the $\mathcal{W}$-algebras. This should implement the BPS/CFT (aka AGT, 2d/4d) correspondence \cite{Alday:2009aq,Wyllard:2009hg}.

Algebraically, the quiver Yangians associated to the toric CY threefolds without compact divisors can be studied with the help of their underlying Kac-Moody algebras, namely $\widehat{\mathfrak{gl}}(m|n)$ and $\widehat{D}(2,1;\alpha)$ (in the toric phase). In this paper, we shall consider the quivers and their Yangians associated to general affine Dynkin diagrams. We expect that these supersymmetric gauge theories can be constructed geometrically similar to the 4d $\mathcal{N}=2$ quivers associated to the corresponding affine Dynkin diagrams. Without the toric setting, it is not clear whether these quiver Yangians would still play the role as the BPS algebras of the gauge theories. Nevertheless, given the similar constructions (either algebraically or geometrically), it is natural to conjecture that they are still intimately related to the BPS states, and recover the BPS algebras possibly with some modifications. In fact, it was proposed recently in \cite{Li:2023zub} that the quiver Yangians for any quivers should still give rise to the BPS algebras. Moreover, the representations called the poset representations were also studied therein.

As the definition of the quiver Yangian can be applied to any quiver, to obtain the algebras, we just need to identify the gauge theories whose quivers are associated to the affine Dynkin diagrams. It turns out that all these quivers are non-chiral/symmetric although the affine Dynkin diagrams could be non-simply laced. Moreover, as we will see, there can be at most one pair of opposite arrows between two (distinct) nodes. In fact, the multiplicities and the information of the long and short roots in the affine Dynkin diagrams are encoded by the weights of the arrows in the quivers which are important in the definition of the quiver Yangians.

Recall that the affine Dynkin diagrams can have both bosonic and fermionic nodes. For quiver Yangians, this $\mathbb{Z}_2$-grading is reflected by (the parity of) the numbers of adjoint loops of the quiver nodes. Unlike the toric quivers where a node can have at most one adjoint loop\footnote{Of course, for $\mathbb{C}^3$, the quiver has one node with three loops, but this gives rise to a bosonic node as opposed to the non-isotropic fermionic nodes.}, we find that there could also be quiver nodes with two self-loops due to the existence of non-isotropic fermionic nodes in the underlying affine Dynkin diagrams. All the above information of the quivers is further supported by the superpotentials and dualities as we shall now briefly explain.

For non-toric quiver gauge theories, the superpotentials may not be uniquely determined. However, we will discuss how the superpotentials can be fixed if we would like to associate the quivers and their Yangians to the affine Dynkin diagrams. In fact, the superpotential terms are in correspondence with the Serre relations of the underlying Kac-Moody algebras.

For the super cases, there can be multiple affine Dynkin diagrams for a given Kac-Moody algebra. These affine Dynkin diagrams are related by odd reflections. We will show that they are actually Seiberg duality of the quivers and in most cases correspond to the isomorphisms of the quiver Yangians. However, as we will discuss, the algebras involve quiver nodes with two adjoint loops are slightly different, and they might give some interesting consequences on the possible BPS story. This extends the discussions for the generalized conifolds in the toric cases in \cite{Bao:2022jhy}. As we will see, the quivers, the edge weights and the superpotentials would be transformed consistently under Seiberg duality.

As the structures of these quiver Yangians (with some subtleties for the phases involving non-isotropic odd nodes) resemble the ones for generalized conifolds, it would be straightforward to obtain certain results in a similar manner, including the minimalistic presentation, the $J$ presentation and the coproduct. We will also mention how the algebras can be related by foldings. Moreover, it would be natural to expect that they can give rise to certain $\mathcal{W}$-algebras with different symmetries under truncations similar to the toric cases for generalized conifolds.

It is known that there exist twisted Yangians that yield $\mathcal{W}$-algebras in the finite cases \cite{ragoucy2001twisted,brown2009twisted}. In fact, an example of the twisted affine Yangian and the surjection to some rectangular $\mathcal{W}$-algebra was given in \cite{ueda2021twisted}. Therefore, we will also introduce the twisted quiver Yangians here analogous to the construction/theorem of the twisted Yangians in the finite cases. The twisted quiver Yangians are by construction associative subalgebras of the quiver Yangians. We will argue that they are actually coideals of the quiver Yangians. We conjecture that their truncations would give rise to certain $\mathcal{W}$-algebras as well, and it would be interesting to study their representations and possible connections to BPS states in future.

Although the precise maps between the (twisted) quiver Yangians and the $\mathcal{W}$-algebras are still not clear, we shall digress slightly and consider the free field realization of the $\mathcal{W}_{m|n\times\infty}$ algebras (whose connections to the quiver Yangians for generalized conifolds are known) as an example. In particular, we determine the screening currents such that the generators of the $\mathcal{W}$-algebras lie in the intersection of their kernels.

Given the quiver Yangians associated to the affine Dynkin diagrams, one can also consider the toroidal and elliptic versions similar to the toric cases. Again, it is straightforward to write down the algebras from the definitions. Here, we will not only fixate on the affine Dynkin cases but consider any general quivers. We will give a free field realization of the toroidal and elliptic algebras.

The paper is orgainzed as follows. In \S\ref{affine}, after reviewing the definition of the quiver Yangians, we will determine the quivers and their Yangians for those associated to the affine Dynkin diagrams in the non-super case. In \S\ref{twisted}, we will introduce the twisted quiver Yangians. Then in \S\ref{supercases}, we shall consider those for the super cases and discuss the Seiberg duality of them. In \S\ref{Walgebras}, we will mention some aspects of the $\mathcal{W}$-algebras. We will consider the toroidal and elliptic algebras for any quivers in \S\ref{toroidalelliptic}. In \S\ref{outlook}, we have some discussions on the outlook. A complete list of the phases of the quivers (that are not given in the main context) can be found in Appendix \ref{SDquivers}.

\section{Affine Dynkin Quiver Yangians}\label{affine}
We shall start with the quiver Yangians arised from the (non-super) affine Dynkin diagrams. Before we contemplate this particular family, let us first introduce the general definition of the quiver Yangians for any quivers \cite{Li:2020rij}.

\subsection{Quiver Yangians}\label{QY}
Given a quiver $Q$ and its superpotential $W$, the quiver Yangian $\mathcal{Y}$ is generated by three types of modes $\uppsi^{(a)}_i$, $\mathtt{e}^{(a)}_j$, $\mathtt{f}^{(a)}_j$, where $i\in\mathbb{Z}$, $j\in\mathbb{Z}_{\geq0}$ and $a$ labels the nodes in the quiver. They satisfy the following defining relations:
\begin{align}
	&\left[\uppsi^{(a)}_n,\uppsi^{(b)}_m\right]=0,\\
	&\left[\mathtt{e}^{(a)}_n,\mathtt{f}^{(b)}_m\right\}=\delta_{ab}\uppsi^{(a)}_{m+n},\\
	&\sum_{k=0}^{|b\rightarrow a|}(-1)^{|b\rightarrow a|-k}\sigma_{|b\rightarrow a|-k}^{b\rightarrow a}\left[\uppsi^{(a)}_n\mathtt{e}^{(b)}_m\right]_k=\sum_{k=0}^{|a\rightarrow b|}\sigma_{|a\rightarrow b|-k}^{a\rightarrow b}\left[\mathtt{e}^{(b)}_m\uppsi^{(a)}_n\right]^k,\\
	&\sum_{k=0}^{|b\rightarrow a|}(-1)^{|b\rightarrow a|-k}\sigma_{|b\rightarrow a|-k}^{b\rightarrow a}\left[\mathtt{e}^{(a)}_n\mathtt{e}^{(b)}_m\right]_k=(-1)^{|a||b|}\sum_{k=0}^{|a\rightarrow b|}\sigma_{|a\rightarrow b|-k}^{a\rightarrow b}\left[\mathtt{e}^{(b)}_m\mathtt{e}^{(a)}_n\right]^k,\\
	&\sum_{k=0}^{|b\rightarrow a|}(-1)^{|b\rightarrow a|-k}\sigma_{|b\rightarrow a|-k}^{b\rightarrow a}\left[\mathtt{f}^{(b)}_m\uppsi^{(a)}_n\right]^k=\sum_{k=0}^{|a\rightarrow b|}\sigma_{|a\rightarrow b|-k}^{a\rightarrow b}\left[\uppsi^{(a)}_n\mathtt{f}^{(b)}_m\right]_k,\\
	&\sum_{k=0}^{|b\rightarrow a|}(-1)^{|b\rightarrow a|-k}\sigma_{|b\rightarrow a|-k}^{b\rightarrow a}\left[\mathtt{f}^{(b)}_m\mathtt{f}^{(a)}_n\right]^k=(-1)^{|a||b|}\sum_{k=0}^{|a\rightarrow b|}\sigma_{|a\rightarrow b|-k}^{a\rightarrow b}\left[\mathtt{f}^{(a)}_n\mathtt{f}^{(b)}_m\right]_k.
\end{align}
Let us explain the notations used here. The bracket $[\text{-},\text{-}\}$ is the super bracket, that is, $[x,y\}=xy-(-1)^{|x||y|}yx$ where $|x|$ denotes the $\mathbb{Z}_2$-grading of the element. In a quiver, a node with an odd (resp.~even) number of adjoint loop(s) is bosonic (resp.~fermionic) such that $|a|=0$ (resp.~$|a|=1$). Then $\left|\mathtt{e}^{(a)}_j\right|=\left|\mathtt{f}^{(a)}_j\right|=|a|$  while $\uppsi^{(a)}_i$ are always bosonic. We use $a\rightarrow b$ to denote the set of arrows from $a$ to $b$, and the total number is $|a\rightarrow b|$. For each edge $X$ in the quiver, we assign a weight/charge $\widetilde{h}_X$ to it, and $\sigma^{a\rightarrow b}_k$ is the $k^\text{th}$ symmetric sum of $\widetilde{h}_X$ for all $X\in a\rightarrow b$. For brevity, when it would not cause any confusions, we shall omit the arrows in the labels such as $\sigma^{ab}_{|ab|-k}$. Moreover, we have
\begin{equation}
	[A_nB_m]_k:=\sum_{l=0}^k(-1)^l\binom{k}{l}A_{n+k-l}B_{m+l},\quad[B_mA_n]^k:=\sum_{l=0}^k(-1)^l\binom{k}{l}B_{m+l}A_{n+k-l}.
\end{equation}
To recover the BPS spectrum correctly, one also needs the Serre relations. We will discuss such relations in more detail later.

The parameters $\widetilde{h}_X$ are not completely independent. Every monomial term $\mathcal{M}$ in the superpotential gives a loop constraint while each node $a$ gives a vertex constraint:
\begin{equation}
	\sum_{X\in\mathcal{M}}\widetilde{h}_X=0,\quad\sum_{X\in a}\text{sgn}_a(X)\widetilde{h}_X=0.
\end{equation}
Here, $X\in a$ stands for arrows that are connected to $a$, and $\text{sgn}_a(X)=\pm1$ indicates whether $X$ starts from or ends at $a$. Recall that for a toric quiver gauge theory whose superpotential can be uniquely determined, the quiver Yangian realizes its BPS algebra. In such case, the above constraints yield two free parameters, say $h_1$ and $h_2$. They parametrize the periodic quiver lattice. Together with the R-symmetry coordinate, they can be identified with the U(1)$^3$ isometry of the toric CY threefold.

It would also be instructive to introduce the currents
\begin{equation}
	\uppsi^{(a)}(z)=\sum_{n=-\infty}^{\infty}\frac{\uppsi^{(a)}_n}{z^{n+1}},\quad\mathtt{e}^{(a)}(z)=\sum_{n=0}^{\infty}\frac{\mathtt{e}^{(a)}}{z^{n+1}},\quad\mathtt{f}^{(a)}(z)=\sum_{n=0}^{\infty}\frac{\mathtt{f}^{(a)}}{z^{n+1}}
\end{equation}
so that the defining relations can be written in terms of these currents:
\begin{align}
	&\uppsi^{(a)}(z)\uppsi^{(b)}(w)=\uppsi^{(b)}(w)\uppsi^{(a)}(z),\\
	&\left[\mathtt{e}^{(a)}(z),\mathtt{f}^{(b)}(w)\right\}\simeq-\delta_{ab}\frac{\uppsi^{(a)}(z)-\uppsi^{(a)}(w)}{z-w},\\
	&\uppsi^{(a)}(z)\mathtt{e}^{(b)}(w)\simeq\varphi^{a\Leftarrow b}(z-w)\mathtt{e}^{(b)}(w)\uppsi^{(a)}(z),\\
	&\mathtt{e}^{(a)}(z)\mathtt{e}^{(b)}(w)\simeq(-1)^{|a||b|}\varphi^{a\Leftarrow b}(z-w)\mathtt{e}^{(b)}(w)\mathtt{e}^{(a)}(z),\\
	&\uppsi^{(a)}(z)\mathtt{f}^{(b)}(w)\simeq\varphi^{a\Leftarrow b}(z-w)^{-1}\mathtt{f}^{(b)}(w)\uppsi^{(a)}(z),\\
	&\mathtt{f}^{(a)}(z)\mathtt{f}^{(b)}(w)\simeq(-1)^{|a||b|}\varphi^{a\Leftarrow b}(z-w)^{-1}\mathtt{f}^{(b)}(w)\mathtt{f}^{(a)}(z).\\
\end{align}
Here, ``$\simeq$'' indicates equality up to some $z^mw^n$ terms. The bond factor $\varphi^{a\Leftarrow b}$ is defined as\footnote{The use of $\zeta$ seems to be redundant here as the function is simply $z$. However, it would be more convenient when we generalize $\zeta$ to different functions leading to other types of quiver BPS algebras.}
\begin{equation}
	\varphi^{a\Leftarrow b}(z):=\frac{\prod\limits_{X\in a\rightarrow b}\zeta\left(\widetilde{h}_X+z\right)}{\prod\limits_{X\in b\rightarrow a}\zeta\left(\widetilde{h}_X-z\right)}\quad\text{with}\quad\zeta(z)=z.\label{bondfactor}
\end{equation}
Since $\zeta(z)=-\zeta(-z)$, we have
\begin{equation}
	\varphi^{a\Leftarrow b}(z)\varphi^{b\Leftarrow a}(-z)=1.
\end{equation}

The quiver Yangian has the crystal melting model as its representation. In short, a crystal model is a 3d uplift of the periodic quiver, where different ``atoms'' with different ``colours'' correspond to different gauge nodes in the periodic quiver with the ``chemical bonds'' specified by the arrows. To recover the BPS spectrum, the crystal ``melts'' in terms of the melting rule, which requires an atom $\mathfrak{a}$ to be in the molten crystal configuration $\mathfrak{C}$ whenever there exists an arrow $X$ such that $X\cdot\mathfrak{a}\in\mathfrak{C}$. In other words, the complements of the molten crystal are specific (but not all) ideals of the Jacobi algebra $\mathbb{C}Q/\langle\partial W\rangle$ \cite{Szendroi:2007nu}. The molten crystal configurations are in one-to-one correspondence with the BPS states as the counting problem of DT invariants and quiver representations can be recasted into enumerating ideals of the Jacobi algebra. The actions of the currents on the crystal modules can be found in \cite[(6.45)]{Li:2020rij}. For quivers with different framings, one can consider the shifted quiver Yangians and certain subcrystals as studied in \cite{Galakhov:2021xum}. This would also cause shifts in the expansion of the $\uppsi^{(a)}(z)$ currents below. However, we shall not further mention the shifts in this paper.

For non-chiral/symmetric quivers, that is, $|a\rightarrow b|=|b\rightarrow a|$ for any $a,b$, the analysis on the actions of $\uppsi$ shows that the negative modes are trivial due to the homogeneity of the bond factor. More specifically, we have
\begin{equation}
	\uppsi^{(a)}(z)=1+\sum_{n=0}^{\infty}\frac{\uppsi^{(a)}_n}{z^{n+1}},
\end{equation}
where $\uppsi^{(a)}_{-1}=1$ and $\uppsi^{(a)}_{n<-1}=0$. In this paper, we will mainly focus on quivers that are non-chiral.

\subsection{Affine Dynkin Cases}\label{affinecases}
It would be convenient to introduce a few notions for our discussions. For any quiver $\widehat{Q}$, which will mainly be some (affine) Dynkin diagram with each edge $X$ given any orientation here, we may consider its doubled quiver $\overline{Q}$. By a doubled quiver, we mean that an edge $X^*$ in the opposite direction is added to the quiver for each $X$. This gives rise to the preprojective algebra $\Pi_{\widehat{Q}}:=\mathbb{C}\overline{Q}/\sum[X,X^*]$. We can then construct the tripled quiver $Q$ where a self-loop $\omega$ is further added to each node\footnote{Later, we will slightly generalize/modify the concept of the ``tripled'' quivers when we have odd nodes in the super cases.}. The (super)potential\footnote{Strictly speaking, the physical superpotential should be $\text{tr}W$. However, we shall not make this difference here for simplicity.} is given by $W=\sum\left(X_{ab}X_{ba}\omega_a-X_{ba}X_{ab}\omega_b\right)$, and we have the Jacobi algebra  $\mathbb{C}Q/\langle\partial W\rangle$, that is, the path algebra $\mathbb{C}Q$ modulo the F-term relations.

It is well-known that for affine ADE (tripled) quivers, the 4d gauge theories that preserve $\mathcal{N}=2$ supersymmetry can be obtained from D-branes probing the singularities $\mathbb{C}\times\mathbb{C}^2/\Gamma$ with $\Gamma$ being finite subgroups of SU(2). In \cite{Bershadsky:1996nh,Cecotti:2012gh}, such geometric construction was extended to the non-simply laced cases. In particular, the BCFG cases can arise from non-split singular elliptic fibrations. This should correspond to the Slodowy correspondence \cite{slodowy2006simple,stekolshchik2005notes} generalizing the McKay correspondence \cite{McKay1980}.

The quiver is again the tripled quiver of the corresponding affine Dynkin diagram/quiver, where any non-simply laced edge indicating long and short roots is still treated as a single edge/arrow. However, the non-simply lacedness, based on the Cartan matrix $A$, changes the above superpotential to
\begin{equation}
	W=\sum\left(X_{ab}X_{ba}\omega_a^{|A_{ab}|}-X_{ba}X_{ab}\omega_b^{|A_{ba}|}\right),\label{superpotential}
\end{equation}
where we will always omit the trace for brevity. The affine Dynkin diagrams, the Cartan matrices and the quivers (with weights associated to the arrows) are then given by\footnote{For simplicity, we will not consider the toric quivers associated to $\mathbb{C}^3$ ($\widehat{\mathfrak{gl}}(1)$), $\mathbb{C}\times\mathbb{C}^2/\mathbb{Z}_2$ ($\widehat{\mathfrak{gl}}(2)$) and the conifold ($\widehat{\mathfrak{gl}}(1|1)$) in this paper.}
\begin{equation}
	\includegraphics[width=15cm]{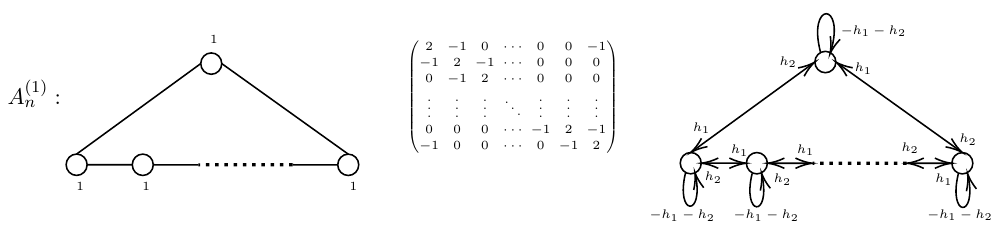},
\end{equation}
\begin{equation}
	\includegraphics[width=15cm]{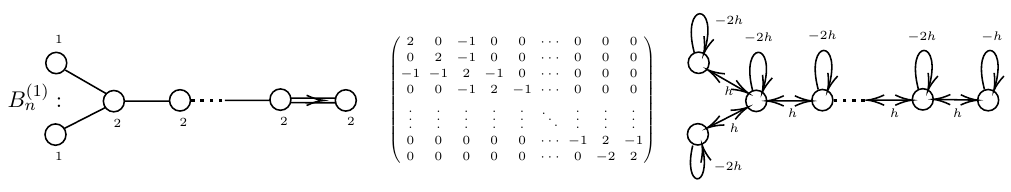},
\end{equation}
\begin{equation}
	\includegraphics[width=15cm]{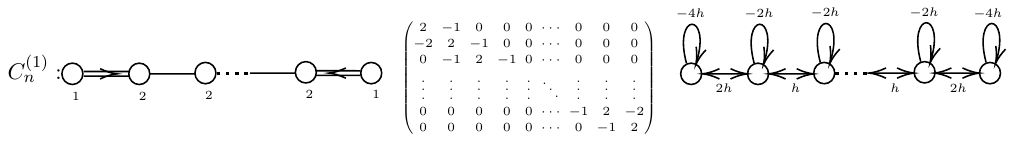},
\end{equation}
\begin{equation}
	\includegraphics[width=15cm]{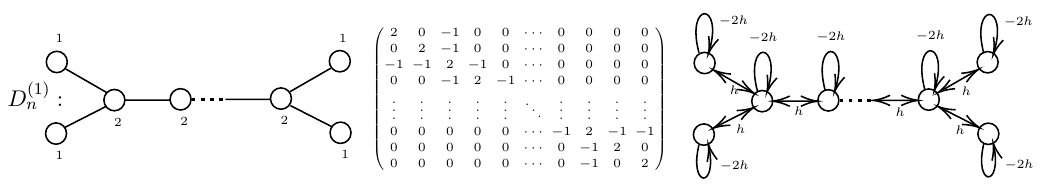},
\end{equation}
\begin{equation}
	\includegraphics[width=15cm]{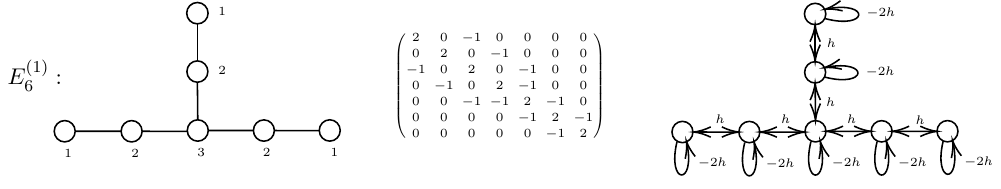},
\end{equation}
\begin{equation}
	\includegraphics[width=15cm]{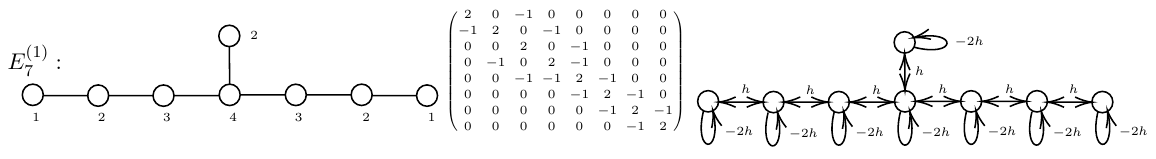},
\end{equation}
\begin{equation}
	\includegraphics[width=15cm]{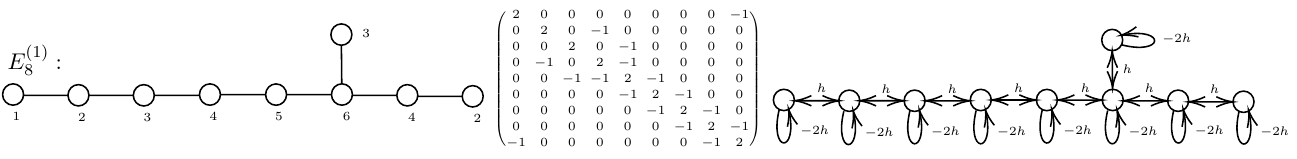},
\end{equation}
\begin{equation}
	\includegraphics[width=15cm]{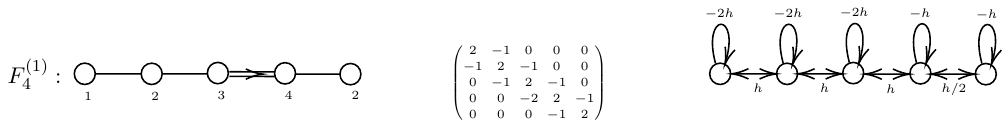},
\end{equation}
\begin{equation}
	\includegraphics[width=15cm]{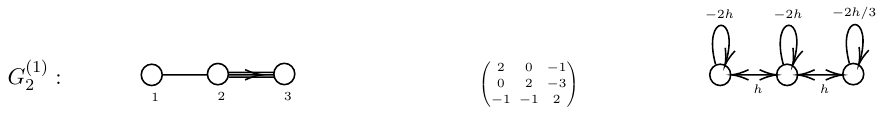}.
\end{equation}
Each quiver has $(n+1)$ nodes when the subscript is $n$, and the numbering of the nodes and the Cartan matrices follow the convention in \texttt{SageMath} \cite{sagemath}. On the other hand, the labels in the affine Dynkin diagrams are the Dynkin-Kac/dual Coxeter numbers.
As we can see, except the A-type cases which are toric, all the non-toric cases here have only one free parameter $h$.

\begin{remark}
	As a digression, it is worth noting that the above affine Dynkin diagrams also appear as magnetic quivers \cite{Cabrera:2019izd} which give the elementary transverse slices in the study of geometric structures of the Higgs branches for certain theories with eight supercharges in various dimensions \cite{Bourget:2019aer}. There are also various brane realizations for (most of) them. However, we should emphasize that the (untwisted) affine Dynkin diagrams/magnetic quivers in this context describe the Coulomb branches/spaces of dressed monopole operators. These symplectic singularities are the closures of the minimal nilpotent orbits of the corresponding simple Lie algebras. A most up-to-date list of known elementary transverse slices can be found in \cite{Bourget:2021siw} (see also \cite{Bourget:2022ehw}).
\end{remark}

Although the geometric constructions may not be clear, from the perspective of defining a Yangian type algebra, nothing prevents us to consider more general Dynkin diagrams with the superpotential chosen following the rule of \eqref{superpotential}. For instance, the twisted affine cases are
\begin{equation}
	\includegraphics[width=10cm]{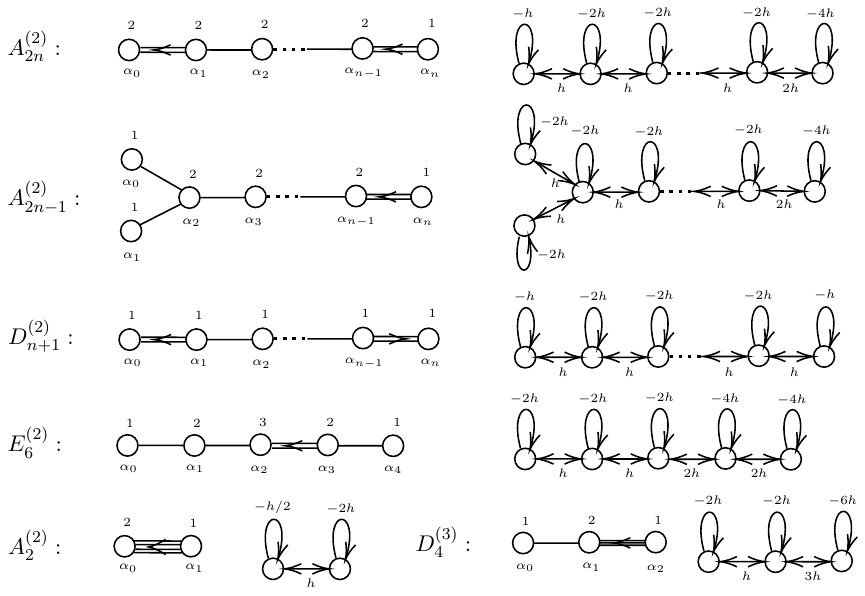}.
\end{equation}
Again, all these quiver Yangians would depend only on one free parameter. Of course, this is not always the case. For example, the compact hyperbolic Dynkin diagram $H_{23}^{(3)}$ (which corresponds to the elementary slice $\mathfrak{ag}_2$ in the above remark) gives a two-parameter algebra:
\begin{equation}
	\includegraphics[width=5cm]{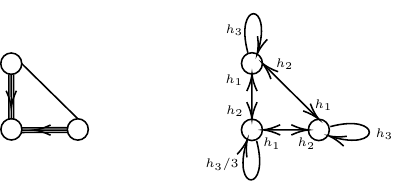},
\end{equation}
where $h_3=-h_1-h_2$. Nevertheless, in this paper, we shall mainly focus on those from the affine Dynkin diagrams. It would also be natural to conjecture that there could be similar geometric constructions for these quiver gauge theories with the chosen superpotentials associated to the twisted affine Dynkin diagrams.

Since there is at most one pair of opposite arrows between any two nodes in the quiver, only $\sigma^{ab}_1$ can be non-trivial. The defining relations of the quiver Yangian can then be written as
\begin{align}
	&\left[\psi^{(a)}_n,\psi^{(b)}_m\right]=0,\label{psipsi}\\
	&\left[e^{(a)}_n,f^{(b)}_m\right\}=\delta_{ab}\psi^{(a)}_{m+n},\\
	&\left[\psi^{(a)}_0,e^{(b)}_m\right]=\left(\alpha^{(a)},\alpha^{(b)}\right)e^{(b)}_m,\\
	&\left[\psi^{(a)}_0,f^{(b)}_m\right]=-\left(\alpha^{(a)},\alpha^{(b)}\right)f^{(b)}_m,\\
	&\left[\psi^{(a)}_{n+1},e^{(b)}_m\right]-\left[\psi^{(a)}_n,e^{(b)}_{m+1}\right]=\sigma^{ba}_1\psi^{(a)}_ne^{(b)}_m+\sigma^{ab}_1e^{(b)}_m\psi^{(a)}_n,\\
	&\left[\psi^{(a)}_{n+1},f^{(b)}_m\right]-\left[\psi^{(a)}_n,f^{(b)}_{m+1}\right]=-\sigma^{ab}_1\psi^{(a)}_nf^{(b)}_m-\sigma^{ba}_1f^{(b)}_m\psi^{(a)}_n,\\
	&\left[e^{(a)}_{n},e^{(b)}_m\right\}=\left[f^{(a)}_{n},f^{(b)}_m\right\}=0\qquad(\sigma_1^{ab}=0),\\
	&\left[e^{(a)}_{n+1},e^{(b)}_m\right\}-\left[e^{(a)}_n,e^{(b)}_{m+1}\right\}=\sigma^{ba}_1e^{(a)}_ne^{(b)}_m+(-1)^{|a||b|}\sigma^{ab}_1e^{(b)}_me^{(a)}_n\qquad(\sigma_1^{ab}\neq0),\\
	&\left[f^{(a)}_{n+1},f^{(b)}_m\right\}-\left[f^{(a)}_n,f^{(b)}_{m+1}\right\}=-\sigma^{ab}_1f^{(a)}_nf^{(b)}_m-(-1)^{|a||b|}\sigma^{ba}_1f^{(b)}_mf^{(a)}_n\qquad(\sigma_1^{ab}\neq0)\label{ff},
\end{align}
where we have rescaled the generators as
\begin{equation}
	\psi^{(a)}_n=-\frac{1}{h_1+h_2}\uppsi^{(a)}_n,\quad e^{(a)}_n=-\frac{1}{\left(h_1+h_2\right)^{1/2}}\mathtt{e}^{(a)}_n\quad f^{(a)}_n=-\frac{1}{\left(h_1+h_2\right)^{1/2}}\mathtt{f}^{(a)}_n
\end{equation}
for convenience (assuming that $h_1+h_2\neq0$)\footnote{In particular, this means that $\psi^{(a)}_{-1}=-1/(h_1+h_2)$. Notice that there can be different conventions on the signs, coming from the conventions in the Cartan matrices and in the edge weights. Nevertheless, the algebras are always isomorphic.}. For non-toric cases which depend on one single parameter, we take $h_1=h_2=h$. Notice that here all the nodes are bosonic. Hence, $|a|=0$ for any $a$ and all superbrackets are just commutators.

The above (symmetric) bracket $(\text{-},\text{-})$ is the invariant inner product on the Kac-Moody algebra $\mathfrak{g}$ associated to the underlying affine Dynkin diagram. By looking at the zero modes, we find that they satisfy the same relations of the Chevalley generators\footnote{We shall comment on the Serre relations shortly.} $\psi^{(a)},e^{(a)},f^{(a)}$ with $\left(e^{(a)},f^{(a)}\right)=1$ and $\left[e^{(a)},f^{(a)}\right]=\psi^{(a)}$. In other words, there is a natural embedding of the universal enveloping algebra of (the derived algebra of) $\mathfrak{g}$ to the quiver Yangian. Therefore, we shall use the zero modes $x^{(a)}_0$ and the chevalley generators $x^{(a)}$ ($x=\psi,e,f$) interchangeably. Write the roots as $\alpha$ with root spaces $\mathfrak{g}_{\alpha}$, and the simple roots are $\alpha^{(a)}$. Recall that the set of roots $\varDelta$ has the decomposition $\varDelta=\varDelta_+\cup\varDelta_-$. Then the sets of positive real and imaginary roots are given by $\varDelta^\text{re}_+=\mathring{\varDelta}_+\cup\left\{n\delta+\alpha|n\in\mathbb{Z}_+,\alpha\in\mathring{\varDelta}\right\}$ and $\varDelta^\text{im}_+=\{n\delta|n\in\mathbb{Z}_+\}$, where $\mathring{\varDelta}$ is the set of roots of the underlying Lie algebra with the zeroth vertex removed in the affine Dynkin diagram and $\delta$ is the minimal positive imaginary root of $\mathfrak{g}$. For each positive root $\alpha$, we can choose a basis $\left\{e^{(\alpha,k)}\right\}$ of $\mathfrak{g}_\alpha$ with a dual basis $\left\{f^{(\alpha,k)}\right\}$ of $\mathfrak{g}_{-\alpha}$, where $k=1,\dots,\dim\mathfrak{g}_\alpha$, such that $\left(e^{(\alpha,k)},f^{(\alpha,l)}\right)=\delta_{kl}$. When $\alpha$ is a real root, $\dim\mathfrak{g}_\alpha=1$ and we shall simply write $e^{(\alpha)}=e^{(\alpha,1)}$, $f^{(\alpha)}=f^{(\alpha,1)}$. In particular, given a simple root $\alpha^{(a)}$, we have $e^{(a)}_0=e^{(a)}=e^{\left(\alpha^{(a)}\right)}$ and $f^{(a)}_0=f^{(a)}=f^{\left(\alpha^{(a)}\right)}$.

With the above notations prepared, it is not hard to see that the affine Dynkin quiver Yangians are basically the affine Yangians asscociated to symmetrizable Kac-Moody algebras studied in \cite{guay2018coproduct}. Therefore, the methods/results therein can be directly applied to the quiver Yangians here (with some slight changes in the coefficients/parameters for certain expressions)\footnote{More precisely, this is the Yangian associated to the derived algebra $\mathfrak{g}'=[\mathfrak{g},\mathfrak{g}]$ instead of the Kac-Moody algebra $\mathfrak{g}$ itself as can be seen from the relations of the zero modes. The Yangian $\mathcal{Y}(\mathfrak{g})$ is given by $\mathcal{Y}(\mathfrak{g}')\cup\mathfrak{h}$ where $\mathfrak{h}$ is the Cartan subalgebra of $\mathfrak{g}$. However, we shall not make this difference here as the following discussions will not be affected by these extra Cartan elements. In particular, they would not change the representations in (possible) connection with the BPS states.}.

\paragraph{Serre relations} Let us now comment on the Serre relations of the quiver Yangians. We may directly take the corresponding relations of the affine Yangians in \cite{guay2018coproduct}, that is,
\begin{equation}
	\sum_{\sigma\in\mathfrak{S}_{1+|A_{ab}|}}\left[e^{(a)}_{n_{\sigma(1)}},\left[e^{(a)}_{n_{\sigma(2)}},\dots,\left[e^{(a)}_{n_{\sigma(1+|A_{ab}|)}},e^{(b)}_m\right]\dots\right]\right]=0\quad(a\neq b),\label{Serre1}
\end{equation}
and likewise for $f$, where $\mathfrak{S}_k$ is the symmetric group. On the other hand, similar relations were given in \cite{Negut:2023iia} for certain toroidal algebras associated to any quivers. Such relations were obtained by considering the action of the (non-reduced) toroidal algebra without Serre relations on the highest weight/crystal representations and finding the kernel of the action. In other words, this gives the smallest possible quotient of the non-reduced toroidal algebra such that the action factors through an action of a reduced toroidal algebra. Taking the rational limit, we can get another proposal of the Serre relations for the (rational) quiver Yangians here. For each monomial term in the superpotential, the involved nodes form an ordered list $\{a_0,a_1,\dots,a_{l-1},a_l=a_0\}$, where the same nodes can appear multiple times in the list. For each ordered list, we also consider the (revolving) sequence $(i,i-1,\dots,1,l,\dots,i+1)$ of the subscripts. The relations are then
\begin{equation}
	\begin{split}
		\sum_{i=1}^l&(-1)^{\frac{1}{2}(i(i-1)-\mathfrak{f}(\mathfrak{f}-1))}\frac{\prod\limits_{\text{pos}(j)>\text{pos}(k)}\upzeta_{a_k,a_j}(z_k-z_j)}{\prod\limits_{\text{pos}(j)=\text{pos}(k)+1}\left(z_k-z_j+\widetilde{h}_{a_ka_j}\right)}\\
		&e^{(a_i)}(z_i)e^{(a_{i-1})}(z_{i-1})\dots e^{(a_1)}(z_1)e^{(l)}(z_l)\dots e^{(a_{i+1})}(z_{i+1})\simeq0,\label{Serre2}
	\end{split}
\end{equation}
and likewise for $f$, where the sum is over all the revolving sequences such that $\text{pos}(j)$ is the position of $j$ in each different sequence $(i,i-1,\dots,1,l,\dots,i+1)$, and $\widetilde{h}_{a_k,a_{k+1}}$ denotes the weight of the corresponding arrow appeared in the monomial term. The function $\upzeta_{ab}(z)$ denotes the numerator of $\varphi^{a\Leftarrow b}(z)$. Here, we have used $\mathfrak{f}$ to denote the number of fermionic nodes in the monomial term/ordered listed so as to incorporate the super cases that will be discussed below. Notice that the signs are slightly different from the ones in \cite{Negut:2023iia}. See the comments around \eqref{Serre3} for more detail.

The above relations are either the most natural expressions in terms of a Yangian type algebra or the ``smallest'' ones regarding the action on the highest weight (crystal) modules. In terms of the BPS invariants, the Serre relations should give the constraints in addition to \eqref{psipsi}$\sim$\eqref{ff} so that the correct counting can be recovered. We shall not further explore this point here. For simplicity, we would only assume that the full Serre relations can be inductively obtained (see the following discussions on the minimalistic presentation) by
\begin{equation}
	\text{ad}\left(e^{(a)}_0\right)^{1-A_{ab}}e^{(b)}_0=0,\quad\text{ad}\left(f^{(a)}_0\right)^{1-A_{ab}}f^{(b)}_0=0\quad(a\neq b),
\end{equation}
where $\text{ad}(x)^ky=[x,\dots,[x,y\}\dots\}$ with the (super)commutator appearing $k$ times.

\paragraph{Minimalistic presentation} From the above relations \eqref{psipsi}$\sim$\eqref{ff}, we get
\begin{align}
	&e^{(a)}_{m+1}=\frac{1}{\left(\alpha^{(a)},\alpha^{(b)}\right)}\left[\widetilde{\psi}^{(b)}_1,e^{(a)}_m\right]-\frac{\sigma^{ab}_1-\sigma^{ba}_1}{2\left(\alpha^{(a)},\alpha^{(b)}\right)}\left[\psi^{(b)}_0,e^{(a)}_m\right]=\frac{1}{\left(\alpha^{(a)},\alpha^{(b)}\right)}\left[\widetilde{\psi}^{(b)}_1,e^{(a)}_m\right]-\frac{\sigma^{ab}_1-\sigma^{ba}_1}{2}e^{(a)}_m,\nonumber\\
	&f^{(a)}_{m+1}=-\frac{1}{\left(\alpha^{(a)},\alpha^{(b)}\right)}\left[\widetilde{\psi}^{(b)}_1,f^{(a)}_m\right]+\frac{\sigma^{ab}_1-\sigma^{ba}_1}{2\left(\alpha^{(a)},\alpha^{(b)}\right)}\left[\psi^{(b)}_0,f^{(a)}_m\right]=-\frac{1}{\left(\alpha^{(a)},\alpha^{(b)}\right)}\left[\widetilde{\psi}^{(b)}_1,f^{(a)}_m\right]-\frac{\sigma^{ab}_1-\sigma^{ba}_1}{2}f^{(a)}_m,\nonumber\\
	&\psi^{(a)}_{m+1}=\left[e^{(a)}_{m+1},f^{(a)}_0\right],\nonumber\\
	\label{finitegenerators}
\end{align}
where $\widetilde{\psi}^{(b)}_1:=\psi^{(b)}_1-\frac{h_1+h_2}{2}\left(\psi^{(b)}_0\right)^2$, and the node $b$ can be taken as $a$ or $a\pm1$. In particular, for the one-parameter non-toric cases, the second parts in $e^{(a)}_{m+1}$ and $f^{(a)}_{m+1}$ with $\left(\sigma^{ab}_1-\sigma^{ba}_1\right)$ vanish. This shows that the higher modes can be inductively obtained by $\psi^{(a)}_{0,1}$, $e^{(a)}_0$, $f^{(a)}_0$. In fact, the algebra is not only finitely generated, but also finitely presented. In other words, it suffices to consider the following minimalistic presentation with finitely many relations \cite{guay2018coproduct}:
\begin{align}
	&\left[\psi^{(a)}_r,\psi^{(b)}_s\right]=0,\label{psipsimin}\\
	&\left[e^{(a)}_0,f^{(b)}_0\right\}=\delta_{ab}\psi^{(a)}_0,\quad\left[e^{(a)}_1,f^{(b)}_0\right\}=\left[e^{(a)}_0,f^{(b)}_1\right\}=\delta_{ab}\psi^{(a)}_1,\\
	&\left[\psi^{(a)}_0,e^{(b)}_r\right]=\left(\alpha^{(a)},\alpha^{(b)}\right)e^{(b)}_r,\\
	&\left[\psi^{(a)}_1,e^{(b)}_0\right]=\left(\alpha^{(a)},\alpha^{(b)}\right)e^{(b)}_1+\sigma^{ba}_1\psi^{(a)}_0e^{(b)}_0+\sigma^{ab}_1e^{(b)}_0\psi^{(a)}_0,\\
	&\left[\psi^{(a)}_0,f^{(b)}_r\right]=-\left(\alpha^{(a)},\alpha^{(b)}\right)f^{(b)}_r,\\
	&\left[\psi^{(a)}_1,f^{(b)}_0\right]=-\left(\alpha^{(a)},\alpha^{(b)}\right)f^{(b)}_1+\sigma^{ab}_1\psi^{(a)}_0f^{(b)}_0+\sigma^{ba}_1f^{(b)}_0\psi^{(a)}_0,\label{psi1fmin}\\
	&\left[e^{(a)}_0,e^{(b)}_0\right\}=\left[f^{(a)}_0,f^{(b)}_0\right\}=0\qquad(\sigma_1^{ab}=0),\\
	&\left[e^{(a)}_1,e^{(b)}_0\right\}-\left[e^{(a)}_0,e^{(b)}_1\right\}=\sigma^{ba}_1e^{(a)}_0e^{(b)}_0+(-1)^{|a||b|}\sigma^{ab}_1e^{(b)}_0e^{(a)}_0,\\
	&\left[f^{(a)}_1,f^{(b)}_0\right\}-\left[f^{(a)}_0,f^{(b)}_1\right\}=-\sigma^{ab}_1f^{(a)}_0f^{(b)}_0-(-1)^{|a||b|}\sigma^{ba}_1f^{(b)}_0f^{(a)}_0,\label{ffmin}
\end{align}
where $r,s\in\{0,1\}$. All the relations involving higher modes can be derived from these relations using the expressions for $x^{(a)}_{m+1}$ ($x=\psi,e,f$). The above discussions on the Serre relations are also expected to be in line with this minimalistic presentation.

\paragraph{Coproduct} Like many Yangian algebras, the quiver Yangians also have a coproduct structure. Using the $J$ presentation discussed below, one can find such coassociative homomorphism $\Delta:\mathcal{Y}\rightarrow\mathcal{Y}\widehat{\otimes}\mathcal{Y}$, where $\mathcal{Y}\widehat{\otimes}\mathcal{Y}$ denotes the completion of $\mathcal{Y}\otimes\mathcal{Y}$. Due to the minimalistic presentation, this coproduct is uniquely determined by
\begin{equation}
	\begin{split}
		&\Delta\left(x^{(a)}_0\right)=\square\left(x^{(a)}_0\right)\quad(x=\psi,e,f),\\
		&\Delta\left(\psi^{(a)}_1\right)=\square\left(\psi^{(a)}_1\right)+(h_1+h_2)\psi^{(a)}_0\otimes\psi^{(a)}_0-(h_1+h_2)\sum_{\alpha\in\varDelta_+^\text{re}}\left(\alpha^{(a)},\alpha\right)f^{(\alpha)}\otimes e^{(\alpha)},
	\end{split}\label{coprod}
\end{equation}
where $\square(x)=x\otimes1+1\otimes x$. Again, we have $h_1=h_2=h$ for the one-parameter algebras.

\paragraph{Representations and counting} Since the quivers are not toric except for the A-type cases, it is still not clear whether the quiver Yangians are precisely the BPS algebras of the gauge theories. As mentioned in \S\ref{intro}, it was proposed in \cite{Li:2023zub} that the quiver Yangians still play some role in the BPS algebras for any quivers, especially for the 4d $\mathcal{N}=2$ theories, where the ADE cases were studied explicitly therein. Indeed, at least for the (A)DE cases, the generalized DT invariants were checked to satisfy certain box counting in \cite{gholampour2009counting}\footnote{In fact, it was conjectured that the coloured box counting can give the generalized DT invariants for the orbifolds $\mathbb{C}^3/G$, where $G$ is a finite subgroup with $G<\text{SU}(2)<\text{SU}(3)$ or $G<\text{SO}(3)<\text{SU}(3)$ \cite{Joyce:2008pc}. When $G$ is an abelian subgroup of $\text{SO}(3)$, this was proven in \cite{Young:2008hn}.}. Regardless of whether the quiver Yangian is the desired BPS algebra or certain modifications/generalizations are required, it is always important to study its representations. Similar to the crystal melting model, the actions of the currents on the states in the highest weight representations can be found in \cite[(3.74)]{Li:2023zub}, where the modules are called poset representations due to the poset structure of the states. In short, the vectors in the modules are eigenstates of $\psi$, and $e$ (resp.~$f$) can be viewed as raising (resp.~lowering) operators.

\subsection{The $J$ Presentation and Foldings}\label{Jpresentation}
It would be useful to introduce another presentation called the $J $ presentation for the affine Dynkin quiver Yangians. It can be thought of the affine extension of Drinfeld's $J$ presentation for the Yangians associated to finite Lie algebras \cite{drinfeld1985hopf}.

The $J$ presentation of the quiver Yangian here is given by the zero modes $x^{(a)}_0$ ($x=\psi,e,f$) and $J\left(x^{(a)}_0\right)$ defined as\footnote{Again, the infinite sums are well-defined as $e^{(\alpha,k)}$ annihilates a state of a module in the category $\mathcal{O}$ for $\alpha$ with sufficiently large height. Alternatively, we can consider the completion of the algebra as in the above discussion of the coproduct.}
\begin{align}
	&J\left(\psi^{(a)}_0\right)=\psi^{(a)}_1+\frac{h_1+h_2}{2}\sum_{\alpha\in\varDelta_+}\left(\alpha,\alpha^{(a)}\right)\sum_{k=1}^{\dim\mathfrak{g}_\alpha}f^{(\alpha,k)}e^{(\alpha,k)}-\frac{h_1+h_2}{2}\left(\psi^{(a)}_0\right)^2,\\
	&J\left(e^{(a)}_0\right)=e^{(a)}_1+\frac{h_1+h_2}{2}\sum_{\alpha\in\varDelta_+}\sum_{k=1}^{\dim\mathfrak{g}_\alpha}f^{(\alpha,k)}\left[e^{(\alpha,k)},e^{(a)}_0\right]-\frac{h_1+h_2}{2}\psi^{(a)}_0e^{(a)}_0,\\
	&J\left(f^{(a)}_0\right)=f^{(a)}_1-\frac{h_1+h_2}{2}\sum_{\alpha\in\varDelta_+}\sum_{k=1}^{\dim\mathfrak{g}_\alpha}\left[f^{(a)}_0,f^{(\alpha,k)}\right]e^{(\alpha,k)}-\frac{h_1+h_2}{2}f^{(a)}_0\psi^{(a)}_0,
\end{align}
where we recall that the generators $e^{(\alpha,k)}$ and $f^{(\alpha,k)}$ were introduced on Page \hyperlink{page.10}{10}. Again, we have $h_1=h_2=h$ for the one-parameter algebras.

Then the quiver Yangian is defined by the following finitely many relations:
\begin{align}
	&\left[\psi^{(a)}_0,J\left(x^{(b)}_0\right)\right]=J\left(\left[\psi^{(a)}_0,x^{(b)}_0\right]\right)\qquad(x=\psi,e,f),\label{JxJx}\\
	&\left[J\left(\psi^{(a)}_0\right),e^{(b)}_0\right]=\left(\alpha^{(a)},\alpha^{(b)}\right)J\left(e^{(b)}_0\right)+\frac{\sigma^{ba}_1-\sigma^{ab}_1}{2}\left(\alpha^{(a)},\alpha^{(b)}\right)e^{(b)}_0,\\
	&\left[J\left(\psi^{(a)}_0\right),f^{(b)}_0\right]=-\left(\alpha^{(a)},\alpha^{(b)}\right)J\left(f^{(b)}_0\right)-\frac{\sigma^{ba}_1-\sigma^{ab}_1}{2}\left(\alpha^{(a)},\alpha^{(b)}\right)f^{(b)}_0,\\
	&\left[J\left(e^{(a)}_0\right),f^{(b)}_0\right]=\left[e^{(a)}_0,J\left(f^{(b)}_0\right)\right]=\delta_{ab}J\left(\psi^{(a)}_0\right),\\
	&\left[J\left(e^{(a)}_0\right),e^{(b)}_0\right]-\left[e^{(a)}_0,J\left(e^{(b)}_0\right)\right]=\frac{1}{2}\left(\sigma^{ba}_1-\sigma^{ab}_1\right)\left[e^{(a)}_0,e^{(b)}_0\right],\\
	&\left[J\left(f^{(a)}_0\right),f^{(b)}_0\right]-\left[f^{(a)}_0,J\left(f^{(b)}_0\right)\right]=-\frac{1}{2}\left(\sigma^{ba}_1-\sigma^{ab}_1\right)\left[f^{(a)}_0,f^{(b)}_0\right],\\
	&\left[J\left(e^{(a)}_0\right),e^{(b)}_0\right]=\left[J\left(f^{(a)}_0\right),f^{(b)}_0\right]=0\qquad(\sigma_1^{ab}=0)\label{JeeJff}.
\end{align}
In particular, some of the relations would get simplified when $\sigma^{ab}_1=\sigma^{ba}_1$ for the one-parameter algebras.

\begin{figure}[h]
	\centering
	\includegraphics[width=10cm]{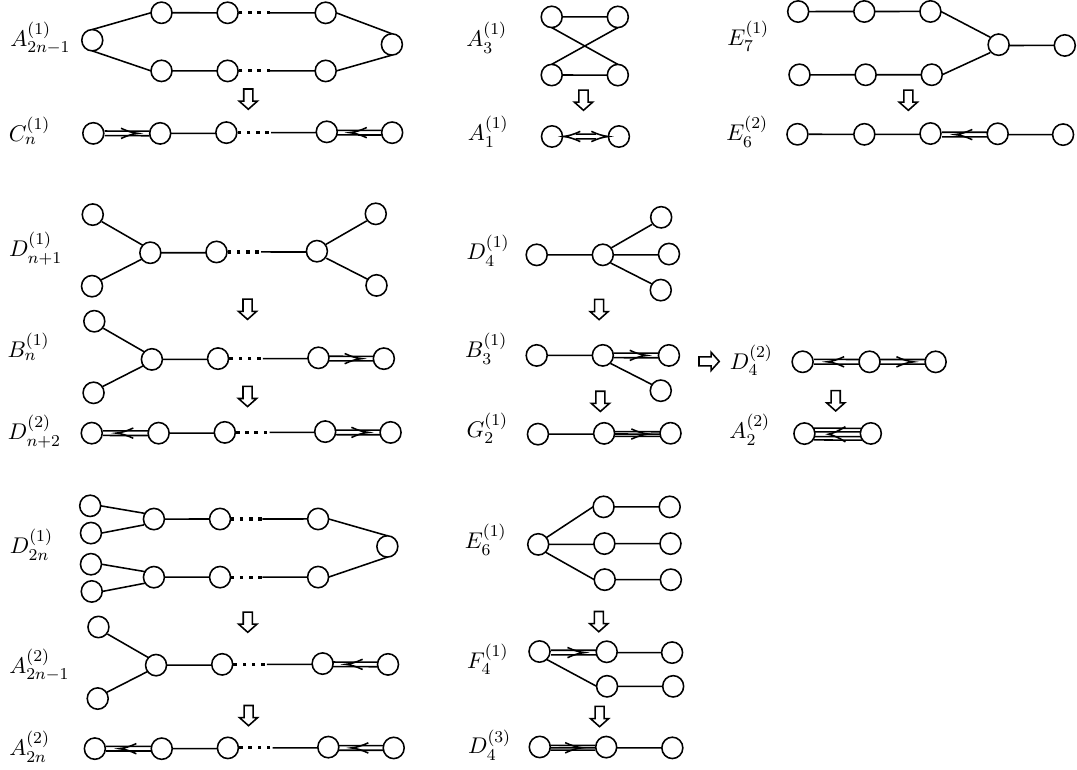}
	\caption{Folding the affine Dynkin diagrams. The quiver Yangians also follow this manipulation.}\label{folding}
\end{figure}

For Kac-Moody algebras, some of the subalgebra structures can be conveniently encoded by folding the (affine) Dynkin diagrams as shown in Figure \ref{folding}.
Suppose the Kac-Moody algebra $\mathfrak{g}_0$ can be folded to $\mathfrak{g}_1$. Then the Chevalley generators simply follow
\begin{equation}
	x'^{(a)}_0=x^{(a_1)}_0+x^{(a_2)}_0\quad(x=\psi,e,f),
\end{equation}
where $a_1$, $a_2$ are the corresponding nodes that are merged in the affine Dynkin diagram for $\mathfrak{g}_0$ and $x'$ denote the generators of $\mathfrak{g}_1$.

In terms of the minimalistic presentation with the extra $\psi^{(a)}_1$, it is not easy to see how the quiver Yangians are related by folding. However, by virtue of the $J$ presentation, the foldings of the quiver Yangians are simply
\begin{equation}
	J\left(x'^{(a)}_0\right)=J\left(x^{(a_1)}_0+x^{(a_2)}_0\right)\quad(x=\psi,e,f),
\end{equation}
revealing their subalgebra structures. Of course, since the A-type cases have two free parameters, when folding to other one-parameter cases, we first need to take $h_1=h_2=h$.

In Figure \ref{folding}, we have also included $A_3^{(1)}\rightarrow A_1^{(1)}$ although the quiver Yangian associated to the latter is not our focus in this paper. Nevertheless, let us still make a comment on this folding. As can be seen from the affine Dynkin diagram, the quiver has two pairs of opposite arrows connecting the two nodes (say 0 and 1), with an adjoint loop for each node. The weights associated to $X_{01,(1)}$ and $X_{10,(1)}$ (resp.~$X_{01,(2)}$ and $X_{10,(2)}$) are $h_1$ (resp.~$h_2$), and both of the adjoint loops have weights $-h_1-h_2$. Therefore, $\sigma^{ab}_2$ is also non-trivial, and the minimalistic presentation requires further analysis. In particular, more relations involving modes at level 2 would be required. Similarly, the above $J$ presentation needs to be expanded as well. However, as shown in \cite[Appendix A]{Bao:2022jhy}, the modes $\psi^{(a)}_{0,1}$, $e^{(a)}_0$, $f^{(a)}_0$ are still sufficient to generate all the higher modes.

Therefore, let us mention how these modes following the folding procedure. Again, the zero modes satisfy $x'^{(a)}_0=x^{(a_1)}_0+x^{(a_2)}_0$. For $\psi'^{(a)}_1$, we find that
\begin{equation}
	\psi'^{(a)}_1=\psi^{(a_1)}_1+\psi^{(a_2)}_1.
\end{equation}
However, this requires $h_3'=-h_1=-h_2$ where $h_3'=-h_1'-h_2'$. Here, the primed symbols still indicate the ones after folding. In other words, only the sum of the two parameters can be determined for $\mathcal{Y}\left(A_1^{(1)}\right)$ while the information of the individual $h_1'$ and $h_2'$ is ``lost''. This is because for the relations at level 1 ($\psi'_1e'_0$ and $\psi'_1f'_0$), we only have $\sigma'^{ab}_1=\sigma'^{ba}_1=-h_3'$.

\section{Twisted Quiver Yangians}\label{twisted}
Whether the quiver Yangians for non-toric cases fully describe the exact BPS algebras would still require further study. It could also be possible that we need some slight modifications or generalizations of them. Here, we shall introduce a twisted version giving subalgebras of the quiver Yangians analogous to the twisted Yangians associated with finite Lie algebras \cite{olshanskii2006twisted,Molev:1994rs,Molev:1997wp,guay2016twisted}. Similar to the usual Yangians (of finite type), they have intimate relations with certain integrable systems (see for example \cite{Gerrard:2017igy,DeLeeuw:2019ohp} for some recent developments). Moreover, finite $\mathcal{W}$-algebras associated to orthogonal and symplectic Lie algebras can be obtained from the truncations of the twisted Yangians \cite{ragoucy2001twisted,brown2009twisted}. Therefore, besides the BPS aspect, the twisted quiver Yangians could also have close connections to integrability and VOAs. A first evidence relating some twisted affine Yangians and rectangular $\mathcal{W}$-algebras was given in \cite{ueda2021twisted}.

The original definition of twisted Yangians associated to finite Lie algebras uses the $RTT$ formalism. A theorem in \cite{Belliard:2014uja} allows one to construct the twisted Yangians using Drinfeld's $J$ presentation. Here, we shall take this as the definition of the twisted Yangians.

Given an involutive automorphism $\rho$, that is $\rho^2=\text{Id}$, of a finite-dimensional simple complex Lie algebra $\mathfrak{g}$, we can decompose the Lie algebra into $\mathfrak{g}=\mathfrak{l}\oplus\mathfrak{m}$. Here, $\mathfrak{l}$ is generated by the elements satisfying $\rho(y)=y$ while $\mathfrak{m}$ is composed of elements having a negative eigenvalue with $\rho(y)=-y$. This symmetric pair decomposition satisfies $[\mathfrak{l},\mathfrak{l}]\subseteq\mathfrak{l}$, $[\mathfrak{l},\mathfrak{m}]=\mathfrak{m}$ and $[\mathfrak{m},\mathfrak{m}]=\mathfrak{l}$. In this positive/negative decomposition, the Casimir element of the (universal enveloping) algebra can be decomposed into $C_{\mathfrak{g}}=C_{\mathfrak{l}}+C_{\mathfrak{m}}$. Let $\{L_i\}$ (resp.~$\{M_j\}$) be a basis of $\mathfrak{l}$ (resp.~$\mathfrak{m}$). Then the twisted Yangian $\widetilde{\mathcal{Y}}(\mathfrak{g},\mathfrak{l})$ is a subalgebra of $\mathcal{Y}(\mathfrak{g})$ topologically generated by $L_i$ and $\mathcal{B}(M_j)$, where $\mathcal{B}(M_j)=J(M_j)+\frac{\hbar}{4}[M_j,C_{\mathfrak{l}}]$. Here, $\hbar$ is the parameter of the Yangian $\mathcal{Y}(\mathfrak{g})$.

\paragraph{Example} Let us illustrate this with the so-called type BDI as an example\footnote{A classification of the symmetric pairs for classical Lie algebras can be found in \cite[Chapter X]{helgason1979differential}.}. We take the Lie algebra $\mathfrak{g}$ to be $\mathfrak{so}(2n)$, and we consider the involution $\rho(x)=gxg^{-1}$, where\footnote{Of course, the type BDI also includes $\mathfrak{g}$ being $\mathfrak{so}(2n+1)$ (with a different $g$), but here we shall just pick one example as an illustration.}
\begin{equation}
	g=\sum_{i=1}^{(p-q)/2}(E_{ii}+E_{-i,-i})+\sum_{i=(p-q)/2+1}^n(E_{-i,i}+E_{i,-i})
\end{equation}
with $p+q=2n$ and $p\geq q>0$. Here, $E_{ij}$ is the elementary matrix with 1 at the entry $(i,j)$ and 0 otherwise. In this convention where $i,j\in\{-n,\dots,-1,1,\dots,n\}$, the algebra $\mathfrak{so}(2n)$ is spanned by $E_{ij}-E_{ji}$. Notice that $g^{-1}=g^{\text{T}}$.

It turns out that $\mathfrak{l}=\mathfrak{so}(p)\oplus\mathfrak{so}(q)$. Let us focus on $p=2n-1$ and $q=1$ here as $\mathfrak{l}$ in this case can be folded from $\mathfrak{g}$. Indeed, $\mathfrak{l}=\mathfrak{so}(p)=\mathfrak{so}(2n-1)$ with $\mathfrak{so}(q)$ being the zero Lie algebra. More explicitly,
\begin{equation}
	g=\sum_{i=1}^{n-1}(E_{ii}+E_{-i,-i})+(E_{-n,n}+E_{n,-n})=\begin{pmatrix}
		0 & \rvline & 0 & \rvline & 1 \\ \hline
		0 & \rvline & I_{2n-2} & \rvline & 0 \\ \hline
		 1 & \rvline & 0 & \rvline & 0
	\end{pmatrix},
\end{equation}
where $I_k$ is the identity matrix of size $k$. Then the positive part with $gxg^{-1}=x$ is spanned by $L_{\alpha}=E_{ij}-E_{-j,-i}$ for $i,j\neq\pm n$ (and $\alpha$ labels the elements in the basis) while the negative part with $gxg^{-1}=-x$ is spanned by $E_{n,\pm n}-E_{\mp n,-n}$. Therefore, the twisted Yangian is generated by $L_{\alpha}$ in $\mathfrak{l}=\mathfrak{so}(2n-1)$ together with $J(E_{n,\pm n}-E_{\mp n,-n})+\frac{\hbar}{4}\left[E_{n,\pm n}-E_{\mp n,-n},\sum\limits_{\alpha,\beta}(\upkappa_{\mathfrak{l}})^{\alpha\beta}\{L_\alpha,L_\beta\}\right]$, where $\upkappa$ denotes the Killing form on $\mathfrak{l}$.

In general, for any simle complex Lie algebra $\mathfrak{g}$ and a given involution, the invariant subalgebra $\mathfrak{l}$ is semisimple or reductive. It can be decomposed into $\mathfrak{l}_1\oplus\mathfrak{l}_2\oplus\mathfrak{l}_0$ where $\mathfrak{l}_{1,2}$ are simple and $\mathfrak{l}_0$ is one-dimensional.

\paragraph{Twisted quiver Yangians} For the quiver Yangians $\mathcal{Y}$ associated to affine Kac-Moody algebras $\widehat{\mathfrak{g}}$, we shall define the twisted Yangians $\widetilde{\mathcal{Y}}\left(\widehat{\mathfrak{g}},\widehat{\mathfrak{l}}\right)$ as their subalgebras in an analogous way using the $J$ presentation in \S\ref{Jpresentation}. The parameter $\hbar$ would be taken as $h_1+h_2$ (or $2h$). In particular, they are generated by $\widehat{\mathfrak{l}}$ and $\mathcal{B}(y)=J\left(y\right)+\frac{\hbar}{4}\left[y,C_{\widehat{\mathfrak{l}}}\right]$ for $y\in\widehat{\mathfrak{m}}$. Here, the Casimir element of $\widehat{\mathfrak{l}}$ is given by\footnote{Notice that this is not the same as (but part of) the Casimir operator of $\widehat{\mathfrak{l}}$ in \cite{kac1990infinite,musson2012lie}.}
\begin{equation}
	C_{\widehat{\mathfrak{l}}}=2\sum_{\alpha\in\varDelta_{\widehat{\mathfrak{l}},+}}\sum_{k=1}^{\dim\mathfrak{g}_{\alpha}}f^{(\alpha,k)}e^{(\alpha,k)}.
\end{equation}
For simplicity, we shall also refer to the twisted Yangian $\widetilde{\mathcal{Y}}$ as a twisted Yangian $\widetilde{\mathcal{Y}}$ associated to $\widehat{\mathfrak{l}}$.

In this paper, we shall always consider the cases satisfying the following two assumptions:
\begin{itemize}
	\item The positive part $\widehat{\mathfrak{l}}$ would have a corresponding affine Dynkin quiver Yangian $\mathcal{Y}\left(\widehat{\mathfrak{l}}\right)$. Then there should be an embedding from (the universal enveloping algebra of) $\widehat{\mathfrak{l}}$ into the quiver Yangian, and we shall simply use the generators of (the full) $\mathcal{Y}\left(\widehat{\mathfrak{l}}\right)$. In particular, using the generators of $\mathcal{Y}\left(\widehat{\mathfrak{l}}\right)$ as (part of) the generators of the twisted quiver Yangian should still be consistent with the possible connections to $\mathcal{W}$-algebras that will be discussed later. By thinking of $\mathcal{Y}\left(\widehat{\mathfrak{l}}\right)$, we can also see that the twisted quiver Yangian recovers the usual quiver Yangian when the involution is trivially the identity map.
	\item The remaining generators of the twisted quiver Yangians can be written as $\mathcal{B}\left(x^{(a)}_0-x^{(b)}_0\right)$ ($x\in\psi,e,f$) with $x^{(a)}-x^{(b)}\in\widehat{\mathfrak{m}}$. It is worth noting that these elements in $\widehat{\mathfrak{m}}$ are either $\psi^{(a)}_0-\psi^{(b)}_0$ or can be written as the commutators of $\psi^{(a)}_0-\psi^{(b)}_0$ and some element from $\widehat{\mathfrak{l}}$.
\end{itemize}

As the twisted quiver Yangians are subalgebras of the quiver Yangians, their actions on the highest weight/poset modules can be naturally induced from those of the quiver Yangians. Of course, this does not mean that the poset representations of $\mathcal{Y}$ automatically become representations of $\widetilde{\mathcal{Y}}$, but it might still be possible to obtain some highest weight representations of $\widetilde{\mathcal{Y}}$ whose states are subsets/combinations of those in the representations of $\mathcal{Y}$. With a better understanding of the Fock modules of the twisted quiver Yangians, it might also lead to some interesting results in integrability.

In the finite cases, the twisted Yangian is a left coideal of the correpsonding Yangian, that is, $\Delta\left(\widetilde{\mathcal{Y}}\right)\subseteq\mathcal{Y}\widehat{\otimes}\widetilde{\mathcal{Y}}$. Here, we expect that a twisted quiver Yangian is a coideal of the quiver Yangian. In other words, $\Delta\left(\widetilde{\mathcal{Y}}\right)\subseteq\mathcal{Y}\widehat{\otimes}\widetilde{\mathcal{Y}}+\widetilde{\mathcal{Y}}\widehat{\otimes}\mathcal{Y}$. This is automatic for the elements from $\widehat{\mathfrak{l}}$. Using Lemma 18.4.1 in \cite{musson2012lie} and the expressions for the coproduct, we have
\begin{align}
	&\Delta\left(J\left(\psi^{(a)}_0\right)\right)=\square\left(J\left(\psi^{(a)}_0\right)\right)+\frac{\hbar}{2}\sum_{\substack{\alpha\in\varDelta_+\\k}}\left(\alpha^{(a)},\alpha\right)\left((-1)^{|\alpha|}e^{(\alpha,k)}\otimes f^{(\alpha,k)}-f^{(\alpha,k)}\otimes e^{(\alpha,k)}\right),\\
	&\Delta\left(J\left(e^{(a)}_0\right)\right)=\square\left(J\left(e^{(a)}_0\right)\right)-\frac{\hbar}{2}\sum_{\substack{\alpha\in\varDelta_+\\k}}\left[e^{(a)}_0,f^{(\alpha,k)}\right\}\otimes e^{(\alpha,k)},\\
	&\Delta\left(J\left(f^{(a)}_0\right)\right)=\square\left(J\left(f^{(a)}_0\right)\right)+\frac{\hbar}{2}\sum_{\substack{\alpha\in\varDelta_+\\k}}f^{(\alpha,k)}\otimes\left[e^{(\alpha,k)},f^{(a)}_0\right\},
\end{align}
where we have included the parity of the roots to incorporate the super cases below. It would also be convenient to write
\begin{equation}
	\Omega_{\widehat{\mathfrak{l}}}=2\sum_{\alpha\in\varDelta_{\widehat{\mathfrak{l}},+}}\sum_{k=1}^{\dim\mathfrak{g}_{\alpha}}f^{(\alpha,k)}\otimes e^{(\alpha,k)}.
\end{equation}
Then
\begin{equation}
		\begin{split}
			\Delta\left(\mathcal{B}\left(\psi^{(a)}_0-\psi^{(b)}_0\right)\right)=&\square\left(\mathcal{B}\left(\psi^{(a)}_0-\psi^{(b)}_0\right)\right)\\
			&+\frac{\hbar}{2}\sum_{\substack{\alpha\in\varDelta_+\\k}}\left(\alpha^{(a)}-\alpha^{(b)},\alpha\right)\left((-1)^{|\alpha|}e^{(\alpha,k)}\otimes f^{(\alpha,k)}-f^{(\alpha,k)}\otimes e^{(\alpha,k)}\right)
		\end{split}
\end{equation}
as the commutators of $\Omega_{\widehat{\mathfrak{l}}}$ and $\square\left(\psi^{(a)}_0\right)$ vanish. For the second line, we notice that
\begin{equation}
	\left(\alpha^{(a)}-\alpha^{(b)},\alpha\right)x^{(\pm\alpha)}\otimes x^{(\mp\alpha)}=\pm\left[\psi^{(a)}_0-\psi^{(b)}_0,x^{(\pm\alpha)}\right]\otimes x^{(\mp\alpha)}=\mp x^{(\pm\alpha)}\otimes\left[\psi^{(a)}_0-\psi^{(b)}_0,x^{(\mp\alpha)}\right].
\end{equation}
Therefore, $\Delta\left(\mathcal{B}\left(\psi^{(a)}_0-\psi^{(b)}_0\right)\right)$ is actually in $\widetilde{\mathcal{Y}}\widehat{\otimes}\widetilde{\mathcal{Y}}$. Moreover,
\begin{equation}
	\begin{split}
		\Delta\left(\mathcal{B}\left(e^{(a)}_0-e^{(b)}_0\right)\right)=&\square\left(\mathcal{B}\left(e^{(a)}_0-e^{(b)}_0\right)\right)-\frac{\hbar}{2}\sum_{\alpha\in\varDelta_{\widehat{\mathfrak{m}},+}}\left[e^{(a)}_0-e^{(b)}_0,f^{(\alpha,k)}\right\}\otimes e^{(\alpha,k)}\\
		&-\frac{\hbar}{2}\sum_{\substack{\alpha\in\varDelta_{\widehat{\mathfrak{l}},+}\\k}}f^{(\alpha,k)}\otimes \left[e^{(\alpha,k)},e^{(a)}_0-e^{(b)}_0\right\}.
	\end{split}
\end{equation}
Using the fact that $[y_1,y_2\}\in\widehat{\mathfrak{l}}$ for $y_1,y_2\in\widehat{\mathfrak{m}}$ and $[y_1,y_2\}\in\widehat{\mathfrak{m}}$ for $y_1\in\widehat{\mathfrak{l}},y_2\in\widehat{\mathfrak{m}}$, we find that $\Delta\left(\mathcal{B}\left(e^{(a)}_0-e^{(b)}_0\right)\right)$ is in $\widetilde{\mathcal{Y}}\widehat{\otimes}\mathcal{Y}$. Likewise, $\Delta\left(\mathcal{B}\left(f^{(a)}_0-f^{(b)}_0\right)\right)$ is in $\mathcal{Y}\widehat{\otimes}\widetilde{\mathcal{Y}}$ following a similar calculation. Notice that as a concept in the coalgebra, we also need $\varepsilon\left(\widetilde{\mathcal{Y}}\right)=0$ for the counit $\varepsilon:\mathcal{Y}\rightarrow\mathbb{C}$. This should be automatic as the counit sends all the modes of $\mathcal{Y}$ to zero.

\paragraph{Examples from foldings} With the above construction, we can consider the twisted quiver Yangians given any involutions. Let us start with some examples that also feature the property of foldings.

We shall first consider the affine extension of the above (finite) example. In fact, it is most conveniently to work with the Chevalley basis as this is what we used for quiver Yangians. Following the numbering as indicated by the Cartan matrices in \S\ref{affinecases}, the involution on $D_n^{(1)}$ is given by
\begin{equation}
	\rho\left(x^{(a)}_0\right)=\begin{cases}
		x^{(a)}_0,&a=2,\dots,n-2\\
		x^{(a+1)}_0,&a=0,n-1\\
		x^{(a-1)}_0,&a=1,n
	\end{cases}
\end{equation}
with $x=\psi,e,f$. Then the positive part has $x^{(a)}_0$ for $a=2,\dots,n-2$ together with $x^{(0)}_0+x^{(1)}_0$ and $x^{(n-1)}_0+x^{(n)}_0$. This is exactly $B_{n-1}^{(1)}$ from folding. On the other hand, the negative part has $x^{(0)}_0-x^{(1)}_0$ and $x^{(n-1)}_0-x^{(n)}_0$. Alternatively, we may also consider $J\left(\psi^{(a)}_0-\psi^{(a+1)}_0\right)$ and $J\left(\left[\psi^{(a)}_0-\psi^{(a+1)}_0,x^{(a)}_0+x^{(a+1)}_0\right]\right)$  ($a=0,n-1$) since
\begin{equation}
	\frac{1}{2}\left[\psi^{(a)}_0-\psi^{(a+1)}_0,e^{(a)}_0+e^{(a+1)}_0\right]=e^{(a)}_0-e^{(a+1)}_0,
\end{equation}
and likewise for $f$.

Let us now consider folding the toric quiver Yangian associated to $A_{2n-1}^{(1)}$. The involution $\rho$ swaps $x^{(a)}_0$ and $x^{(2n-2-a)}_0$ for $a\in\{0,1,\dots,n-2,n,\dots,2n-2\}$ while leaving $x^{(n-1)}_0$ and $x^{(2n-1)}_0$ invariant. As a result, we have a twisted quiver Yangian associated to $C_n^{(1)}$, and the negative part has $x^{(a)}_0-x^{(2n-2-a)}_0$ ($a\neq n-1,2n-1$). We should also make a comment that the usual quiver Yangian associated to $C_n^{(1)}$ is a one-parameter algebra. However, by definition of the twisted quiver Yangian, since the quiver Yangian associated to the A-type has two parameters, this twisted version of C-type would also have two parameters.

As another example, we shall consider the folding not just with the $\mathbb{Z}_2$ symmetry. From Figure \ref{folding}, we can see that $D_4^{(1)}$ can be folded to $G_2^{(1)}$. To obtain this twisted quiver Yangian associated to $G_2^{(1)}$ from the quiver Yangian for $D_4^{(1)}$, we can take two steps. Let us label the node in the centre by 2 and other nodes by 0, 1, 3, 4 in $D_4^{(1)}$. Then we can get a twisted Yangian associated to $B_3^{(1)}$ by merging the nodes 3 and 4. In other words, the $\mathcal{B}$ generators contain $J\left(x^{(4)}_0-x^{(5)}_0\right)$.

Then from this twisted quiver Yangian, we can further fold the node 2 into the above merged node. This yields a twisted quiver Yangian associated to $G_2^{(1)}$ generated by elements from the current algebra/quiver Yangian for $G_2^{(1)}$ and $\mathcal{B}\left(x^{(4)}_0-x^{(5)}_0-x^{(2)}_0\right)$. Here, $x^{(a)}_0$ still refers to the modes from the original (quiver Yangian associated to) $D_4^{(1)}$.

All the twisted quiver Yangians related to the foldings as in Figure \ref{folding} can be obtained in a similar manner. Comparing them with the usual quiver Yangians obtained from foldings, we can see that the folded part contains $J\left(x^{(a_1)}_0-x^{(a_2)}_0\right)$ instead of $J\left(x^{(a_1)}_0+x^{(a_2)}_0\right)$.

\paragraph{More examples} Of course, the twisted quiver Yangians do not always have to be obtained from foldings. Let us mention some examples here as well. Consider the quiver Yangian associated to $A_{2n}^{(1)}$ ($n\geq2$). First, we take the involution $\rho_1\left(x^{(a)}_0\right)=x^{(2n-1-a)}_0$ ($a=0,1,\dots,2n-1$). Then the positive (resp.~negative) part has $x^{(a)}_0+x^{(2n-1-a)}_0$ (resp.~$x^{(a)}_0-x^{(2n-1-a)}_0$). This gives a twisted quiver Yangian associated to $A_1^{(1)}$ for $n=2$ and $D_{n+1}^{(2)}$ for $n>2$.

Let us also consider the involution $\rho_2\left(x^{(a)}_0\right)=x^{(n+a)}_0$. Then the positive (resp.~negative) part has $x^{(a)}_0+x^{(n+a)}_0$ (resp.~$x^{(a)}_0-x^{(n+a)}_0$). This gives a twisted quiver Yangian associated to $A_n^{(1)}$. This is different from the usual quiver Yangian for $\mathbb{C}\times\mathbb{C}^2/\mathbb{Z}_{n+1}$.

\section{Super Cases}\label{supercases}
A natural extension of the above discussions would be considering the affine Dynkin diagrams for superalgebras. For the toric cases, namely the generalized conifolds with $A(m-1,n-1)^{(1)}$, details can be found in \cite{Li:2020rij,Bao:2022jhy} (see also \cite{ueda2019affine,ueda2022surjectivity,ueda2022affine} for similar affine super Yangians). Although the other affine Dynkin cases are not toric and the possible connections to the BPS spectra might need further analysis, it is still straightforward to obtain the quiver Yangians. Therefore, the main job is to find out how to get these $\mathcal{N}=1$ quivers. We shall find the superpotentials such that the quiver Yangians arise from the corresponding affine Dynkin diagrams and Kac-Moody superalgebras in the same sense as the non-super untwisted affine cases. Hence, we expect that there could exist similar geometric constructions for these $\mathcal{N}=1$ quiver gauge theories.

In these affine Dynkin diagrams, there are also odd nodes. Therefore, we need to generalize/modify the notion of the ``tripled'' quivers. More specifically, when the odd node $a$ is isotropic (resp.~non-isotropic) with $A_{aa}=0$ (resp.~$A_{aa}=2$), we instead add 0 (resp.~2) self-loops to the node. As a result, in the ``tripled'' quiver, $a$ is still fermionic with $|a|=1$.

When drawing the affine Dynkin diagrams, we shall adopt the common notation in literature. A white (grey, resp.~black) node is associated to an even (odd isotropic, resp.~odd non-isotropic) simple root. A small black dot is used to denote a node that could be either white or grey, and this will be denoted by a node with a dashed arrow in the ``tripled'' quiver. Explicitly, we have
\begin{equation}
	\includegraphics[width=12cm]{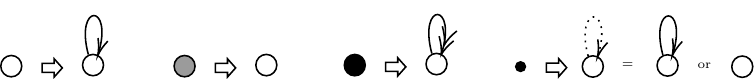}.
\end{equation}

In the affine Dynkin diagrams, the non-simply lacedness could give multiple edges either with arrows or without arrows between two nodes. It turns out that any of these edges would give only one pair of opposite arrows in the quiver just like the simply laced ones. This is due to Seiberg duality as we will see later. In fact, the non-simply lacedness is encoded by the edge weights.

Therefore, the only non-trivial symmetric sum of the edge weights would be $\sigma^{ab}_1$ (except $\sigma^{aa}_2$ for black nodes). As a result, most of the discussions for the non-super cases can be directly applied/extended to the super cases. Before we analyze the explicit construction of the quivers and superpotentials for each case, let us first summarize the results here. Again, let us use the rescaled generators:
\begin{equation}
	\psi^{(a)}_n=-\frac{1}{h_1+h_2}\uppsi^{(a)}_n,\quad e^{(a)}_n=-\frac{1}{\left(h_1+h_2\right)^{1/2}}\mathtt{e}^{(a)}_n\quad f^{(a)}_n=-\frac{1}{\left(h_1+h_2\right)^{1/2}}\mathtt{f}^{(a)}_n
\end{equation}
(assuming that $h_1+h_2\neq0$), and we still have $h_1=h_2=h$ for the one-parameter cases\footnote{Of course, one can assign different values to the edges as long as the ratios of the weights/two parameters remain invariant among different choices. This would trivially give isomorphic algebras with a rescaling of the parameter(s).}. Then the defining relations remain to be \eqref{psipsi}$\sim$\eqref{ff}, but with non-trivial signs and supercommutators for those involving fermionic generators. Notice that this does not fully include the cases with black nodes, and we will comment on this shortly.

All the higher modes can still be inductively derived from $\psi^{(a)}_{0,1},e^{(a)}_0,f^{(a)}_0$. Notice that for a grey node $a$, we need to take $b=a\pm1$ (but not $b=a$) in \eqref{finitegenerators}. Moreover, we have the same minimalistic presentation and $J$ presentation with supercommutators in \eqref{psipsimin}$\sim$\eqref{ffmin} and \eqref{JxJx}$\sim$\eqref{JeeJff}. Likewise, the coproduct is given by \eqref{coprod}. Given the $J$ presentation, we can also define the twisted quiver Yangians for the super cases in the same manner. Hence, we will not give more examples here.

Now, let us have a brief discussion on the cases whose quivers have nodes with two adjoint loops. As can be seen from the lists below in this section and in Appendix \ref{SDquivers}, for a given affine Dynkin diagram, there can be at most one black node locating at the end of the diagram, and it always belongs to the set of nodes with the shortest roots among all the simple roots. In the quiver, the two self-loops always have weights (proportional to) $h$ and $-2h$ respectively. In fact, all the relations are still the same as those without black nodes, except the $\psi e$, $ee$ relations with the two modes both from this black node (the ones with the $f$ modes are completely similar, and hence we shall not repeat them in the discussions here). The only different relations read
\begin{align}
	&[\psi_{n+2},e_m]-2[\psi_{n+1},e_{m+1}]+[\psi_n,e_{m+2}]=2h^2[\psi_n,e_m]-h\{\psi_{n+1},e_m\}+h\{\psi_n,e_{m+1}\},\\
	&\{e_{n+2},e_m\}-2\{e_{n+1},e_{m+1}\}+\{e_n,e_{m+2}\}=2h^2\{e_n,e_m\}-h[e_{n+1},e_m]+h[e_n,e_{m+1}],\\
	&[\psi_{n+2},f_m]-2[\psi_{n+1},f_{m+1}]+[\psi_n,f_{m+2}]=2h^2[\psi_n,f_m]+h\{\psi_{n+1},f_m\}-h\{\psi_n,f_{m+1}\},\\
	&\{f_{n+2},f_m\}-2\{f_{n+1},f_{m+1}\}+\{f_n,f_{m+2}\}=2h^2\{f_n,f_m\}+h[f_{n+1},f_m]-h[f_n,f_{m+1}],
\end{align}
where we have omitted the superscripts on the modes as there is only one node involved. Let us focus on the $\psi e$ relation, and the discussion on the $ee$, $\psi f$, $ff$ relations would be similar. By taking $n=-2,-1,0$ respectively and $m=0$, we have\footnote{Notice that one cannot take $n=-2,-1$ in the $ee$ relations.}
\begin{align}
	&[\psi_0,e_0]=-2he_0,\\
	&[\psi_1,e_0]=e_1-h\{\psi_0,e_1\},\\
	&[\psi_2,e_0]-2[\psi_1,e_1]+[\psi_0,e_2]=2h^2[\psi_0,e_0]-h\{\psi_1,e_0\}+h\{\psi_0,e_1\}.
\end{align}
These relations should still be able to derive all the relations with higher $m$ and $n$. In other words, the minimalistic presentation in this case needs to be extended by including the third line (as well as the similar ones for $ee$, $\psi f$ and $ff$). From the first two lines, we can also see that they remain the same pattern as in the other cases, and $\psi_{0,1},e_0,f_0$ can still generate all the higher modes\footnote{In fact, by considering the relations such as $\psi^{(a-1)}e^{(a)}$, where $a$ denotes the node with two adjoint loops, one can also see that $\psi_{0,1},e_0,f_0$ are sufficient to generate all the modes.}. Therefore, the quiver Yangian in such case can actually be thought of as the usual algebra same as the other cases, but further quotiented by four relations involving some modes at level 2. The generators in terms of $J$ and the coproduct above should be compatible with the extra relations although they need to be included in the complete $J$ presentation.

\paragraph{Serre relations} Let us again make some comments on the Serre relations. A natural possibility would be some relations similar to \eqref{Serre1} for the non-super cases. For instance, we may write
\begin{equation}
	\left[e^{(a)}_{n_1},\left[e^{(a+1)}_{m_1},\left[e^{(a)}_{n_2},e^{(a-1)}_{m_2}\right\}\right\}\right\}+\left[e^{(a)}_{n_2},\left[e^{(a+1)}_{m_1},\left[e^{(a)}_{n_1},e^{(a-1)}_{m_2}\right\}\right\}\right\}=0
\end{equation}
when $a$ is a (generic) node without adjoint loops. As pointed out in \cite{Bao:2022jhy}, with these relations, there are some subleties when writing the minimalistic presentation for the $\widehat{\mathfrak{gl}}(2|1)$ quiver and one toric phase of the $\widehat{\mathfrak{gl}}(2|2)$ quiver. On the other hand, we may also consider the relations given by \eqref{Serre2}. Here, for simplicity, we shall always assume that the full Serre relations (involving the higher modes) can be derived from the Serre relations of the underlying Kac-Moody algebra. In fact, as we will see, these Serre relations are intimately related to the superpotentials we will take, and this was already observed for the toric cases in \cite{Li:2020rij,Galakhov:2021vbo}.

\subsection{Distinguished Cases}\label{distinguished}
A different feature in the super cases is that there can be multiple affine Dynkin diagrams for the same Kac-Moody superalgebra. A classification of all possible diagrams was given in \cite{frappat1989structure} (see also \cite{Frappat:1996pb}). In this subsection, let us explicitly construct the quivers and superpotentials for those assoicated to the distinguished Dynkin diagrams for the untwisted affine cases. The other phases\footnote{The way to obtain the quivers associated to the twisted affine cases classified in \cite[Table 11]{frappat1989structure} is completely the same. Hence, we shall not list them explicitly. Here, ``twisted'' refers to the underlying affine Lie superalgebras and should not be confused with the twisted quiver Yangians.} can be found in \S\ref{seiberg} and in Appendix \ref{SDquivers}.

We find that the terms in the superpotential can be obtained from the Serre relations of the underlying Kac-Moody algebra. For instance, in the above non-super cases, we have the correspondence\footnote{In the followings, we shall always omit the Serre relations of the same forms for $f$.}
\begin{equation}
	\left[e^{(a)}_0,\dots,\left[e^{(a)}_0,e^{(b)}_0\right]\dots\right]=0\quad\leftrightarrow\quad X_{aa}^kX_{ab}X_{ba}\in W.
\end{equation}
These relations and  superpotential terms would still appear in the super cases. However, due to the existence of fermionic nodes, these Serre relations are not sufficient to recover the Kac-Moody algebra. The extra Serre relations can be found in \cite{Frappat:1996pb,yamane1999defining}. They would yield more terms in the superpotential. As (almost) every affine Dynkin diagram has at least one grey node, a common case would be
\begin{equation}
	\includegraphics[width=10cm]{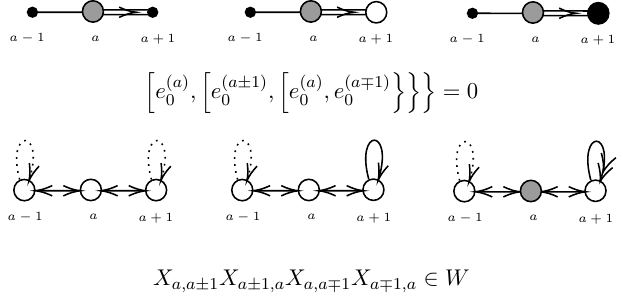}.
\end{equation}
We will give the other cases when we meet them in the following discussions. As the couplings of the superpotential terms are not of particular importance here, we shall always omit them and only give the relevant monomial factors for each case.

\paragraph{A-type and $D(2,1;\alpha)^{(1)}$} These quivers are toric and have been well-studied in various literature. Hence, we shall only give the weight assignments to the edges for completeness:
\begin{equation}
	\includegraphics[width=10cm]{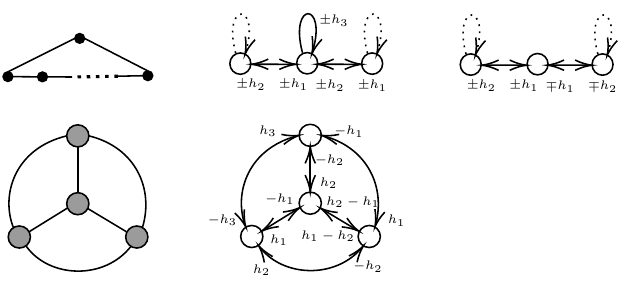},\label{toriccases}
\end{equation}
where $h_3=-h_1-h_2$. Here, we give all the toric dual phases for $A(m-1|n-1)^{(1)}$ (with $m+n$ nodes). The distinguished case has two fermionic nodes at position 0 and $m$. All the other phases can be obtained by dualizing an (isotropic) odd node as will be discussed in the next subsection. The ``$\pm$'' signs depend on the convention just like the signs in the Cartan matrix. For $D(2,1;\alpha)^{(1)}$, there is another phase which is not toric. We will analyze this case in more detail shortly.

\paragraph{B-type $B(m,n)^{(1)}$} For the distinguished case of $\mathfrak{osp}(2m+1|2n)^{(1)}$, all the superpotential terms follow the above rules. We have
\begin{equation}
	\includegraphics[width=10cm]{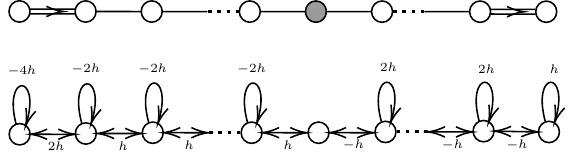}.
\end{equation}

\paragraph{B-type $B(0,n)^{(1)}$} For the distinguished case of $\mathfrak{osp}(1|2n)^{(1)}$, there is a non-isotropic fermionic node, but (almost) all the superpotential terms follow the above rules. We have
\begin{equation}
	\includegraphics[width=7cm]{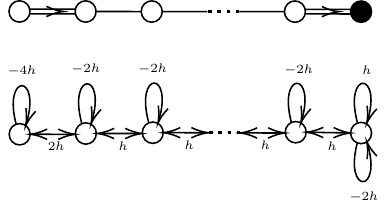}.
\end{equation}
The superpotential terms involving the rightmost node $n$ (where the labelling starts from 0 for the leftmost node) are $X_{n,n-1}X_{n-1,n}X_{nn,(2)}$ and $X_{nn,(1)}^2X_{nn,(2)}$. As we can see, the terms involving the black node are different from the other terms we have seen before\footnote{Notice that the second term does not correspond to any Serre relations, but we still have $\left[e^{(a)}_0,\left\{e^{(a)}_0,e^{(a)}_0\right\}\right]=0$ as the same mode should anticommute with itself for fermionic $a$.}. This will be sharpened when we discuss Seiberg duality in \S\ref{seiberg}.

\paragraph{C-type $C(n+1)^{(1)}$} To discuss the distinguished case of $\mathfrak{osp}(2|2n)^{(1)}$, we first need to consider a new general configuration:
\begin{equation}
	\includegraphics[width=5cm]{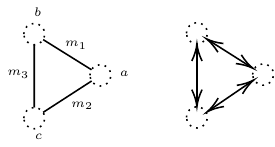},
\end{equation}
where the multiplicities of the edges are given by $m_1=|(\alpha_a,\alpha_b)|$ etc. Here, the dashed nodes can be any of the three types. Moreover, the sum of the three inner products should be zero, and $|a||b|+|a||c|+|b||c|\equiv1$. The corresponding Serre relation is
\begin{equation}
	(-1)^{|a||c|}(\alpha_a,\alpha_c)\left[\left[e^{(a)}_0,e^{(b)}_0\right\},e^{(c)}_0\right\}=(-1)^{|a||b|}(\alpha_a,\alpha_b)\left[\left[e^{(a)}_0,e^{(c)}_0\right\},e^{(b)}_0\right\}.
\end{equation}
Therefore, there are two corresponding terms in the superpotential given by the two loops of length 3 (one clockwise and one counterclockwise). Notice that regardless of the multiplicities of the edges in the affine Dynkin diagram, there is always one pair of opposite arrows connecting any two of the three nodes in the quiver. This will also be verified in the next subsection. We then have
\begin{equation}
	\includegraphics[width=7cm]{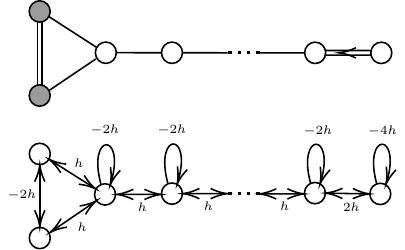}.
\end{equation}
As we can see, the multiplicities of the edges are in fact reflected by the weights of the arrows.

\paragraph{D-type $D(m,n)^{(1)}$} For the distinguished case of $\mathfrak{osp}(2m|2n)^{(1)}$, all the superpotential terms follow the above rules. We have
\begin{equation}
	\includegraphics[width=10cm]{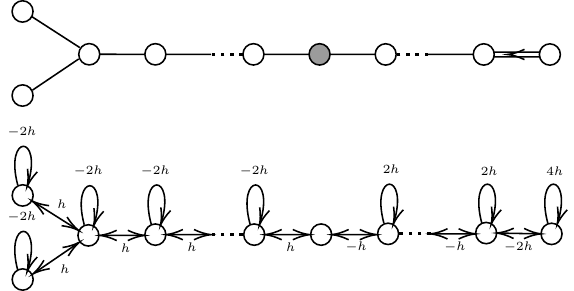}.
\end{equation}

\paragraph{F-type $F(4)^{(1)}$} For the distinguished case of $F(4)^{(1)}$, we have
\begin{equation}
	\includegraphics[width=5cm]{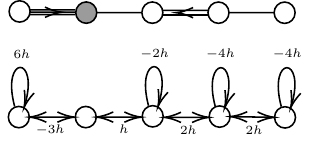}.
\end{equation}
Label the nodes by $0,1,\dots,4$ from left to right. There is an extra Serre relation given by
\begin{equation}
	\left[\left[\left[\left[\left[e^{(0)}_0,e^{(1)}_0\right],e^{(2)}_0\right],e^{(3)}\right],e^{(1)}_0\right],e^{(2)}_0\right]=2\left[\left[\left[\left[\left[e^{(0)}_0,e^{(1)}_0\right],e^{(2)}_0\right],e^{(3)}\right],e^{(2)}_0\right],e^{(1)}_0\right].
\end{equation}
This corresponds to the superpotential term $X_{01}X_{12}X_{23}X_{32}X_{21}X_{10}$, which is consistent with the weight assignment to the arrows.

\paragraph{G-type $G(3)^{(1)}$} For the distinguished case of $G(3)^{(1)}$, we have
\begin{equation}
	\includegraphics[width=4cm]{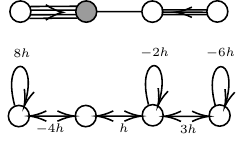}.
\end{equation}
Label the nodes by $0,1,2,3$ from left to right. There is an extra Serre relation given by
\begin{equation}
	2\left[\left[\left[\left[\left[e^{(0)}_0,e^{(1)}_0\right],e^{(2)}_0\right],e^{(3)}\right],e^{(1)}_0\right],e^{(2)}_0\right]=3\left[\left[\left[\left[\left[e^{(0)}_0,e^{(1)}_0\right],e^{(2)}_0\right],e^{(3)}\right],e^{(2)}_0\right],e^{(1)}_0\right].
\end{equation}
This corresponds to the superpotential term $X_{01}X_{12}X_{23}X_{32}X_{21}X_{10}$, which is consistent with the weight assignment to the arrows.

As we can see, all the cases except $A(m-1|n-1)^{(1)}$ and $D(2,1;\alpha)^{(1)}$ have only one free parameter. This would also be the case for the other phases which are related by Seiberg duality as we will discuss below.

\subsection{Seiberg Duality and Odd Reflections}\label{seiberg}
Given an affine Lie superalgebra, it can have multiple distinct affine Dynkin diagrams. As a result, they would give rise to different quivers. For the affine Lie algebras, these diagrams are related by odd reflections on the grey nodes ($(\alpha,\alpha)=0$) \cite{serganova2011kac}:
\begin{equation}
	r_{\alpha}(\alpha')=\begin{cases}
		-\alpha',&\alpha'=\alpha\\
		\alpha+\alpha',&(\alpha,\alpha')\neq0\\
		\alpha',&\alpha'\neq\alpha\text{ and }(\alpha,\alpha')=0
	\end{cases}.
\end{equation}
Physically, the quiver gauge theories should be related by Seiberg duality. In particular, the odd reflection on a grey node correspond to dualizing the gauge node without any adjoint loops\footnote{In principle, one may also consider dualizing a node with adjoints in a quiver. However, we shall not discuss this here as they are not associated to the affine Dynkin diagrams and odd reflections.}.

Let us briefly recall how Seiberg duality \cite{Seiberg:1994pq,Franco:2003ja,Feng:2000mi,Beasley:2001zp,Feng:2001bn,Feng:2002zw} can be performed for quivers. We first choose a node $\digamma$ to dualize so that the other gauge nodes decouple. Then we reverse all the arrows connected to $\digamma$ as the quarks would be transformed into the conjugate (flavour) representations. For every pair of reversed arrows $(X_{i\digamma},X_{\digamma j})\rightarrow(X'_{\digamma i},X'_{j\digamma})$, we add a meson $M_{ij}$. As all the flavours are ``gauged back'', all the arrows are promoted to bifundamentals and adjoints. In the superpotential, we should replace all the 2-paths $X_{i\digamma}X_{\digamma j}$ with the mesons $M_{ij}$ and add the cubic terms $X'_{j\digamma}X'_{\digamma i}M_{ij}$. We also need to discard the fields that become massive in terms of their equations of motions. This would remove the corresponding arrows in the quiver. In fact, for the quivers associated to affine Dynkin diagrams disscussed here, the nodes connected to the dualized node would change parity under Seiberg duality.

As the quivers are Seiberg duals, it would be natural to wonder whether the quiver Yangians are isomorphic. For the toric CYs without compact divisors, this was proven in \cite{Bao:2022jhy}. For the quiver Yangians discussed in this paper, they all have underlying Kac-Moody algebras, and the quivers are non-chiral. Let us first exclude the phases involving black nodes. Then there can be at most one pair of opposite arrows between any two nodes. In other words, $\sigma^{ab}_1$ are the only non-trivial symmetric sums of the edge weights. Therefore, the isomorphic map in \cite{Bao:2022jhy} between two Seiberg dual quiver Yangians can be directly applied here. It would be most conveniently to write it in the $J$ presentation:
\begin{align}
	&\psi'^{(a)}_0=\begin{cases}
		-\psi^{(a)}_0,&a=\digamma,\\
		\psi^{(a)}_0+\psi^{(\digamma)}_0,&(\alpha^{(a)},\alpha^{(\digamma)})\neq0,\\
		\psi^{(a)}_0,&\text{otherwise};
	\end{cases}\\
	&e'^{(a)}_0=\begin{cases}
		f^{(a)}_0,&a=\digamma,\\
		\left[e^{(\digamma)}_0,e^{(a)}_0\right],&(\alpha^{(a)},\alpha^{(\digamma)})\neq0,\\
		e^{(a)}_0,&\text{otherwise};
	\end{cases}\\
	&f'^{(a)}_0=\begin{cases}
		-e^{(a)}_0,&a=\digamma,\\
		-\frac{1}{(\alpha^{(a)},\alpha^{(\digamma)})}\left[f^{(a)}_0,f^{(\digamma)}_0\right],&(\alpha^{(a)},\alpha^{(\digamma)})\neq0,\\
		f^{(a)}_0,&\text{otherwise};
	\end{cases}\\
	&J\left(\psi'^{(a)}_0\right)=\begin{cases}
		-J\left(\psi^{(a)}_0\right),&a=\digamma,\\
		J\left(\psi^{(a)}_0\right)+J\left(\psi^{(\digamma)}_0\right)-\frac{1}{2}\left(\sigma_1^{a\digamma}-\sigma_1^{\digamma a}\right)\psi^{(\digamma)}_0,&(\alpha^{(a)},\alpha^{(\digamma)})\neq0,\\
		J\left(\psi^{(a)}_0\right),&\text{otherwise};
	\end{cases}\\
	&J\left(e'^{(a)}_0\right)=\begin{cases}
		J\left(f^{(a)}_0\right),&a=\digamma,\\
		\left[e^{(\digamma)}_0,J\left(e^{(a)}_0\right)\right],&(\alpha^{(a)},\alpha^{(\digamma)})\neq0,\\
		J\left(e^{(a)}_0\right),&\text{otherwise};
	\end{cases}\\
	&J\left(f'^{(a)}_0\right)=\begin{cases}
		-J\left(e^{(a)}_0\right),&a=\digamma,\\
		-\frac{1}{(\alpha^{(a)},\alpha^{(\digamma)})}\left[J\left(f^{(a)}_0\right),f^{(\digamma)}_0\right],&(\alpha^{(a)},\alpha^{(\digamma)})\neq0,\\
		J\left(f^{(a)}_0\right),&\text{otherwise}.
	\end{cases}
\end{align}
Notice that the transformation of the zero modes is exactly the transformation of the Chevalley basis of the affine Lie superalgebras under the odd reflection.

If we consider the phases with black nodes, then there would be some subtleties. From a quiver Yangian without the black node, this map would give rise to the algebra that do not have the extra constraints with the level 2 modes. In other words, we further need to quotient it by these extra relations as mentioned above. Therefore, this gives an algebra that is slightly larger than the quiver Yangian. We then have the following two possibilities:
\begin{itemize}
	\item If the quiver Yangians still describe the BPS algebras, then Seiberg dual theories can have non-isomorphic BPS algebras.
	\item If the BPS algebras should be isomorphic under Seiberg duality, then (at least some of) the non-toric quiver Yangians would need further modifications to get the BPS algebras.
\end{itemize}
Notice that for a quiver with a black node, if a grey node not connected to the black one is dualized, then the two quiver Yangians would still be isomorphic.

Let us now illustrate the procedure of Seiberg duality with some examples. Recall that the superpotential terms can be obtained from the corresponding Serre relations of the underlying affine Lie superalgebras. Here, we shall omit these Serre relations and only write the superpotentials. Indeed, we find that the superpotentials can be reproduced under Seiberg duality.

\paragraph{Example 1: $D(2,1;\alpha)^{(1)}$} The quiver associated to $D(2,1;\alpha)^{(1)}$ in \eqref{toriccases} is toric. This is the only toric phase for $\mathbb{C}^3/(\mathbb{Z}_2\times\mathbb{Z}_2)$. However, there is another phase which is Seiberg dual of the quiver. This is expected not only from the underlying affine Dynkin diagrams but also from the fact that we can dualize any of the four fermionic nodes (which are equivalent due to the symmetry of the quiver). Let us reproduce the toric phase here:
\begin{equation}
	\includegraphics[width=7cm]{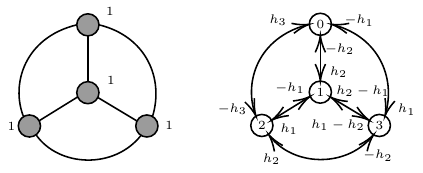},
\end{equation}
where the numbers in the affine Dynkin diagram indicate the Dynkin-Kac labels while the numbers in the quiver are just labelling the nodes. The superpotential is
\begin{equation}
	\begin{split}
		W=&X_{13}X_{32}X_{21}+X_{20}X_{01}X_{12}+X_{31}X_{10}X_{03}+X_{02}X_{23}X_{30}\\
		&-X_{31}X_{12}X_{23}-X_{02}X_{21}X_{10}-X_{13}X_{30}X_{01}-X_{20}X_{03}X_{32}.
	\end{split}
\end{equation}
Let us dualize the node 1 so that all the arrows connected to this node would be reversed. This leads to
\begin{equation}
	\begin{split}
		W=&M_{23}X_{32}+X_{20}M_{02}+M_{30}X_{03}+{\color{red}X_{02}}{\color{orange}X_{23}}{\color{violet}X_{30}}-M_{32}X_{23}-X_{02}M_{20}-M_{03}X_{30}-{\color{olive}X_{20}}{\color{cyan}X_{03}}{\color{blue}X_{32}}\\
		&-{\color{blue}M_{23}}X'_{31}X'_{12}-{\color{olive}M_{02}}X'_{21}X'_{10}-{\color{cyan}M_{30}}X'_{01}X'_{13}+{\color{orange}M_{32}}X'_{21}X'_{13}+{\color{red}M_{20}}X'_{01}X'_{12}+{\color{violet}M_{03}}X'_{31}X'_{10}\\
		&+M_{33}X_{31}X_{13}+M_{22}X_{12}X_{21}+M_{00}X_{01}X_{10}.
	\end{split}
\end{equation}
The quadratic terms are massive terms and hence the coloured fields should be integrated out. For instance, the F-term relation $\partial W/\partial M_{23}=0$ gives $X_{32}=X'_{31}X'_{12}$. This yields the term $-X_{20}X_{03}X'_{31}X'_{12}$. Keep this procedure, and we get the quiver
\begin{equation}
	\includegraphics[width=7cm]{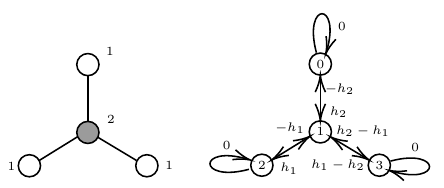}
\end{equation}
with the superpotential
\begin{equation}
	W=X_{01}X_{12}X_{21}X_{13}X_{31}X_{10}-X_{01}X_{13}X_{31}X_{12}X_{21}X_{10}+M_{33}X_{31}X_{13}+M_{22}X_{12}X_{21}+M_{00}X_{01}X_{10},
\end{equation}
where we have relabelled all the fields (including $M$ and $X'$) by $X$.

\paragraph{Example 2: $\mathfrak{osp}(3|6)^{(1)}$} Let us consider an example having a fermionic node with two adjoint loops:
\begin{equation}
	\includegraphics[width=10cm]{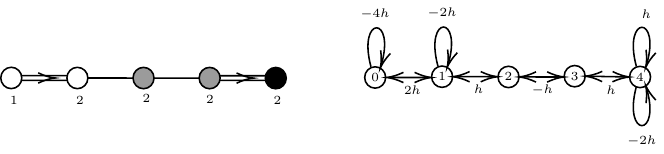}.\label{osp361}
\end{equation}
Let us only list the superpotential terms that would be changed when dualizing the node 3:
\begin{equation}
	X_{23}X_{34}X_{43}X_{32},\quad X_{44,(1)}X_{44,(1)}X_{44,(2)},\quad X_{43}X_{34}X_{44,(2)}.
\end{equation}
Under Seiberg duality, they become
\begin{equation}
	M_{24}M_{42},\quad X_{44,(1)}X_{44,(1)}{\color{blue} X_{44,(2)}},\quad M_{44}X_{44,(2)},
\end{equation}
and we have the new terms
\begin{equation}
	{\color{red} M_{24}}X'_{43}X'_{32},\quad {\color{red} M_{42}}X'_{23}X'_{34},\quad M_{22}X'_{23}X'_{32},\quad {\color{blue} M_{44}}X'_{43}X'_{34}.
\end{equation}
Integrating out the massive fields in terms of their equations of motions, these terms become
\begin{equation}
	X_{44,(1)}X_{44,(1)}X_{43}X_{34},\quad X_{43}X_{32}X_{23}X_{34},
\end{equation}
where we have relabelled all the fields (including $M$ and $X'$) by $X$, with the quiver
\begin{equation}
	\includegraphics[width=10cm]{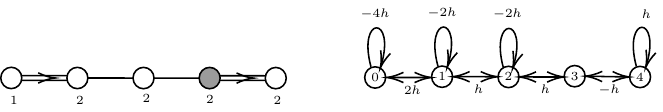}
\end{equation}
as expected.

Let us now dualize the node 2 in \eqref{osp361}. In particular, the superpotential has the terms $X_{10}X_{01}X_{11}X_{11}$ and $X_{12}X_{21}X_{11}$. They become $X_{10}X_{01}X_{11}X_{11}$ and $M_{11}X_{11}$. Together with the new term $M_{11}X'_{12}X'_{21}$, we have $\partial W/\partial M_{11}=0$ giving $X_{11}=X'_{12}X'_{21}$. Therefore, we are left with the term $X_{10}X_{01}X_{12}X_{21}X_{12}X_{21}$, where we have relabelled all the fields (including $M$ and $X'$) by $X$. The other terms in the superpotential are changed in the same way as those we have in the toric cases. The quiver reads
\begin{equation}
	\includegraphics[width=10cm]{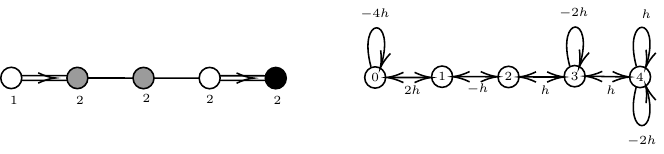}.
\end{equation}

Then we can dualize the node 1. The terms $X_{10}X_{01}X_{12}X_{21}X_{12}X_{21}$ and $X_{01}X_{10}X_{00}$ in the superpotential become $M_{02}M_{22}M_{20}$ and $M_{00}X_{00}$. We also have the new terms $M_{20}X'_{01}X'_{12}$, $M_{02}X'_{21}X'_{10}$ and $M_{00}X'_{01}X'_{10}$. The last term become the mass term under the F-term relation $\partial W/\partial M_{00}=0$. The remaining two cubic terms are exactly the superpotential terms for the triangle configuration in the quiver
\begin{equation}
	\includegraphics[width=10cm]{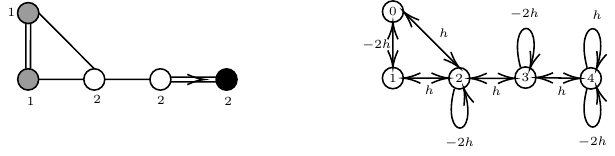}.
\end{equation}

\paragraph{Example 3: $F(4)^{(1)}$} Let us analyze all the phases for $F(4)^{(1)}$. Recall that the distinguished case has the quiver
\begin{equation}
	\includegraphics[width=10cm]{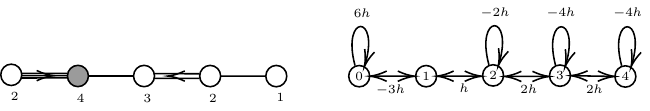}
\end{equation}
with the superpotential
\begin{equation}
	\begin{split}
		W=&g_1X_{01}X_{12}X_{23}X_{32}X_{21}X_{10}+g_2X_{01}X_{10}X_{00}+g_3X_{21}X_{12}X_{22}+g_4X_{23}X_{32}X_{22}^2\\
		&+g_5X_{32}X_{23}X_{33}+g_6X_{34}X_{43}X_{33}+g_7X_{43}X_{34}X_{44}.
	\end{split}
\end{equation}
The couplings $g_i$ are not really important here. Dualizing the node 1, the superpotential becomes
\begin{equation}
	\begin{split}
		W=&g_1X_{01}X_{12}X_{23}X_{32}X_{21}X_{10}+g_2M_{00}X_{00}+g_3M_{22}X_{22}+g_4X_{23}X_{32}X_{22}^2+g_5X_{32}X_{23}X_{33}\\
		&+g_6X_{34}X_{43}X_{33}+g_7X_{43}X_{34}X_{44}+g_8M_{00}X'_{01}X'_{10}+g_9M_{22}X'_{21}X'_{12}+g_{10}M_{20}X'_{01}X'_{12}\\
		&+g_{11}M_{02}X'_{21}X'_{10}.
	\end{split}
\end{equation}
In particular, $\partial W/\partial M_{00}=0$ (resp.~$\partial W/\partial M_{22}=0$) gives $X_{00}=X'_{01}X'_{01}$ (resp.~$X_{22}=X'_{21}X'_{12}$). This yields the quiver
\begin{equation}
	\includegraphics[width=10cm]{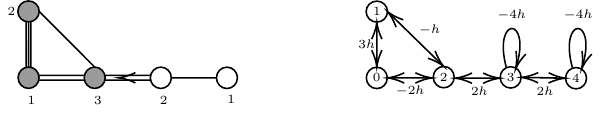}\label{F42}
\end{equation}
with the superpotential
\begin{equation}
	\begin{split}
		W=&g_1X_{02}X_{23}X_{32}X_{20}+g_2X_{23}X_{32}X_{21}X_{12}X_{21}X_{12}+g_3X_{32}X_{23}X_{33}+g_4X_{34}X_{43}X_{33}\\
		&+g_5X_{43}X_{34}X_{44}+g_6X_{01}X_{12}X_{20}+g_7X_{21}X_{10}X_{02},
	\end{split}
\end{equation}
where we have renamed the fields and the couplings. Now we can either dualize the node 0 or dualize the node 2. Let us first dualize the node 0, and the superpotential becomes
\begin{equation}
	\begin{split}
		W=&g_1M_{22}X_{23}X_{32}+g_2X_{23}X_{32}X_{21}X_{12}X_{21}X_{12}+g_3X_{32}X_{23}X_{33}+g_4X_{34}X_{43}X_{33}\\
		&+g_5X_{43}X_{34}X_{44}+g_6M_{21}X_{12}+g_7X_{21}M_{12}+g_8M_{11}X'_{10}X'_{01}+g_9M_{22}X'_{20}X'_{02}\\
		&+g_{10}M_{21}X'_{10}X'_{02}+g_{11}M_{12}X'_{20}X'_{01}.
	\end{split}
\end{equation}
In particular, $\partial W/\partial M_{12}=0$ (resp.~$\partial W/\partial M_{21}=0$) gives $X_{21}=X'_{20}X'_{01}$ (resp.~$X_{12}=X'_{10}X'_{02}$). This yields the quiver
\begin{equation}
	\includegraphics[width=10cm]{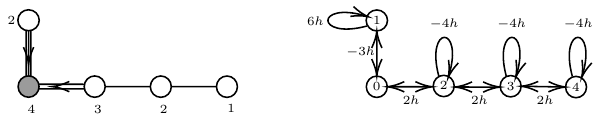}
\end{equation}
with the superpotential
\begin{equation}
	\begin{split}
		W=&g_1X_{22}X_{23}X_{32}+g_2X_{23}X_{32}X_{20}X_{01}X_{10}X_{02}X_{20}X_{01}X_{10}X_{02}+g_3X_{32}X_{23}X_{33}\\
		&+g_4X_{34}X_{43}X_{33}+g_5X_{43}X_{34}X_{44}+g_6X_{10}X_{01}X_{11}+g_7X_{20}X_{02}X_{22},
	\end{split}
\end{equation}
where we have renamed the fields and the couplings. Let us now dualize the node 2 in \eqref{F42}. The superpotential becomes
\begin{equation}
	\begin{split}
		W=&g_1M_{03}M_{30}+g_2M_{13}M_{31}M_{11}+g_3M_{33}X_{33}+g_4X_{34}X_{43}X_{33}+g_5X_{43}X_{34}X_{44}\\
		&+g_6X_{01}M_{10}+g_7M_{01}X_{10}+g_8M_{00}X'_{02}X'_{20}+g_9M_{11}X'_{12}X'_{21}+g_{10}M_{33}X'_{32}X'_{23}\\
		&+g_{11}M_{01}X'_{12}X'_{20}+g_{12}M_{10}X'_{02}X'_{21}+g_{13}M_{03}X'_{32}X'_{20}+g_{14}M_{30}X'_{02}X'_{23}\\
		&+g_{15}M_{13}X'_{32}X'_{21}+g_{16}M_{31}X'_{12}X'_{23}.
	\end{split}
\end{equation}
Integrating out the massive fields yields the quiver
\begin{equation}
	\includegraphics[width=10cm]{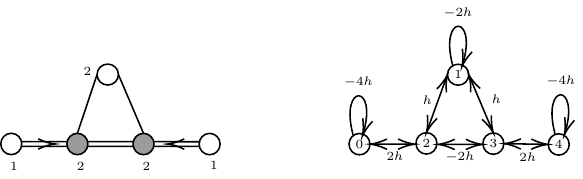}
\end{equation}
with the superpotential
\begin{equation}
	\begin{split}
		W=&g_1X_{13}X_{31}X_{11}+g_2X_{34}X_{43}X_{32}X_{23}+g_3X_{43}X_{34}X_{44}+g_4X_{02}X_{20}X_{00}+g_5X_{12}X_{21}X_{11}\\
		&+g_6X_{32}X_{20}X_{02}X_{23}+g_7X_{32}X_{21}X_{13}+g_8X_{12}X_{23}X_{31},
	\end{split}
\end{equation}
where we have renamed the fields and the couplings. Overall, there are four phases under dualizing the grey nodes.

\paragraph{Example 4: $G(3)^{(1)}$}  Let us analyze all the phases for $G(3)^{(1)}$. Recall that the distinguished case has the quiver
\begin{equation}
	\includegraphics[width=9cm]{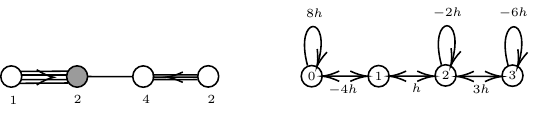}
\end{equation}
with the superpotential
\begin{equation}
	\begin{split}
		W=&g_1X_{01}X_{10}X_{00}+g_2X_{01}X_{12}X_{23}X_{32}X_{21}X_{10}+g_3X_{21}X_{12}X_{22}\\
		&+g_4X_{23}X_{32}X_{22}X_{22}X_{22}+g_5X_{32}X_{23}X_{33}.
	\end{split}
\end{equation}
Dualizing the node 1, the superpotential becomes
\begin{equation}
	\begin{split}
		W=&g_1M_{00}X_{00}+g_2M_{02}X_{23}X_{32}M_{20}+g_3M_{22}X_{22}+g_4X_{23}X_{32}X_{22}X_{22}X_{22}\\
		&+g_5X_{32}X_{23}X_{33}+g_6M_{00}X'_{01}X'_{10}+g_7M_{22}X'_{21}X'_{12}+g_8M_{20}X'_{01}X'_{12}+g_9M_{02}X'_{21}X'_{10}.
	\end{split}
\end{equation}
Integrating out the massive fields yields the quiver
\begin{equation}
	\includegraphics[width=9cm]{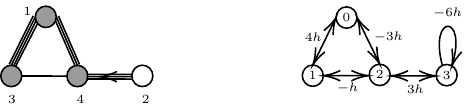}\label{G32}
\end{equation}
with the superpotential
\begin{equation}
	\begin{split}
		W=&g_1X_{02}X_{23}X_{32}X_{20}+g_2X_{23}X_{32}X_{21}X_{12}X_{21}X_{12}X_{21}X_{12}+g_3X_{32}X_{23}X_{33}\\
		&+g_4X_{01}X_{12}X_{20}+g_5X_{21}X_{10}X_{02},
	\end{split}
\end{equation}
where we have renamed the fields and the couplings. Now we can either dualize the node 0 or dualize the node 2. Let us first dualize the node 0, and the superpotential becomes
\begin{equation}
	\begin{split}
		W=&g_1M_{22}X_{23}X_{32}+g_2X_{23}X_{32}X_{21}X_{12}X_{21}X_{12}X_{21}X_{12}+g_3X_{32}X_{23}X_{33}+g_4M_{21}X_{12}\\
		&+g_5X_{21}M_{12}+g_6M_{11}X'_{10}X'_{01}+g_7M_{22}X'_{20}X'_{02}+g_8M_{21}X'_{10}X'_{02}+g_9M_{12}X'_{20}X'_{01}.
	\end{split}
\end{equation}
Integrating out the massive fields yields the quiver
\begin{equation}
	\includegraphics[width=9cm]{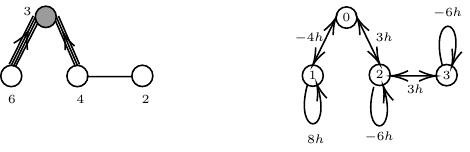}
\end{equation}
with the superpotential
\begin{equation}
	\begin{split}
		W=&g_1X_{22}X_{23}X_{32}+g_2X_{23}X_{32}X_{20}X_{01}X_{10}X_{02}X_{20}X_{01}X_{10}X_{02}X_{20}X_{01}X_{10}X_{02}\\
		&+g_3X_{32}X_{23}X_{33}+g_4X_{10}X_{01}X_{11}+g_5X_{20}X_{02}X_{22},
	\end{split}
\end{equation}
where we have renamed the fields and the couplings. Let us now dualize the node 2 in \eqref{G32}. The superpotential becomes
\begin{equation}
	\begin{split}
		W=&g_1M_{03}M_{30}+g_2M_{13}M_{31}M_{11}M_{11}+g_3M_{33}X_{33}+g_4X_{01}M_{10}+g_5M_{01}X_{10}\\
		&g_6M_{00}X'_{02}X'_{20}+g_7M_{11}X'_{12}X'_{21}+g_8M_{33}X'_{32}X'_{23}+g_9M_{30}X'_{02}X'_{23}+g_{10}M_{03}X'_{32}X'_{20}\\
		&+g_{11}M_{31}X'_{12}X'_{23}+g_{12}M_{13}X'_{32}X'_{21}+g_{13}M_{10}X'_{02}X'_{21}+g_{14}M_{01}X'_{12}X'_{20}.
	\end{split}
\end{equation}
Integrating out the massive fields yields the quiver
\begin{equation}
	\includegraphics[width=9cm]{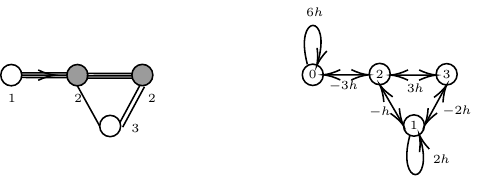}
\end{equation}
with the superpotential
\begin{equation}
	\begin{split}
		W=&g_1X_{13}X_{31}X_{11}X_{11}+g_2X_{02}X_{20}X_{00}+g_3X_{12}X_{21}X_{11}+g_4X_{02}X_{23}X_{32}X_{20}\\
		&+g_5X_{12}X_{23}X_{31}+g_6X_{32}X_{21}X_{13},
	\end{split}
\end{equation}
where we have renamed the fields and the couplings. Then we can dualize the node 3, and superpotential becomes
\begin{equation}
	\begin{split}
		W=&g_1M_{11}X_{11}X_{11}+g_2X_{02}X_{20}X_{00}+g_3X_{12}X_{21}X_{11}+g_4X_{02}M_{22}X_{20}+g_5X_{12}M_{21}\\
		&+g_6X_{21}M_{12}+g_7M_{11}X'_{13}X'_{31}+g_8M_{22}X'_{23}X'_{32}+g_9M_{21}X'_{13}X'_{32}+g_{10}M_{12}X'_{23}X'_{31}.
	\end{split}
\end{equation}
Integrating out the massive fields yields the quiver
\begin{equation}
	\includegraphics[width=9cm]{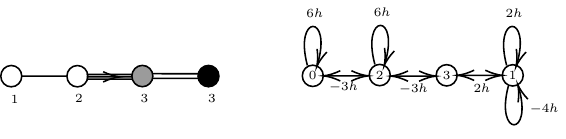}
\end{equation}
with the superpotential
\begin{equation}
	\begin{split}
		W=&g_1X_{11,(2)}X_{11,(1)}X_{11,(1)}+g_2X_{02}X_{20}X_{00}+g_3X_{13}X_{32}X_{23}X_{31}X_{11,(1)}+g_4X_{02}X_{22}X_{20}\\
		&+g_5X_{13}X_{31}X_{11,(2)}+g_6X_{23}X_{32}X_{22},
	\end{split}
\end{equation}
where we have renamed the fields and the couplings. Overall, there are five phases under dualizing the grey nodes.

\section{Connections to $\mathcal{W}$-Algebras}\label{Walgebras}
It is expected that the quiver Yangians are closely related to the VOAs. More specifically, the $\mathcal{W}$-algebras\footnote{More precisely, it should be the universal enveloping algebras $U(\mathcal{W})$ of the $\mathcal{W}$-algebras as the $\mathcal{W}$-algebras are non-associative (with respect to the normal ordered products). The name universal enveloping algebra originates from the Borcherds Lie algebra	of $\mathcal{W}$. See \cite{frenkel1992vertex,matsuo2010quasi}. For brevity, we shall make a slight abuse here.} should be truncations of the quiver Yangians. This was proven in the case of the quiver Yangian for $\mathbb{C}^3$ and the $\mathcal{W}_{1+\infty}$-algebra in \cite{schiffmann2013cherednik,Gaberdiel:2017dbk,Prochazka:2019dvu}. For quiver Yangians associated to (most) $A(m-1|n-1)^{(1)}$, which are known to be the BPS algebras for gauge theories on generalized conifolds, it was shown that they can be truncated to $\mathcal{W}$-algebras of type $\widehat{\mathfrak{gl}}(m|n)$ \cite{Bao:2022jhy}. It is natural to conjecture that the quiver Yangians associated to other affine Dynkin diagrams would also give rise to $\mathcal{W}$-algebras of certain types.

As discussed in \cite{Li:2023zub}, one may compare the characters of the two types of algebras as a check. It could be possible that the truncations arised from these quiver Yangians can be constructed directly from the Miura operators and the underlying Kac-Moody algebras similar to the $\widehat{\mathfrak{gl}}(m|n)$ case as we will mention below. However, there could also be more general $\mathcal{W}$-algebras whose generators do not completely follow this construction. This is already known for the finite cases through the quantum Drinfeld-Sokolov reduction \cite{Bouwknegt:1992wg}. See also \cite{Keller:2011ek} for some explicit constructions of the generators, as well as the comparison between the norms of the Gaiotto-Whittaker vectors and certain one-instanton partition functions.

Indeed, there exist twisted Yangians in the finite cases which can be truncated to different $\mathcal{W}$-algebras as studied in \cite{ragoucy2001twisted,brown2009twisted}. Analogously, we expect that the twisted quiver Yangians discussed above can also have truncations that are $\mathcal{W}$-algebras. As mentioned before, given a quiver Yangian, there should exist a surjective homomorphism to the universal enveloping algebra of the underlying current algebra. Following the A-type cases which were found in \cite{guay2007affine,kodera2021guay,ueda2019affine,ueda2022surjectivity}, we shall call this map the evaluation map ev. As a twisted quiver Yangian is constructed as a subalgebra of the quiver Yangian, we conjecture that there is a surjective homomorphism from the twisted quiver Yangian $\widetilde{\mathcal{Y}}\left(\widehat{\mathfrak{l}}\right)$ given by
\begin{equation}
	\Phi=\left(\bigotimes_{k=1}^l\text{ev}\right)\cdot\Delta^l
\end{equation}
such that the image is a $\mathcal{W}$-algebra with symmetry $\mathfrak{l}$. Here, $l$ indicates the level of the truncation.

Let us make some comments. This surjection $\Phi$ actually maps elements to the completion $\widehat{U}\left(\widehat{\mathfrak{g}}\right)^{\otimes l}$. However, there should be an injective homomorphism from the $\mathcal{W}$-algebra embedded into it induced by the Miura transformation \cite{arakawa2017explicit,arakawa2017introduction,ueda2021twisted}. Moreover, the $\mathcal{W}$-algebra should have its spin 1 generatos being the generators of $\widehat{\mathfrak{l}}$. However, the higher spin generators could be more intricate.

Following the BPS/CFT correspondence, finding the connections between (twisted) quiver Yangians and $\mathcal{W}$-algebras could give us more hints on the study of BPS states. Of course, this does not mean that these Yangian algebras have to be the BPS algebras for certain gauge theories (and we have mentioned some subtleties in the previous section). For instance, the truncations of the quiver Yangians for generalized conifolds have the construction similar to the ones for some affine super Yangians in \cite{ueda2019affine,ueda2022affine}. However, the two types of Yangians are not isomorphic. Likewise, the twisted affine Yangians in \cite{ueda2021twisted} are slightly different from the definition here. Nevertheless, different Yangian algebras may still give rise to various $\mathcal{W}$-algebras. It could also be possible that one needs to modify the above (twisted) quiver Yangians to obtain the $\mathcal{W}$-algebras as truncations and/or to find the BPS algebras.

As aforementioned, the generators of the $\mathcal{W}$-algebras may not follow directly from the expansion of the product of Miura operators. Nevertheless, one may always construct the generators using the definition of $\mathcal{W}$-algebras. Recall that a $\mathcal{W}$-algebra is defined via the BRST cohomology of some larger vertex algebra. The elements that are annihilated by the odd derivation then form the $\mathcal{W}$-algebra.

In the rest of this section, we shall consider the screening currents of the $\mathcal{W}$-algebras such that the generators give the intersection of the kernels of these screening currents \cite{Litvinov:2016mgi}. For the case of $\mathbb{C}^3$, this was studied in \cite{Prochazka:2018tlo}. Here, we shall focus on the generalized conifold cases as an illustration. In particular, the generators are already known in such cases. We shall exploit these generators by considering their free field realization, and we will find the screening currents of the $\mathcal{W}$-algebras. We hope that  this could shed light on our understanding of any general $\mathcal{W}$-algebras and help us learn more about their connections to (twisted) quiver Yangians.

As the $\mathcal{W}$-algebra is associated to the generalized conifold $xy=z^mw^n$ (and has the symmetry of A-type), we shall refer to it as the $\mathcal{W}_{m|n\times l}$-algebra for brevity. The construction from the odd derivation can actually be found for example in \cite{arakawa2017explicit,ueda2022affine}. Nevertheless, it would be more convenient to use the Miura operators \cite{Rapcak:2019wzw,Eberhardt:2019xmf}
\begin{equation}
	\mathcal{L}^{(x)}_r=\kappa\partial+J^{(x)}_r
\end{equation}
to write down the generators. Here, the superscript $x$ indicates the type of the truncation of $\mathcal{W}_{m|n\times\infty}$ as we will discuss shortly, and $J^{(x)}$ is the supermatrix with $m$ bosonic and $n$ fermionic rows/columns. The entries of the matrix are the currents generating the $\widehat{\mathfrak{gl}}(m|n)_{\kappa}$ algebra:
\begin{equation}
	\begin{split}
		J^{(x)}_{r,ab}(z)J^{(x)}_{s,cd}(w)\sim&\frac{(-1)^{p(b)p(c)}\kappa\delta_{rs}\delta_{ad}\delta_{cb}+\delta_{rs}\delta_{ab}\delta_{cd}}{(z-w)^2}\\
		&+\frac{(-1)^{p(a)p(b)+p(c)p(d)+p(c)p(b)}\delta_{rs}\delta_{ad}J^{(x)}_{cb}(w)-(-1)^{p(b)p(c)}\delta_{rs}\delta_{cb}J^{(x)}_{ad}(w)}{z-w},
	\end{split}
\end{equation}
where we have used $p(a)$ to denote the parity of $a$ so as to distinguish it with $|a|$ in the quiver Yangians. The generators $U_i$ of $\mathcal{W}_{m|n\times l}$ in the Miura basis\footnote{The generators in a different basis, namely the so-called primary basis, can be found in \cite{Creutzig:2019qos}.} can then be obtained from
\begin{equation}
	\mathcal{L}^{(x)}_1\mathcal{L}^{(x)}_2\dots\mathcal{L}^{(x)}_l=(\kappa\partial)^l+U_1(\kappa\partial)^{l-1}+\dots+U_l.
\end{equation}
As shown in \cite{ueda2022affine}, the currents $U_1$, $U_2$ at spin 1 and 2 are sufficient to generate the whole algebra. They have the expressions
\begin{align}
	&U_{1,ab}=\sum_{r=1}^lJ^{(x)}_{r,ab},\label{U1}\\
	&U_{2,ab}=\sum_{r=1}^{l-1}\sum_{s=r+1}^l\sum_c\left(J^{(x)}_{r,ac}J^{(x)}_{s,cb}\right)+\kappa\sum_{r=1}^l(r-1)\partial J^{(x)}_{r,ab},\label{U2}
\end{align}
where the normal ordering is denoted as $(\dots)$. Their OPEs can be found for example in \cite{Rapcak:2019wzw,Eberhardt:2019xmf}.

Such algebra can arise from the junction of supersymmetric interfaces in certain 4d gauge theory. It admits a brane web construction \cite{Prochazka:2017qum,Prochazka:2018tlo,Rapcak:2019wzw}, extending the vertex algebra at the corner for the $\mathbb{C}^3$ case \cite{Gaiotto:2017euk}. In particular, the coupling $\Psi$ of the gauge theory is related to the level of the underlying Kac-Moody algebra by $\Psi=\kappa+\mathtt{h}^{\vee}$, where $\mathtt{h}^{\vee}=m-n$ is the dual Coxeter number. In our following discussions, we shall always assume $\Psi\neq0$. Besides, these $\mathcal{W}$-algebras also have a coset construction from the dual coset CFTs.

As shown in \cite{Bao:2022jhy}, different ways of permuting the bosonic and fermionic rows/columns (which correspond to permutations of parity sequences) give isomorphic $\mathcal{W}$-algebras. Therefore, we shall focus on the distinguished case below for convenience, where $p(a)=0$ for $1\leq a\leq m$ and $p(a)=1$ for $m+1\leq a\leq m+n$. In the same manner as the construction for the $\widehat{\mathfrak{gl}}(m|n)_{\kappa}$ current algebra in \cite{Yang:2007zzb}, the free field realization of $\mathcal{W}_{m|n\times l}$ can be given by the following content:
\begin{itemize}
	\item $\frac{1}{2}l(m(m-1)+n(n-1))$ copies of (bosonic) $\beta\gamma$ systems, $\{\beta_{r,ab},\gamma_{r,ab}\}$ ($1\leq r\leq l$, $1\leq a<b\leq m$), $\{\overline{\beta}_{r,ab},\overline{\gamma}_{r,ab}\}$ ($1\leq r\leq l$, $1\leq a<b\leq n$);
	\item $lmn$ copies of (fermionic) $bc$ systems,  $\{\psi_{r,ab},\psi^{\dagger}_{r,ab}\}$ ($1\leq r\leq l$, $1\leq a\leq m$, $1\leq b\leq n$);
	\item $l(m+n)$ scalars, $\phi_{r,a}$ ($1\leq r\leq l$, $1\leq a\leq m+n$).
\end{itemize}
They have the OPEs
\begin{equation}
	\begin{split}
		&\beta_{r,ab}(z)\gamma_{s,cd}(w)=-\gamma_{s,cd}(z)\beta_{r,ab}(w)\sim\frac{\delta_{rs}\delta_{ac}\delta_{bd}}{z-w},\quad\overline{\beta}_{r,ab}(z)\overline{\gamma}(w)_{s,cd}=-\overline{\gamma}_{s,cd}(z)\overline{\beta}_{r,ab}(w)\sim\frac{\delta_{rs}\delta_{ac}\delta_{bd}}{z-w},\\
		&\psi_{r,ab}(z)\psi^{\dagger}_{s,cd}(w)=\psi^{\dagger}_{s,cd}(z)\psi_{r,ab}(w)\sim\frac{\delta_{rs}\delta_{ac}\delta_{bd}}{z-w},\quad\phi_{r,a}(z)\phi_{s,b}\sim\delta_{rs}\delta_{ab}\log(z-w).
	\end{split}
\end{equation}
Then the free field realization of the $\mathcal{W}$-algebra can be obtained by
\begin{align}
	&J^{(x)}_{r,a,a+1}=-\sum_{b=1}^{a-1}(\gamma_{r,ba}\beta_{s,b,a+1})-\beta_{r,a,a+1},\quad1\leq a\leq m-1,\\
	&J^{(x)}_{r,a,a+1}=\sum_{b=1}^{m-1}(\gamma_{r,bm}\psi_{r,b,1})+\psi_{r,m,1},\quad a=m,\\
    &J^{(x)}_{r,m+a,m+a+1}=\sum_{b=1}^m\left(\psi^{\dagger}_{r,ba}\psi_{r,b,a+1}\right)+\sum_{b=1}^{a-1}\left(\overline{\gamma}_{r,ba}\overline{\beta}_{r,b,a+1}\right)+\overline{\beta}_{r,a,a+1},\quad1\leq a\leq n-1,\\
    &J^{(x)}_{r,aa}=-\sum_{b=1}^{a-1}(\gamma_{r,ba}\beta_{r,ba})+\sum_{b=a+1}^m(\gamma_{r,ab}\beta_{r,ab})+\sum_{b=1}^n\left(\psi^{\dagger}_{ab}\psi_{ab}\right)-\Psi^{1/2}\partial\phi_{r,a},\quad1\leq a\leq m,\\
    &J^{(x)}_{r,m+a,m+a}=\sum_{b=1}^{a-1}(\overline{\gamma}_{r,ba}\overline{\beta}_{r,ba})-\sum_{b=a+1}^n(\overline{\gamma}_{r,ab}\overline{\beta}_{r,ab})+\sum_{b=1}^m\left(\psi^{\dagger}_{ba}\psi_{ba}\right)+\text{i}\Psi^{1/2}\partial\phi_{r,m+a},\quad1\leq a\leq n,\\
    &J^{(x)}_{r,a+1,a}=-\sum_{b=1}^{a-1}(\gamma_{r,b,a+1}\beta_{r,ba})+\sum_{b=a+2}^m(\gamma_{r,ab}\beta_{r,a+1,b})+\sum_{b=1}^n\left(\psi^{\dagger}_{r,ab}\psi_{r,a+1,b}\right)+\sum_{b=a+1}^m(\gamma_{r,a,a+1}\gamma_{r,ab}\beta_{r,ab})\\
    &\qquad\qquad\quad+\sum_{b=1}^n\left(\gamma_{r,a,a+1}\psi^{\dagger}_{r,ab}\psi_{r,ab}\right)-\sum_{b=a+2}^m(\gamma_{r,a,a+1}\gamma_{r,a+1,b}\beta_{r,a+1,b})-\sum_{b=1}^n\left(\gamma_{r,a,a+1}\psi^{\dagger}_{r,a+1,b}\psi_{r,a+1,b}\right)\\
    &\qquad\qquad\quad-\Psi^{1/2}(\gamma_{r,a,a+1}\partial\phi_{r,a})+\Psi^{1/2}(\gamma_{r,a,a+1}\partial\phi_{r,a+1})-(\kappa+a-1)\partial\gamma_{r,a,a+1},\quad1\leq a\leq m-1,\\
    &J^{(x)}_{r,a+1,a}=-\sum_{b=1}^{m-1}\left(\psi^{\dagger}_{r,b,1}\beta_{r,bm}\right)-\sum_{b=2}^n(\psi_{r,mb}\overline{\beta}_{r,1,b})+\sum_{b=2}^n\left(\psi^{\dagger}_{r,m,1}\psi^{\dagger}_{r,mb}\psi_{r,mb}\right)+\sum_{b=2}^n\left(\psi^{\dagger}_{r,m,1}\overline{\gamma}_{r,1,b}\overline{\beta}_{r,1,b}\right)\\
    &\qquad\qquad\quad-\Psi^{1/2}\left(\psi^{\dagger}_{r,m,1}\partial\phi_{r,m}\right)-\text{i}\Psi^{1/2}\left(\psi^{\dagger}_{r,m,1}\partial\phi_{r,m+1}\right)-(\kappa+m-1)\partial\psi^{\dagger}_{r,m,1},\quad a=m,\\
    &J^{(x)}_{r,m+a+1,m+a}=\sum_{b=1}^m\left(\psi^{\dagger}_{r,b,a+1}\psi_{r,ba}\right)+\sum_{b=1}^{a-1}(\overline{\gamma}_{r,b,a+1}\overline{\beta}_{r,ba})-\sum_{b=a+2}^n(\overline{\gamma}_{r,ab}\overline{\beta}_{r,a+1,b})-\sum_{b=a+1}^n(\overline{\gamma}_{r,a,a+1}\overline{\gamma}_{r,ab}\overline{\beta}_{r,ab})\\
    &\qquad\qquad\quad+\sum_{b=a+2}^n(\overline{\gamma}_{r,a,a+1}\overline{\gamma}_{r,a+1,b}\overline{\beta}_{r,a+1,b})+\text{i}\Psi^{1/2}(\overline{\gamma}_{r,a,a+1}\partial\phi_{r,m+a})-\text{i}\Psi^{1/2}(\overline{\gamma}_{r,a,a+1}\partial\phi_{r,m+a+1})\\
    &\qquad\qquad\quad-(\kappa+m+a+1)\partial\overline{\gamma}_{r,a,a+1},\quad1\leq a\leq n-1.
\end{align}
The other currents $J^{(x)}_{ab}$ can be directly obtained from the OPEs of these currents. For instance, $J^{(x)}_{a,a\pm2}(w)$ is the residue of the OPE $-(-1)^{p(a\pm1)}J^{(x)}_{a,a\pm1}(z)J^{(x)}_{a\pm1,a\pm2}(w)$, and this procedure can be done inductively. It is then straightforward to write down the generators in terms of the free fields using \eqref{U1} and \eqref{U2}.

One can also express the stress tensor\footnote{This agrees with the stress tensor for $n=0$ given in \cite{Eberhardt:2019xmf}.}
\begin{equation}
	T(z)=\sum_{a,b}\frac{(-1)^{p(a)}}{2\Psi}(U_{1,ab}U_{1,ba})+\sum_a\frac{(-1)^{p(a)}\kappa(l-1)}{2\Psi}\partial U_{1,aa}-\sum_a\frac{(-1)^{p(a)}}{\Psi}U_{2,aa}
\end{equation}
satisfying
\begin{align}
		&T(z)T(w)\sim\frac{c/2}{(z-w)^4}+\frac{2T(w)}{(z-w)^2}+\frac{\partial T(w)}{z-w},\\
		&T(z)U_{1,ab}(w)\sim\frac{U_{1,ab}(w)}{(z-w)^2}+\frac{\partial U_{1,ab}}{z-w},\\
		&T(z)U_{2,ab}(w)\sim\dots+\frac{2U_{2,ab}}{(z-w)^2}+\frac{\partial U_{1,ab}}{z-w}
\end{align}
in terms of these free fields. The central charge is given by
\begin{equation}
	c=\frac{(m-n)l}{\Psi}\left(1+\kappa(m-n)-\left(l^2-1\right)\kappa^2\right).
\end{equation}

The screening currents $S$ of the $\mathcal{W}$-algebra satisfy \cite{Litvinov:2016mgi}
\begin{equation}
	[S,U_{i,ab}(z)]:=\oint_{C_z}V(w)U_{i,ab}(z)\text{d}w=0,
\end{equation}
where $C_z$ is a closed contour encircling $z$. In other words, the $\mathcal{W}$-algebra can be defined as the intersection of the kernels of all such screening currents. Therefore, we would like to find the vertex operators $V(w)$ such that the residues vanish in the OPEs of $U_{i,ab}(z)V(w)$. However, unlike the $\mathcal{W}$-algebras studied in \cite{Litvinov:2016mgi}, we find that the vertex operators satisfying this do not necessarily have conformal dimension 1, that is, $\mathtt{\Delta}$ may not be 1 in the OPE
\begin{equation}
	T(z)V(w)\sim\frac{\mathtt{\Delta}V(w)}{(z-w)^2}+\frac{\partial V(w)}{z-w}.
\end{equation}

To find such vertex operators, it would be convenient to consider the bosonization of the above free fields. We have
\begin{equation}
	\beta=-\left(\partial \text{e}^{-\text{i}\chi}\text{e}^{-\xi}\right),\quad\gamma=\left(\text{e}^{\text{i}\chi+\xi}\right),\quad\overline{\beta}=-\left(\partial \text{e}^{-\text{i}\overline{\chi}}\text{e}^{-\overline{\xi}}\right),\quad\overline{\gamma}=\left(\text{e}^{\text{i}\overline{\chi}+\overline{\xi}}\right),\quad\psi=\left(\text{e}^{\text{i}\eta}\right),\quad\psi^{\dagger}=\left(\text{e}^{-\text{i}\eta}\right),
\end{equation}
where we have suppressed the subscripts for brevity. They satisfy the OPEs\footnote{In fact, we may further reduce the number of the free scalars by taking $\eta$ to be $\chi$ or $\overline{\chi}$. However, we shall not make this identification here.}
\begin{equation}
	\mathtt{x}(z)\mathtt{x}(w)\sim-\log(z-w)\quad(\mathtt{x}=\chi,\xi,\overline{\chi},\overline{\xi},\eta).
\end{equation}
As mentioned above, we only need to compute the OPEs of the vertex operators with $U_1$ and $U_2$. Moreover, it suffices to consider $l=2$ as higher $l$ with more $J^{(x)}_r$ would not give new conditions. Alternatively, similar to \cite{Prochazka:2018tlo}, we may think of the Miura operators as nodes ordered on a line with insertions of the corresponding screening charges between two neighbouring nodes.

\begin{remark}
	For $(m,n)=(2,0),(1,1)$, the corresponding quiver Yangians are not our main focus in this paper. In particular, the minimalistic presentations may involve modes at level $2$. Likewise, the minimal sets of generators of the $\mathcal{W}$-algebras may include those of spin greater than $2$. The surjective maps from the quiver Yangian to the $\mathcal{W}$-algebras would therefore require further study. Nevertheless, we shall not address these points here and just consider the results from $U_1$, $U_2$. Similarly, for $(m,n)=(1,0)$, we need higher mode/spin generators as known in literature. However, we find that the screening current below in such case is consistent with that in \cite{Prochazka:2018tlo}.
\end{remark}

With the above bosonization, the vertex operator can be written as\footnote{Of course, we would rule out the trivial case with all $k_{\mathtt{x}}=0$.}
\begin{equation}
	V(z)=\left(\exp\left(\sum_{\mathtt{x}}\text{i}k_{\mathtt{x}}\mathtt{x}\right)\right)
\end{equation}
for $\mathtt{x}=\chi,\xi,\overline{\chi},\overline{\xi},\eta$ with various subscripts. To write its conformal dimension, let us introduce the shorthand notation $\widetilde{\mathtt{\Delta}}(\underline{\mathtt{x}})$ which is equal to
\begin{itemize}
	\item \begin{equation}
		\Psi^{1/2}\sum_{a=1}^{m+n}k_{\phi_{1,a}}^2+\sum_{a=1}^m(\Psi-2a+1)k_{\phi_{1,a}}+\text{i}\sum_{a=1}^n(\Psi-2n+2a-1)k_{\phi_{1,m+a}}
	\end{equation}
    when $\underline{\mathtt{x}}=\{\phi_{1,a}|1\leq a\leq m+n\}$;
    \item \begin{equation}
    	\Psi^{1/2}\sum_{a=1}^{m+n}k_{\phi_{2,a}}^2-\sum_{a=1}^m(\Psi+2a-1)k_{\phi_{2,a}}-\text{i}\sum_{a=1}^n(\Psi+2n-2a+1)k_{\phi_{2,m+a}}
    \end{equation}
    when $\underline{\mathtt{x}}=\{\phi_{2,a}|1\leq a\leq m+n\}$;
    \item \begin{equation}
    	\Psi^{1/2}\sum_{\mathtt{x}}\left(k_{\mathtt{x}}^2+k_{\mathtt{x}}\right)
    \end{equation}
    when $\underline{\mathtt{x}}=\{\chi_{r,ab}|r=1,2,~1\leq a<b\leq m\}$ or $\underline{\mathtt{x}}=\{\overline{\chi}_{r,ab}|r=1,2,~1\leq a<b\leq n\}$;
    \item \begin{equation}
    	\Psi^{1/2}\sum_{\mathtt{x}}\left(k_{\mathtt{x}}^2-\text{i}k_{\mathtt{x}}\right)
    \end{equation}
    when $\underline{\mathtt{x}}=\{\xi_{r,ab}|r=1,2,~1\leq a<b\leq m\}$ or $\underline{\mathtt{x}}=\{\overline{\xi}_{r,ab}|r=1,2,~1\leq a<b\leq n\}$;
    \item \begin{equation}
    	\Psi^{1/2}\sum_{\mathtt{x}}\left(k_{\mathtt{x}}^2-k_{\mathtt{x}}\right)
    \end{equation}
    when $\underline{\mathtt{x}}=\{\eta_{r,ab}|r=1,2,~1\leq a\leq m,~1\leq b\leq n\}$.
\end{itemize}
Then the conformal dimension is
\begin{equation}
	\mathtt{\Delta}=\frac{1}{2\Psi^{1/2}}\sum_{\underline{\mathtt{x}}}\widetilde{\mathtt{\Delta}}(\underline{\mathtt{x}}).
\end{equation}

To get the screening currents, the coefficients $k_{\mathtt{x}}$ need to satisfy the following conditions. We find that
\begin{align}
	&\sum_{a=1}^mk_{\phi_{r,a}}+\text{i}\sum_{a=m+1}^{m+n}k_{\phi_{r,a}}=0\quad(r=1,2),\\
	&-\text{i}\sum_{b=1}^{a-1}k_{\xi_{r,ba}}+\text{i}\sum_{b=a+1}^{m-1}k_{\xi_{r,ab}}+\sum_{b=1}^nk_{\eta_{r,ab}}-\Psi^{1/2}k_{\phi_{r,a}}=0\quad(r=1,2,~1\leq a\leq m),\\
	&\text{i}\sum_{b=1}^{a-m-1}k_{\overline{\xi}_{r,b,a-m}}-\text{i}\sum_{b=a-m+1}^nk_{\overline{\xi}_{r,a-m,b}}+\sum_{b=1}^nk_{\eta_{r,a-m,b}}+\text{i}\Psi^{1/2}k_{\phi_{r,a}}=0,\quad(r=1,2,~m+1\leq a\leq m+n)
\end{align}
for generic $k_{\mathtt{x}}$. In particular, there is no constraint on $\chi$ and $\overline{\chi}$ from them. Notice that these conditions are not all linearly independent. It is also worth noting that the terms in the second and third lines actually correspond to the terms in $J^{(x)}_{r,aa}$ ($r=1,2,~1\leq a\leq m+n$).

Let us explain what ``generic'' means for these coeffcients. It turns out that the coefficients for $\mathtt{x}$ other than $\phi$ should satisfy some extra conditions. They are actually encoded by terms involving these $\mathtt{x}$ of degree no greater than 2 (resp.~3) in $U_{1,ab}$ ($a\neq b$) (resp.~$U_{2,ab}$). For instance\footnote{We have checked this up to $m+n=4$ using the \texttt{Mathematica} package \texttt{math.ope} \cite{Fujitsu:1994np}.},
\begin{align}
	&\pm\partial\gamma_{ab},\pm\partial\gamma_{ab}\beta_{ab}:~\pm(k_{\chi_{r,ab}}-\text{i}k_{r,\xi_{r,ab}})\neq0,-1,\\
	&\pm\partial\psi^{\dagger}_{ab},\pm\psi^{\dagger}_{ab}\psi_{ab}:~\pm(k_{\eta_{r,ab}})\neq0,-1\\
	&\pm\gamma_{ab}\beta_{cd}:~\pm(k_{\chi_{r,ab}}-\text{i}k_{r,\xi_{r,ab}})\mp(k_{\chi_{r,cd}}-\text{i}k_{r,\xi_{r,cd}})\neq0,-1,\\
	&\pm\psi^{\dagger}_{cd}\beta_{ab}:~\pm(k_{\chi_{r,ab}}-\text{i}k_{\xi_{r,ab}})\mp k_{\eta_{r,ab}}\neq0,-1,\\
	&\pm\psi_{cd}\overline{\beta}_{ab}:~\pm(k_{\chi_{r,ab}}-\text{i}k_{\xi_{r,ab}})\mp k_{\eta_{r,ab}}\neq0,-1,\\
	&\pm\gamma_{ab}\psi_{cd}:~\pm(k_{\chi_{r,ab}}-\text{i}k_{\xi_{r,ab}})\mp k_{\eta_{r,ab}}\neq0,-1.
\end{align}

Let us make some more comments here. As studied in \cite{Rapcak:2019wzw}, there can be different types of truncations of $\mathcal{W}_{m|n\times\infty}$ corresponding to different divisors in the CY$_3$/web diagram similar to the $\mathbb{C}^3$ case. We have discussed the $x$-algebras while there are $y$-, $z$- and $w$-algebras as well. Together they play the role as building blocks of the general truncations $x^{N_1}y^{N_2}z^{N_3}w^{N_4}$-algebras.

For the $y$-algebra, we have $J^{(y)}$ with the OPE
\begin{equation}
	\begin{split}
		J^{(y)}_{r,ab}(z)J^{(y)}_{s,cd}(w)\sim&\frac{(-1)^{p(b)p(c)}\widetilde{\kappa}\delta_{rs}\delta_{ad}\delta_{cb}+\frac{\widetilde{\kappa}}{\kappa}\delta_{rs}\delta_{ab}\delta_{cd}}{(z-w)^2}\\
		&+\frac{(-1)^{p(a)p(b)+p(c)p(d)+p(c)p(b)}\delta_{rs}\delta_{ad}J^{(y)}_{cb}(w)-(-1)^{p(b)p(c)}\delta_{rs}\delta_{cb}J^{(y)}_{ad}(w)}{z-w}
	\end{split}
\end{equation}
with a different normalization of the diagonal $\widehat{\mathfrak{gl}}(1)$. Here, the level is given by $\widetilde{\kappa}=-\Psi=-\kappa-\mathtt{h}^{\vee}$. The free field realization should be similar to the one of the $x$-algebra. For the $z$-algebra, one considers the $\beta\gamma$-systems $\{\beta_{r,a},\gamma_{r,a}\}$ ($1\leq a\leq m$) and the $bc$-systems $\{\psi_{r,a},\psi^{\dagger}_{r,a}\}$ ($m+1\leq a\leq m+n$). Then
\begin{equation}
	J^{(z)}_{r,ab}=(\mathtt{x}_{r,a}\mathtt{y}_{r,b})+\frac{1}{\kappa+1}\delta_{ab}J_r,
\end{equation}
where $\mathtt{x}=\beta,\psi$, $\mathtt{y}=\gamma,\psi^{\dagger}$ and $J$ satisfies
\begin{equation}
	J_r(z)J_s(w)\sim\frac{\kappa\delta_{rs}+\delta_{rs}}{(z-w)^2}.
\end{equation}
Likewise, for the $w$-algebra, we have
\begin{equation}
	J^{(w)}_{r,ab}=-(\mathtt{x}_{r,a}\mathtt{y}_{r,b})+\frac{\text{i}}{\kappa+1}\delta_{ab}J_r.
\end{equation}

The Miura operator reads
\begin{equation}
	\mathcal{L}^{(\omega)}=(\kappa\partial)^{\nu_{\omega}}+U^{(\omega)}_1(\kappa\partial)^{\nu_{\omega}-1}+U^{(\omega)}_2(\kappa\partial)^{\nu_{\omega}-2}+\dots,
\end{equation}
where $\omega=x,y,z,w$. The expressions of $U^{(\omega)}_i$ can be found in \cite{Rapcak:2019wzw}. In particular, we now have the pseudo-differential operator that acts as
\begin{equation}
	\partial^{\nu}F(u)=\sum_{k=0}^{\infty}\frac{(-1)^k(-\nu)_k}{k!}\left(\partial^kF(u)\right)\partial^{\nu-k}
\end{equation}
for any function $F(u)$, and $(v)_k=(v+k-1)\dots(v+k)v$ is the Pochhammer symbol. Therefore, unlike the $x$-algebras where the generators terminate at spin $l$, the $y$-/$z$-/$w$-algebras have infinitely many $U_i$ for truncations at any level $l$. Denote the equivariant parameters of the associated CY$_3$ as $\epsilon_1,\epsilon_2$. Then
\begin{equation}
	\nu_x=1,\quad\nu_y=\frac{\widetilde{\kappa}}{\kappa}=-\frac{\epsilon_2}{\mathtt{h}^{\vee}\epsilon_1+\epsilon_2},\quad\nu_z=\frac{\epsilon_1}{\mathtt{h}^{\vee}\epsilon_1+\epsilon_2},\quad\nu_w=-\frac{\epsilon_1}{\mathtt{h}^{\vee}\epsilon_1+\epsilon_2}.
\end{equation}
For the general $x^{N_1}y^{N_2}z^{N_3}w^{N_4}$-algebra, the generators can be obtained from
\begin{equation}
	\mathcal{L}^{(x)}_1\dots\mathcal{L}^{(x)}_{N_1}\mathcal{L}^{(y)}_{N_1+N_2}\dots\mathcal{L}^{(z)}_{N_1+N_2+N_3}\dots\mathcal{L}^{(w)}_{N_1+N_2+N_3+N_4}=(\kappa\partial)^N+U_1(\kappa\partial)^{N-1}+U_2(\kappa\partial)^{N-2}+\dots,
\end{equation}
where $N=N_1\nu_x+N_2\nu_y+N_3\nu_z+N_4\nu_z$. The expressions for $U_1$, $U_2$ in terms of $U^{(\omega)}_i$ can be found in \cite[(5.47)]{Rapcak:2019wzw}.

With the free field realizations, one may obtain the screening currents for general truncations in a similar manner as above. Moreover, it was conjectured in \cite{Rapcak:2019wzw} that $\mathcal{L}^{(x)}\mathcal{L}^{(y)}=\left(\mathcal{L}^{(z)}\right)^m\left(\mathcal{L}^{(w)}\right)^n$, which is in line with the defining equation $xy=z^mw^n$ of the associated generalized conifold. As a preliminary check, we have
\begin{equation}
	\nu_x+\nu_y=m\nu_z+n\nu_z=\frac{\mathtt{h}^{\vee}\epsilon_1}{\mathtt{h}^{\vee}\epsilon_1+\epsilon_2}.
\end{equation}
To verify this conjecture, one way is to compare the OPEs of the generators on both sides. Recall that $U^{(x)}_1$ and $U^{(x)}_2$ are sufficient to generate the whole $x$-algebra. If the other types are also finitely generated and finitely presented, then it could be possible that one only needs to check the OPEs at finitely many orders. Alternatively, one may also consider the free field realizations of these algebras and compare their components. However, this is still non-trivial and we leave this to future work.

In \cite{Creutzig:2019wfe}, some $\mathcal{W}$-algebras with orthosymplectic symmetries were obtained via Hamiltonian reductions. They can be realized as the asymptotic symmetries of higher spin gravities. Moreover, one may also consider the $\mathcal{W}$-algebras associated to the web diagram of $\mathbb{C}^3/(\mathbb{Z}_2\times\mathbb{Z}_2)$ which is related to $D(2,1;\alpha)$. We expect that their free field realizations can be obtained in the same manner as the ones in \cite{Yang:2008hd,Yang:2008ghr,Yang:2008vb,Chen:2011us}.

\section{Toroidal and Elliptic Algebras}\label{toroidalelliptic}
Following \cite{Galakhov:2021vbo}, the (rational) quiver Yangians can be generalized to toroidal and elliptic quiver algebras. This hierarchical construction of the elliptic/trigonometric/rational algebras should also be in line with their expected connections to integrable systems. Physically, these algebras can be realized by 3-/2-/1-dimensional quantum field theories with four supercharges along dimensional reductions. As a result, we have the extra parameters $\beta$ for the radius of $S^1$ and $q=\text{e}^{2\pi i\tau}$ for the squared nome of the torus in the toroidal and elliptic algebras respectively. Taking $q\rightarrow0$ yields the toroidal algebra from an elliptic algebra while further taking the rational limit $\beta\rightarrow0$ gives rise to the rational algebra.

In this subsection, we shall briefly mention such extensions. Given the quivers at hand from the above discussions, it is immediate to write their toroidal and elliptic algebras. After recalling the definitions, our main purpose is to construct a free field realizations for any quivers.

The elliptic/trigonometric/rational hierarchy is captured by the function $\zeta$ in the bond factor \eqref{bondfactor}:
\begin{equation}
	\zeta(z)=\begin{cases}
		z,&\text{rational}\\
		\text{Sin}_{\beta}(z):=2\sinh\frac{\beta z}{2}=Z^{1/2}-Z^{-1/2},&\text{trigonometric}\\
		\Theta_q(z):=-Z^{-1/2}\theta_q(z)=(Z^{1/2}-Z^{-1/2})\prod\limits_{k=1}^{\infty}(1-Z^{-1}q^k)(1-Zq^k),&\text{elliptic}
	\end{cases},
\end{equation}
where $\theta_q(z)=(Z;q)_{\infty}(qZ^{-1};q)_{\infty}$ in terms of the $q$-Pochhammer symbol. Here, $Z=\text{e}^{\beta z}$. In this subsection, the parameters in the upper case will be likewise related to those in the lower cases, such as $(H_i,h_i)$, $(C,c)$, and we shall henceforth use them in the arguments of the functions/currents interchangeably. This function $\zeta$ explains the nomenclature of the algebras.

As $\zeta$ always satisfies $\zeta(z)=-\zeta(-z)$, we have the reciprocity condition $\varphi^{a\Leftarrow b}(z)\varphi^{b\Leftarrow a}(-z)=1$. To get rid of the half-integer powers in the spectral parameters, we will use the balanced bond factor:
\begin{equation}
	\phi^{a\Leftarrow b}(z,w)=(ZW)^{\frac{\mathfrak{t}}{2}\chi_{ab}}\varphi^{a\Leftarrow b}(z-w),
\end{equation}
where $\mathfrak{t}$ is 1 for the toroidal and elliptic cases (but vanishes for the rational case), and $\chi_{ab}=|a\rightarrow b|-|b\rightarrow a|$ is the chirality. From the reciprocity condition, we get
\begin{equation}
	\phi^{a\Leftarrow b}(z,w)\phi^{b\Leftarrow a}(w,z)=1.
\end{equation}
Moreover, we have
\begin{equation}
	\phi^{a\Leftarrow b}(z+s,w)=s^{\mathfrak{t}\chi_{ab}}\phi^{a\Leftarrow b}(z,w-s).
\end{equation}

The relations of the three types of the quiver Yangians can then be written in a unified way as
\begin{align}
	&\uppsi^{(a)}_{\pm}(z)\psi^{(b)}_{\pm}(w)\simeq C^{\pm\mathfrak{t}\chi_{ab}}\uppsi^{(b)}_{\pm}(w)\uppsi^{(a)}_{\pm}(z),\label{psipsi1}\\
	&\uppsi^{(a)}_+(z)\uppsi^{(b)}_-(w)\simeq\frac{\phi^{a\Leftarrow b}(z+c/2,w-c/2)}{\phi^{a\Leftarrow b}(z-c/2,w+c/2)}\uppsi^{(b)}_-(w)\uppsi^{(a)}_+(z),\\
	&\uppsi^{(a)}_{\pm}(z)\mathtt{e}^{(b)}(w)\simeq\phi^{a\Leftarrow b}(z\pm c/2,w)\mathtt{e}^{(b)}(w)\uppsi^{(a)}_{\pm}(z),\\
	&\uppsi^{(a)}_{\pm}(z)\mathtt{f}^{(b)}(w)\simeq\phi^{a\Leftarrow b}(z\mp c/2,w)^{-1}\mathtt{f}^{(b)}(w)\uppsi^{(a)}_{\pm}(z),\label{psif}\\
	&\mathtt{e}^{(a)}(z)\mathtt{e}^{(b)}(w)\simeq(-1)^{|a||b|}\phi^{a\Leftarrow b}(z,w)\mathtt{e}^{(b)}(w)\mathtt{e}^{(a)}(z),\label{ee}\\
	&\mathtt{f}^{(a)}(z)\mathtt{f}^{(b)}(w)\simeq(-1)^{|a||b|}\phi^{a\Leftarrow b}(z,w)^{-1}\mathtt{f}^{(b)}(w)\mathtt{f}^{(a)}(z),\\
	&\left[\mathtt{e}^{(a)}(z),\mathtt{f}^{(b)}(w)\right\}\simeq-\delta_{ab}\left(\delta(z-w-c)\uppsi^{(a)}_+(z-c/2)-\delta(z-w+c)\uppsi^{(a)}_-(w-c/2)\right)\label{ef}.
\end{align}
For the trigonometric and elliptic cases, there can also be a non-trivial central element $c$ (which is 0 for the rational case). Notice that we also have two types of $\uppsi$ currents, $\uppsi_{\pm}$, which is different from the rational algebra where $\uppsi_+=\uppsi_-=\uppsi$. The symbol ``$\simeq$'' has a different meaning in those two cases as well. It means that the Laurent expansion on the two sides should agree, and we shall henceforth simply write it as ``$=$''. Moreover, $\delta(z)=1/z$ for the rational algebra while for the other two cases, it is the formal delta function
\begin{equation}
	\delta(z)=\sum_{k\in\mathbb{Z}}Z^k.
\end{equation}

When working with the current relations, not all of them are independent. As shown in \cite[Appendix B]{Galakhov:2021vbo}, \eqref{psipsi1}$\sim$\eqref{psif} can be derived from \eqref{ee}$\sim$\eqref{ef}. Therefore, it suffices to consider the $\mathtt{e}\mathtt{e}$, $\mathtt{f}\mathtt{f}$ and $\mathtt{e}\mathtt{f}$ current relations.

Some comments are in order:
\begin{itemize}
	\item The mode expansions of the currents are different in each case. Since we are not going to use them here, readers are referred to \cite[(2.17)$\sim$(2.19)]{Galakhov:2021vbo} for the explicit expressions.
	
	\item It is natural to conjecture that the toroidal and elliptic quiver algebras are isomorphic for Seiberg dual quivers (except for the subtleties when there are quiver nodes with two adjoint loops). For the affine Dynkin cases discussed in this paper, this should be proven in a similar manner as in \cite{bezerra2021braid,Bao:2023kkh} for generalized conifolds\footnote{As discussed in \cite{Bao:2023kkh}, it is expected to have a subalgebra structure (whose map can be constructed similar to the transformation under Seiberg duality) for the quivers related by higgsing at least for the one-parameter degeneration in the toric cases. For the non-toric cases discussed here, they are all one-parameter algebras. Hence, we should also have a subalgebra structure which is consistent with the underlying affine Lie algebras. The gauge theories should be related by higgsing as well.}.
	
	\item Similar to the rational quiver Yangians, the toroidal and elliptic cases should also be subject to the Serre relations. As before, we mention two possible choices here. One would be the symmetric sum of permuted modes in the nested supercommutators such as in \eqref{Serre1}. However, the supercommutators should be replaced with the corresponding deformed brackets (see for example \cite{Galakhov:2021vbo}). The other one would be the relations from \cite{Negut:2023iia} (whose rational limit was given in \eqref{Serre2}):
	\begin{equation}
		\begin{split}
			\sum_{i=1}^l&(-1)^{\frac{1}{2}(i(i-1)-\mathfrak{f}(\mathfrak{f}-1))}\frac{\prod\limits_{\text{pos}(j)>\text{pos}(k)}\upzeta_{a_k,a_j}(Z_kZ_j^{-1})\left(Z_j^{1/2}Z_k^{-1/2}\right)^{\delta_{a_j,a_k}\delta_{j<k}}}{\prod\limits_{\text{pos}(j)=\text{pos}(k)+1}\left(1-Z_k^{1/2}Z_j^{-1/2}\widetilde{H}_{a_k,a_j}^{1/2}\right)}\\
		    &\left(\frac{Z_1\widetilde{H}_{a_1,a_2}\dots \widetilde{H}_{a_{i-1},a_i}}{Z_i}\right)^{1/2}\mathtt{e}^{(a_i)}(Z_i)\mathtt{e}^{(a_{i-1})}(Z_{i-1})\dots\mathtt{e}^{(a_1)}(Z_1)\mathtt{e}^{(a_l)}(Z_l)\dots\mathtt{e}^{(a_{i+1})}(Z_{i+1})=0,\label{Serre3}
		\end{split}
	\end{equation}
    and likewise for $\mathtt{f}$. The notations can be found around \eqref{Serre2}. The signs $(-1)^{\frac{1}{2}(i(i-1)-\mathfrak{f}(\mathfrak{f}-1))}$ are slightly different from those in \cite{Negut:2023iia}. This is because the $\mathtt{e}\mathtt{e}$ and $\mathtt{f}\mathtt{f}$ relations of the toroidal algebras therein do not have $(-1)^{|a||b|}$. Again, we shall not fixate on determining the full Serre relations here. We will simply assume that the Serre relations would be consistent with the following discussions.
	
	\item Unlike the rational case, the coproduct is more straightforward for the other two cases. It has relatively simple expressions in terms of the currents:
	\begin{align}
		&\Delta\left(\mathtt{e}^{(a)}(Z)\right)=\mathtt{e}^{(a)}(Z)\otimes1+\uppsi^{(a)}\left(C_1^{1/2}Z\right)\otimes\mathtt{e}^{(a)}(C_1Z),\\
		&\Delta\left(\mathtt{f}^{(a)}(Z)\right)=1\otimes\mathtt{f}^{(a)}(Z)+\mathtt{f}^{(a)}(C_2Z)\otimes\mathtt{f}^{(a)}\left(C_2^{1/2}Z\right),\\
		&\Delta\left(\uppsi^{(a)}_+(Z)\right)=\uppsi^{(a)}_+(Z)\otimes\uppsi^{(a)}_+\left(C_1^{-1}Z\right),\\
		&\Delta\left(\uppsi^{(a)}_-(Z)\right)=\uppsi^{(a)}_-\left(C_2^{-1}Z\right)\otimes\uppsi^{(a)}_-(Z),\\
		&\Delta(C)=C\otimes C.
	\end{align}
	Here, $C_1=C\otimes1$ and $C_2=1\otimes C$ indicate where the $C$ factors should be in the mode expressions. The expressions in terms of modes can be found for example around (2.26)$\sim$(2.29) in \cite{Bao:2023kkh}.
	
	\item When $c=0$, the quiver algebras would still have crystal/poset modules which were studied in \cite{Galakhov:2021vbo}. Mathematically, this is known as the vertical representation in the toroidal case. For non-trivial $c$, one needs to consider the horizontal representation. The so-called (1,0) representation for the toroidal algebras associated to generalized conifolds was found in \cite{bezerra2021quantum}. In particular, the horizontal representations in terms of vertex operators should be useful when studying the connection to deformed VOAs/$\mathcal{W}$-algebras. We hope that the free field realization below would also shed light on this connection.
	
	\item As we have introduced the twisted quiver Yangians above, it is natural to wonder if there could be certain trigonometric and elliptic extensions for them as well. A naive construction would be replacing the modes in the rational algebra with the corresponding cases. However, modifications will be required. Of course, they should still be subalgebras of the toroidal/elliptic quiver Yangians, and one needs to study such extensions of the $J$ presentation. We expect that such twisted algebras would be related to certain deformed $\mathcal{W}$-algebras via truncations. However, this has not been very well-studied even for the usual toroidal/elliptic algebras (see also \S\ref{outlook}). Therefore, we postpone the possible generalizations of the twisted quiver Yangians to future work.
	
	\item One does not have to stop at the torus, and it could be possible to consider more general Riemann surfaces of genus greater than one and even with punctures. This would lead us to the realm of generalized cohomology theories. The rational/trigonometric/elliptic quiver algebras are associated with the ordinary cohomology/K-theory/elliptic cohomology. In general, we can consider the function $\zeta(z)$ given by the inverse function of some formal group law logarithm.. This should also be closely related to (generalized) CoHAs. See \cite{Galakhov:2021vbo,Galakhov:2023aev,Li:2023zub} for some recent developments.
\end{itemize}

Now, let us discuss the free field realization of the toroidal and elliptic algebras. As aforementioned, the (1,0) representation in the case of $A(m-1|n-1)^{(1)}$ for toroidal algebras was found in \cite{bezerra2021quantum}. For quivers without any self-loops (such as toric chiral quivers), a free field realization was given in \cite{Bao:2023kkh}. Here, we shall construct a free field realization for any general quivers\footnote{Notice that this free field representation is different from the crystal/vertical representation. It might be considered as some ``horizontal-like'' representation whose construction is similar to the one in \cite{bezerra2021quantum} for $A(m-1|n-1)^{(1)}$ quantum toroidal algebras.}.

We shall first consider the toroidal quiver algebras. It would be instructive to write $\mathfrak{q}=C$, and we have the $\mathfrak{q}$-number $[n]_{\mathfrak{q}}=\frac{\mathfrak{q}^n-\mathfrak{q}^{-n}}{\mathfrak{q}-\mathfrak{q}^{-1}}$. The OPE of two vertex operators can be written as $V_1(Z)V_2(W)=\langle V_1(Z)V_2(W)\rangle(V_1(Z)V_2(W))$, where $\langle\dots\rangle$ and $(\dots)$ denote the contraction and the normal ordering respectively. This contraction is a rational function understood as a Laurent series that converges in the region $|Z|\gg|W|$. It would also be convenient to write the difference operator:
\begin{equation}
	\partial V(Z)=\frac{V(\mathfrak{q}Z)-V(\mathfrak{q}^{-1}Z)}{(\mathfrak{q}-\mathfrak{q}^{-1})Z}.
\end{equation}

Let us introduce the modes with the commutation relations
\begin{align}
	&\left[k^{(a)}_r,k^{(b)}_s\right]=\delta_{r+s,0}\frac{[r]_{\mathfrak{q}}^2}{r}\sum_{i\in\{b\rightarrow a\}}\mathfrak{q}\widetilde{H}_{ba,i},\\
	&\left[l^{(a)}_r,l^{(b)}_s\right]=\delta_{r+s,0}\frac{[r]_{\mathfrak{q}}^2}{r}\sum_{i\in\{a\rightarrow b\}}\mathfrak{q}\widetilde{H}_{ab,i},\\
	&\left[\gamma^{(a)}_r,\gamma^{(b)}_s\right]=\delta_{ab}\delta_{r+s,0}\frac{[r]_{\mathfrak{q}}^2}{r},\\
	&\left[\lambda^{(a)}_r,\lambda^{(b)}_s\right]=\delta_{ab}\delta_{r+s,0}\frac{[2r]_{\mathfrak{q}}[r]_{\mathfrak{q}}}{r},\\
	&\left[\upsilon^{(a)}_r,\upsilon^{(b)}_s\right]=\delta_{ab}\delta_{r+s,0}\frac{[2r]_{\mathfrak{q}}[r]_{\mathfrak{q}}}{r},
\end{align}
with the other commutators vanishing for $r,s\in\mathbb{Z}$. Consider the currents
\begin{equation}
	\mathtt{x}^{(a)}_{\pm}(Z)=\sum_{r>0}\frac{\mathtt{x}^{(a)}_{\pm r}}{[r]_{\mathfrak{q}}}Z^{\mp r}\quad(\mathtt{x}=k,l,\gamma,\lambda,\upsilon).
\end{equation}
Then the non-trivial contractions among them are
\begin{align}
	&\left\langle\exp\left(k^{(a)}_+(Z)\right)\exp\left(k^{(b)}_-(W)\right)\right\rangle=\prod_i\left(1-\mathfrak{q}\widetilde{H}_{ba,i}\frac{W}{Z}\right),\\
	&\left\langle\exp\left(l^{(a)}_+(Z)\right)\exp\left(l^{(b)}_-(W)\right)\right\rangle=\prod_i\left(1-\mathfrak{q}\widetilde{H}_{ab,i}\frac{W}{Z}\right),\\
	&\left\langle\exp\left(\gamma^{(a)}_+(Z)\right)\exp\left(\gamma^{(b)}_-(W)\right)\right\rangle=\left(1-\frac{W}{Z}\right)^{-\delta_{ab}},\\
	&\left\langle\exp\left(\lambda^{(a)}_+(Z)\right)\exp\left(\lambda^{(b)}_-(W)\right)\right\rangle=\left(1-\mathfrak{q}\frac{W}{Z}\right)^{-\delta_{ab}}\left(1-\mathfrak{q}^{-1}\frac{W}{Z}\right)^{-\delta_{ab}},\\
	&\left\langle\exp\left(\upsilon^{(a)}_+(Z)\right)\exp\left(\upsilon^{(b)}_-(W)\right)\right\rangle=\left(1-\mathfrak{q}\frac{W}{Z}\right)^{-\delta_{ab}}\left(1-\mathfrak{q}^{-1}\frac{W}{Z}\right)^{-\delta_{ab}}.\\
\end{align}
We shall define the vertex operators
\begin{align}
	&K^{(a)}(Z)=\left(Z\exp\left(k^{(a)}_-(\mathfrak{q}^{-1}Z)\right)\exp\left(-k^{(a)}_+(Z)\right)\text{e}^{\overline{k}^{(a)}}Z^{k^{(a)}_0}\right),\\
	&L^{(a)}(Z)=\left(Z\exp\left(l^{(a)}_-(\mathfrak{q}^{-1}Z)\right)\exp\left(-l^{(a)}_+(Z)\right)\text{e}^{-\overline{l}^{(a)}}Z^{-l^{(a)}_0}\right),\\
	&\Gamma^{(a)}_+(Z)=\left(\exp\left(\gamma^{(a)}_-(Z)\right)\exp\left(-\gamma^{(a)}_+(Z)\right)\text{e}^{\overline{\gamma}^{(a)}}Z^{\gamma^{(a)}_0}\right),\\
	&\Gamma^{(a)}_-(Z)=\left(\exp\left(-\gamma^{(a)}_-(Z)\right)\exp\left(\gamma^{(a)}_+(Z)\right)\text{e}^{-\overline{\gamma}^{(a)}}Z^{-\gamma^{(a)}_0}\right),\\
	&\Lambda^{(a)}(Z)=\left(\exp\left(\upsilon^{(a)}_-(Z)\right)\exp\left(-\lambda^{(a)}_+(Z)\right)\text{e}^{\overline{\upsilon}^{(a)}}Z^{\lambda^{(a)}_0}\right),\\
	&\Upsilon^{(a)}(Z)=\left(\exp\left(-\lambda^{(a)}_-(Z)\right)\exp\left(\upsilon^{(a)}_+(Z)\right)\text{e}^{-\overline{\lambda}^{(a)}}Z^{-\upsilon^{(a)}_0}\right),
\end{align}
where we have also introduced elements with
\begin{align}
	&\left\langle\text{e}^{\overline{k}^{(a)}}Z^{k^{(b)}_0}\right\rangle=Z^{|a\rightarrow  b|},\quad\left\langle\text{e}^{\overline{l}^{(a)}}Z^{l^{(b)}_0}\right\rangle=Z^{-|a\rightarrow  b|},\\
	&\left\langle Z^{\gamma^{(a)}_0}\text{e}^{\overline{\gamma}^{(b)}}\right\rangle=Z^{\delta_{ab}},\quad\left\langle Z^{\lambda^{(a)}_0}\text{e}^{\overline{\lambda}^{(b)}}\right\rangle=Z^{2\delta_{ab}},\quad\left\langle Z^{\upsilon^{(a)}_0}\text{e}^{\overline{\upsilon}^{(b)}}\right\rangle=Z^{2\delta_{ab}},\\
	&\left(\text{e}^{\overline{k}^{(a)}}\text{e}^{\overline{k}^{(b)}}\right)=\epsilon(a,b)\left(\text{e}^{\overline{k}^{(a)}+\overline{k}^{(b)}}\right),\quad\left(\text{e}^{\overline{l}^{(a)}}\text{e}^{\overline{l}^{(b)}}\right)=\epsilon(b,a)\left(\text{e}^{\overline{l}^{(a)}+\overline{l}^{(b)}}\right),\\
	&\left(\text{e}^{\overline{k}^{(a)}}\text{e}^{\overline{l}^{(b)}}\right)=\widetilde{\epsilon}(a,b)\left(\text{e}^{\overline{k}^{(a)}+\overline{l}^{(b)}}\right).
\end{align}
The cocycle factors satisfy
\begin{equation}
	\frac{\epsilon(a,b)}{\epsilon(b,a)}=(-1)^{|a||b|}(-1)^{\chi_{ab}}\frac{\prod\widetilde{H}_{ab,i}}{\prod\widetilde{H}_{ba,j}},\quad\frac{\widetilde{\epsilon}(a,b)}{\widetilde{\epsilon}(b,a)}=(-1)^{|a||b|}
\end{equation}
for $a\neq b$, and $\epsilon(a,a)=\widetilde{\epsilon}(a,a)=1$. The OPEs of the vertex operators can then be directly obtained from the OPEs of their building blocks.

Recall that it suffices to consider the $\mathtt{ee}$, $\mathtt{ff}$ and $\mathtt{e}\mathtt{f}$ relations. It is a straightforward check that the followings satisfy the $\mathtt{e}\mathtt{e}$ and $\mathtt{f}\mathtt{f}$ relations:
\begin{align}
	&\mathtt{e}^{(a)}(Z)=\left(K^{(a)}(Z)\Lambda^{(a)}(Z)\right)\quad(|a|=0),\\
	&\mathtt{e}^{(a)}(Z)=\left(K^{(a)}(Z)\partial\Gamma^{(a)}_-(Z)\right)\quad(|a|=1),\\
	&\mathtt{f}^{(a)}(Z)=\left(L^{(a)}(Z)\Upsilon^{(a)}(Z)\right)\quad(|a|=0),\\
	&\mathtt{f}^{(a)}(Z)=\left(L^{(a)}(Z)\Gamma^{(a)}_+(Z)\right)\quad(|a|=1).
\end{align}
We can then use the $\mathtt{e}\mathtt{f}$ relations to construct the $\uppsi_{\pm}$ currents. For bosonic nodes, we have
\begin{equation}
	\begin{split}
		&\left\langle K^{(a)}(Z)\Lambda^{(a)}(Z)L^{(a)}(W)\Upsilon^{(a)}(W)\right\rangle-\left\langle L^{(a)}(W)\Upsilon^{(a)}(W)K^{(a)}(Z)\Lambda^{(a)}(Z)\right\rangle\\
		=&\left(\frac{1}{(Z-\mathfrak{q}W)(Z-\mathfrak{q}^{-1}W)}\right)_{|Z|\gg|W|}-\left(\frac{1}{(Z-\mathfrak{q}W)(Z-\mathfrak{q}^{-1}W)}\right)_{|W|\gg|Z|}\\
		=&-\frac{1}{\mathfrak{q}Z^2(\mathfrak{q}-\mathfrak{q}^{-1})}\delta\left(\mathfrak{q}\frac{Z}{W}\right)+\frac{1}{\mathfrak{q}W^2(\mathfrak{q}-\mathfrak{q}^{-1})}\delta\left(\mathfrak{q}^{-1}\frac{Z}{W}\right),
	\end{split}
\end{equation}
where $(\dots)_{\mathcal{R}}$ denotes the Laurent expansion in the region $\mathcal{R}$. It would be useful to notice that $\delta(Z/W)F(Z)=\delta(Z/W)F(W)$ for any Laurent series $F(Z)$. Therefore,
\begin{align}
	&\uppsi^{(a)}_+(Z)=\frac{1}{\mathfrak{q}^{-1}-\mathfrak{q}}\left( K^{(a)}(\mathfrak{q}^{1/2}Z)\Lambda^{(a)}(\mathfrak{q}^{1/2}Z)L^{(a)}(\mathfrak{q}^{-1/2}Z)\Upsilon^{(a)}(\mathfrak{q}^{-1/2}Z)\right)\quad(|a|=0),\\
	&\uppsi^{(a)}_-(Z)=\frac{1}{\mathfrak{q}^{-1}-\mathfrak{q}}\left( K^{(a)}(\mathfrak{q}^{-1/2}Z)\Lambda^{(a)}(\mathfrak{q}^{-1/2}Z)L^{(a)}(\mathfrak{q}^{1/2}Z)\Upsilon^{(a)}(\mathfrak{q}^{1/2}Z)\right)\quad(|a|=0).
\end{align}
For fermionic nodes, we have
\begin{equation}
	\begin{split}
		&\left( K^{(a)}(Z)\partial\Gamma^{(a)}_-(Z)L^{(a)}(W)\Gamma^{(a)}_+(W)\right)+\left( L^{(a)}(W)\Gamma^{(a)}_+(W)K^{(a)}(Z)\partial\Gamma^{(a)}_-(Z)\right)\\
		=&\left(K^{(a)}(Z)L^{(a)}(W)\partial_Z\left(\frac{1}{W}\delta\left(\frac{W}{Z}\right)\Gamma^{(a)}_+(W)\Gamma^{(a)}_-(Z)\right)\right)\\
		=&\left(K^{(a)}(Z)L^{(a)}(W)\partial_Z\left(\frac{1}{W}\delta\left(\frac{W}{Z}\right)\right)\right)
	\end{split}
\end{equation}
since
\begin{equation}
	\begin{split}
		&\left\langle\Gamma^{(a)}_-(Z)\Gamma^{(a)}_+(W)\right\rangle=\frac{1}{Z-W}\quad(|Z|\gg|W|),\\
		&\left\langle\Gamma^{(a)}_+(W)\Gamma^{(a)}_-(Z)\right\rangle=\frac{1}{W-Z}\quad(|W|\gg|Z|).
	\end{split}
\end{equation}
Therefore,
\begin{align}
	&\uppsi^{(a)}_+(Z)=\frac{1}{\mathfrak{q}^{-1}-\mathfrak{q}}\left( K^{(a)}(\mathfrak{q}^{1/2}Z)L^{(a)}(\mathfrak{q}^{-1/2}Z)\right)\quad(|a|=1),\\
	&\uppsi^{(a)}_-(Z)=\frac{1}{\mathfrak{q}^{-1}-\mathfrak{q}}\left( K^{(a)}(\mathfrak{q}^{-1/2}Z)L^{(a)}(\mathfrak{q}^{1/2}Z)\right)\quad(|a|=1).
\end{align}
One may also check that $\left[\mathtt{e}^{(a)}(Z),\mathtt{f}^{(b)}(W)\right\}=0$ for $a\neq b$ as expected.

For the elliptic case, the free field realization can be directly obtained from the toroidal case. Consider the ``elliptically deformed'' vertex operators with the contractions
\begin{align}
	&\left\langle\exp\left(\chi^{(a)}(Z)\right)\exp\left(\chi^{(b)}(W)\right)\right\rangle=\frac{\prod\left(q\widetilde{H}_{ab,i}^{-1}Z^{-1}W;q\right)_{\infty}}{\prod\left(q\widetilde{H}_{ba,j}^{-1}Z^{-1}W;q\right)_{\infty}},\\
	&\left\langle\exp\left(\xi^{(a)}(Z)\right)\exp\left(\xi^{(b)}(W)\right)\right\rangle=\frac{\prod\left(q\widetilde{H}_{ba,j}^{-1}Z^{-1}W;q\right)_{\infty}}{\prod\left(q\widetilde{H}_{ab,i}^{-1}Z^{-1}W;q\right)_{\infty}}.
\end{align}
Then the currents of the elliptic algebra can be written as
\begin{align}
	&\mathtt{e}^{(a)}(Z)=\left(\exp\left(\chi^{(a)}(Z)\right)\mathtt{E}^{(a)}(Z)\right),\\
	&\mathtt{f}^{(a)}(Z)=\left(\exp\left(\xi^{(a)}(Z)\right)\mathtt{F}^{(a)}(Z)\right),\\
	&\uppsi^{(a)}_{\pm}(Z)=\left(\exp\left(\chi^{(a)}\left(\mathfrak{q}^{\pm1/2}Z\right)\right)\exp\left(\xi^{(a)}\left(\mathfrak{q}^{\mp1/2}Z\right)\right)\mathtt{\Psi}^{(a)}_{\pm}(Z)\right).
\end{align}
In particular, $\mathtt{E}$, $\mathtt{F}$ and $\mathtt{\Psi}$ satisfy the current relations of the toroidal algebra. As a result, this gives the free field realization of the elliptic algebras.

\section{Discussions and Outlook}\label{outlook}
Let us have some discussions on possible future directions. We have considered the non-toric quivers associated to affine Dynkin diagrams in this paper. It is natural to expect that the quiver Yangians in these cases can also encode the Bethe/gauge correspondence \cite{Nekrasov:2009uh,Nekrasov:2009ui}. Similar to the toric cases in \cite{Galakhov:2022uyu,Bao:2022fpk,Litvinov:2020zeq,Chistyakova:2021yyd,Kolyaskin:2022tqi}, one may simply use the bond factor in each case to get the possible Bethe ansatz equations (see for example \cite[(2.53)]{Galakhov:2022uyu}) for some Hitchin integrable system. In the toric cases, these Bethe equations can be obtained by considering the $RTT$ relation and the actions on 2d crystals. For the non-toric cases, the Fock modules one should consider might be related to the poset representations in \cite{Li:2023zub}. In particular, since the quivers associated to the affine Dynkin diagrams are all non-chiral. There should be no obstructions that appear in the cases of chiral quivers as pointed out in \cite{Galakhov:2022uyu}.

In this paper, we have introduced the twisted quiver Yangians. For the twisted Yangians in the finite cases, one has the $RTT$ presentations, and they have connections with certain integrable systems. It would be interesting to see if/how we can construct the $R$-matrix formalism for the twisted quiver Yangians. This would require a more detailed study on their representations. It is also natural to wonder whether there exist any trigonometric and elliptic versions similar to the quiver Yangian cases.

We have mentioned how the foldings of the affine Dynkin diagram could arise in the context of (twisted) quiver Yangians. In the finite cases, the $\mathcal{W}$-algebras can be folded as studied in \cite{ragoucy2001twisted,Frappat:1992xz}. Therefore, we might also consider whether there is a similar folding story for the $\mathcal{W}$-algebras discussed here. This might help us elucidate the relations between the (twisted) quiver Yangians and the $\mathcal{W}$-algebras.

Of course, we may consider the AGT correspondence for the toroidal and elliptic counterparts of the quiver Yangians as well. For the toroidal (resp.~elliptic) cases, we expect their truncations to give rise to $\mathfrak{q}$-deformed (resp.~elliptic deformations of) $\mathcal{W}$-algebras. There is still little known even for those associated to the generalized conifolds, let alone other cases. In \cite{Harada:2021xnm,Noshita:2022dxv}, the 5d AGT correspondence was studied for the deformed VOAs at the corner and the toroidal algebra associated to $\widehat{\mathfrak{gl}}(1)$. Regarding the elliptic case, we need to construct possible elliptic deformations of the $\mathcal{W}$-algebras similar to the finite cases in \cite{Avan:2018pyf}.

Let us make some more comments on the $\mathfrak{q}$-deformed $\mathcal{W}$-algebras of A-type. In \cite{Negut:2019agq,negut2022deformed}, a deformed version associated to $\widehat{\mathfrak{gl}}(m)$ was constructed. As this is a deformed $\mathcal{W}$-algebra, the generators satisfy certain quadratic relation. To write down the quadratic relation, we need the following $R$-matrix:
\begin{equation}
	R_{12}(z)=\sum_{1\leq a,b\leq m+n}E_{aa}\otimes E_{bb}\left(\frac{\mathfrak{q}-z\mathfrak{q}^{-1}}{1-z}\right)^{\delta_{ab}}+\left(\mathfrak{q}-\mathfrak{q}^{-1}\right)\sum_{1\leq a\neq b\leq m+n}(-1)^{p(b)}E_{ab}\otimes E_{ba}\frac{z^{\delta_{a<b}}}{1-z},
\end{equation}
where we are actually considering a naive generalization to the super cases (and the orinigal definition for the non-super cases would not have the factor $(-1)^{p(b)}$). The elementary matrix $E_{ab}$ has 1 at the entry $(a,b)$ and 0 otherwise. Again, we shall focus on the distinguished case. In other words, the directions $1\leq a\leq m$ (resp.~$m+1\leq a\leq m+n$) are bosonic (resp.~fermionic). We will use $R_{21}=\pi R_{12}$ to denote the one with the two components of the tensor product exchanged, where $\pi$ is this permutation operator. As the name suggests, the $R$-matrix satisfies the Yang-Baxter equation
\begin{equation}
	R_{12}\left(\frac{z_1}{z_2}\right)R_{13}\left(\frac{z_1}{z_3}\right)R_{23}\left(\frac{z_2}{z_3}\right)=R_{23}\left(\frac{z_2}{z_3}\right)R_{13}\left(\frac{z_1}{z_3}\right)R_{12}\left(\frac{z_1}{z_2}\right).
\end{equation}
Moreover, we have
\begin{equation}
	R_{12}\left(\frac{z_1}{z_2}\right)R_{21}\left(\frac{z_2}{z_1}\right)=\mathscr{F}\left(\frac{z_1}{z_2}\right)\text{Id},
\end{equation}
where
\begin{equation}
	\mathscr{F}(z)=\frac{\left(1-z\mathfrak{q}^2\right)\left(1-z\mathfrak{q}^{-2}\right)}{(1-z)^2}.
\end{equation}
This allows us to express the inverse of $R_{12}$ using $R_{21}$.

Let us still write the generators as the (super)matrix $U_i=(U_{i,ab})$ with spin $i$. The quadratic relation is then given by
\begin{equation}
	\begin{split}
		&R_{12}\left(\frac{z}{w}\mathfrak{p}^{2(k-k')}\right)(U_k(z)\otimes1)R_{21}\left(\frac{w}{z}\mathfrak{p}^{2k'}\right)(1\otimes U_{k'}(w))\prod_{i=k'-k+1}^{k'-1}\mathscr{F}\left(\frac{z}{w}\mathfrak{p}^{-2i}\right)\\
		&-(1\otimes U_{k'}(w))R_{12}\left(\frac{z}{w}\mathfrak{p}^{2k}\right)(U_k(z)\otimes1)R_{21}\left(\frac{w}{z}\right)\prod_{i=1}^{k-1}\mathscr{F}\left(\frac{w}{z}\mathfrak{p}^{-2i}\right)\\
		=&\sum_{j\in\{-k',\dots,k-k'-1\}\cup\{1,\dots,k\}}\text{sgn}(j)\delta\left(\frac{z}{w}\mathfrak{p}^{2j}\right)\\
		&\left(\mathfrak{q}^{-1}-\mathfrak{q}\right)\pi\left((U_{k'+j}(z)\otimes1)R_{21}\left(\mathfrak{p}^{2k}\right)(1\otimes U_{k-j}(w))\prod_{i=1}^{k-1}\mathscr{F}\left(\mathfrak{p}^{-2i}\right)\right),
	\end{split}
\end{equation}
where $\mathfrak{p}$ is the parameter of the algebra. The left hand side should be understood as follows. The first (resp.~second) line above is expanded in the region $|z|\gg|w|$ (resp.~$|w|\gg|z|$). This is similar to the discussions in the toroidal quiver Yangians.

As this is a deformation of $\mathcal{W}_{m|n\times l}$, the generators can actually be obtained via the deformed Miura transformation:
\begin{equation}
	\left(\Lambda_1\left(z\mathfrak{p}^{2(l-1)}\right)-D\right)\left(\Lambda_2\left(z\mathfrak{p}^{2(l-2)}\right)-D\right)\dots\left(\Lambda_l(z)-D\right)=\sum_{k=0}^l(-D)^{l-k}U_k(z),
\end{equation}
where the difference operator $D$ acts as $DF(z)=F\left(z\mathfrak{p}^2\right)$ for any function $F(z)$. Then
\begin{equation}
	U_k(z)=\sum_{1\leq i_1<\dots<i_k\leq l}\Lambda_{i_1}\left(z\mathfrak{p}^{2(k-1)}\right)\dots\Lambda_{i_{k-1}}\left(z\mathfrak{p}^2\right)\Lambda_{i_k}(z).
\end{equation}
In particular, $U_{k>l}$ vanishes (and $U_0=1$). As we can see, this is actually the deformation of the $x$-algebra truncations, and $\Lambda$ plays the role as $J^{(x)}$. It would also be interesting to study the deformations of more general truncations.

Therefore, to study the possible connections to the toroidal quiver BPS algebras, we can first focus on the building blocks $\Lambda_{ab}$. In terms of $\Lambda$, the quadratic relation reads
\begin{align}
	&\left(\frac{\mathfrak{q}-zw^{-1}\mathfrak{q}^{-1}}{1-zw^{-1}}\right)^{\delta_{ac}}\left(\frac{\mathfrak{q}-wz^{-1}\mathfrak{p}^2\mathfrak{q}^{-1}}{1-wz^{-1}\mathfrak{p}^2}\right)^{\delta_{bc}}\Lambda_{ab}(z)\Lambda_{cd}(w)\nonumber\\
	&+\delta_{bc}\left(\mathfrak{q}-\mathfrak{q}^{-1}\right)\sum_{e\neq c}(-1)^{p(e)}\left(\frac{\mathfrak{q}-zw^{-1}\mathfrak{q}^{-1}}{1-zw^{-1}}\right)^{\delta_{ab}}\frac{\left(wz^{-1}\mathfrak{p}^2\right)^{\delta_{c<e}}}{1-wz^{-1}\mathfrak{p}^2}\Lambda_{ae}(z)\Lambda_{ed}(w)\nonumber\\
	&+\delta_{a\neq c}\left(\mathfrak{q}-\mathfrak{q}^{-1}\right)(-1)^{p(a)p(c)+p(c)p(b)+p(a)p(b)}\frac{\left(zw^{-1}\right)^{\delta_{a<c}}}{1-zw^{-1}}\left(\frac{\mathfrak{q}-wz^{-1}\mathfrak{p}^2\mathfrak{q}^{-1}}{1-zw^{-1}\mathfrak{p}^2}\right)^{\delta_{ab}}\Lambda_{cb}(z)\Lambda_{ad}(w)\nonumber\\
	&+\delta_{ab}\delta_{a\neq c}\left(\mathfrak{q}-\mathfrak{q}^{-1}\right)^2\sum_e(-1)^{p(e)+p(a)}\frac{\left(zw^{-1}\right)^{\delta_{a<c}}}{1-zw^{-1}}\frac{\left(wz^{-1}\mathfrak{p}^2\right)^{\delta_{a<e}}}{1-wz^{-1}\mathfrak{p}^2}\Lambda_{ce}(z)\Lambda_{ed}(w)\nonumber\\
	=&(-1)^{(p(c)+p(d))(p(a)+p(b))}\left(\frac{\mathfrak{q}-zw^{-1}\mathfrak{p}^2\mathfrak{q}^{-1}}{1-zw^{-1}\mathfrak{p}^2}\right)^{\delta_{ad}}\left(\frac{\mathfrak{q}-wz^{-1}\mathfrak{q}^{-1}}{1-wz^{-1}}\right)^{\delta_{bd}}\Lambda_{cd}(w)\Lambda_{ab}(z)\nonumber\\
	&+\delta_{b\neq d}\left(\mathfrak{q}-\mathfrak{q}^{-1}\right)(-1)^{p(c)+(p(c)+p(d))(p(a)+p(b))}\left(\frac{\mathfrak{q}-zw^{-1}\mathfrak{p}^2\mathfrak{q}^{-1}}{1-zw^{-1}\mathfrak{p}^2}\right)^{\delta_{ab}}\frac{\left(wz^{-1}\right)^{\delta_{b<d}}}{1-wz^{-1}}\Lambda_{cb}(w)\Lambda_{ad}(z)\nonumber\\
	&+\delta_{ad}\left(\mathfrak{q}-\mathfrak{q}^{-1}\right)\sum_{e\neq a}(-1)^{p(a)p(b)+p(c)p(a)+p(c)p(b)}\frac{\left(zw^{-1}\mathfrak{p}^2\right)^{\delta_{a<e}}}{1-zw^{-1}\mathfrak{p}^2}\left(\frac{\mathfrak{q}-wz^{-1}\mathfrak{q}^{-1}}{1-zw^{-1}}\right)^{\delta_{ab}}\Lambda_{ce}(w)\Lambda_{eb}(z)\nonumber\\
	&+\delta_{ab}\left(\mathfrak{q}-\mathfrak{q}^{-1}\right)^2\sum_{e\neq a}(-1)^{p(d)+p(b)}\frac{\left(zw^{-1}\mathfrak{p}^2\right)^{\delta_{b<e}}}{1-zw^{-1}\mathfrak{p}^2}\frac{\left(wz^{-1}\right)^{\delta_{a<d}}}{1-wz^{-1}}\Lambda_{ce}(w)\Lambda_{ed}(z).
\end{align}
A possible approach is to consider the free field realization of the currents and compare it with the one for toroidal quiver BPS algebras. One possible free field realization could be some $\mathfrak{q}$-deformation of what we have discussed for $\mathcal{W}_{m|n\times\infty}$. Although it should be straightforward to write down the deformed commutators of the free field modes, this is still not evident since there could be extra terms in the vertex operators that would vanish in the rational limit.

One may also consider finding a deformed version of the surjection from the rational quiver Yangian to $\mathcal{W}_{m|n\times l}$. However, we expect the truncation in the deformed case to be a bit different (if there is one). In particular, the positive (or negative) copy of the toroidal quiver BPS algebra generated by $\mathtt{e}$ (or $\mathtt{f}$) might be sufficient to give rise to such deformed $\mathcal{W}$-algebra. Indeed, it was shown in \cite{negut2022deformed} that there is an embedding from this deformed $\mathcal{W}$-algebra (in the non-super case) to some positive copy of certain quantum toroidal algebra associated to $\widehat{\mathfrak{gl}}(m)$.

When constructing the embedding, two shuffle algebra structures of the deformed $\mathcal{W}$-algebra were found in \cite{negut2022deformed}. On the other hand, a modification of the positive (or negative) part of the toroidal quiver Yangian was discussed in \cite[\S5.2]{Galakhov:2021vbo}. In particular, this is a shuffle algebra that should be associated to the Coulomb branch (cf. \cite{Galakhov:2018lta}). Such shuffle algebra, albeit different from the toroidal quiver Yangian of the Higgs branch, should give the isomorphic Hilbert space of the BPS states following the Higgs-Coulomb duality. Therefore, it could also be possible that the truncation of the algebra associated to the Coulomb branch can give rise to the deformed $\mathcal{W}$-algebras. Despite the resemblance, this is still quite non-trivial as can be seen from the different shuffle products on the two sides.

Of course, it could also be possible to construct the deformed $\mathcal{W}$-algebras similar to the ones discussed here from the toroidal quiver Yangians or the quiver shuffle algebras directly. Then one needs to show that such algebras do have a deformed vertex algebra structure and that they are deformations of the (rational) $\mathcal{W}$-algebras. Moreover, it would be important to find the quadratic relation of the algebras.

\section*{Acknowledgement}
I would like to thank Yang-Hui He for enjoyable discussions. I am grateful to Yuji Tachikawa for his kind and patient correspondences. The research is supported by a CSC scholarship.

\appendix

\section{Affine Dynkin Diagrams and Seiberg Dual Quivers}\label{SDquivers}
Let us list all the possible quivers and their edge weights in the quiver Yangians associated to the affine Lie superalgebras. The distinguished diagrams/quivers for the untwisted cases are depicted in \S\ref{distinguished}, and hence we shall only give the remaining phases for them here. For the twisted affine Dynkin diagrams which were classified in \cite[Table 11]{frappat1989structure}, the quiver Yangians can be obtained in the same manner. In particular, the possible superpotential terms and ways to assign weights are completely the same as those for the untwisted affine cases discussed in this paper. Hence, we will not repeat the procedure for plotting these quivers explicitly here.

All the superpotential terms follow the rules discussed in \S\ref{supercases} unless otherwise specified below. Moreover, we shall not repeat those for $A(m-1|n-1)^{(1)}$ and the exceptional cases ($D(2,1;\alpha)^{(1)}$, $F(4)^{(1)}$, $G(3)^{(1)}$) as all of their phases have already been discussed in \S\ref{supercases}. Again, the ranks of the gauge groups are proportional to the Dynkin-Kac labels in the affine Dynkin diagram given below. We also use $\mathcal{F}$ to denote the parity of the grey nodes in the affine Dynkin diagram, that is, the fermionic nodes without adjoint loops in the quiver.

\paragraph{Case 1: $\mathfrak{osp}(2m|2n)^{(1)}$} There are four types of configurations for $\mathcal{F}=0$:
\begin{equation}
	\includegraphics[width=12cm]{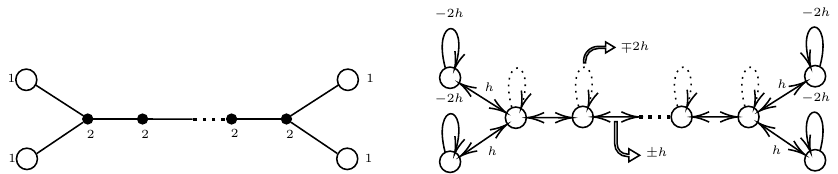},
\end{equation}
\begin{equation}
	\includegraphics[width=12cm]{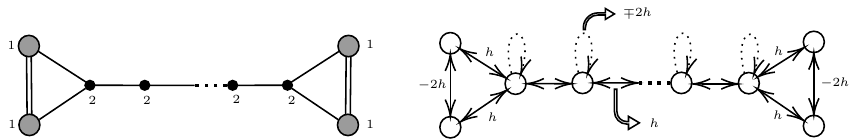},
\end{equation}
\begin{equation}
	\includegraphics[width=12cm]{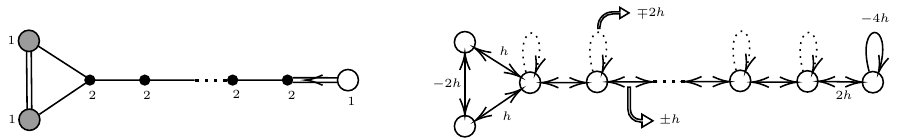},
\end{equation}
\begin{equation}
	\includegraphics[width=12cm]{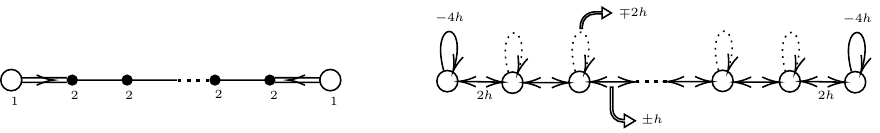}.
\end{equation}
There are two types of configurations for $\mathcal{F}=1$:
\begin{equation}
	\includegraphics[width=12cm]{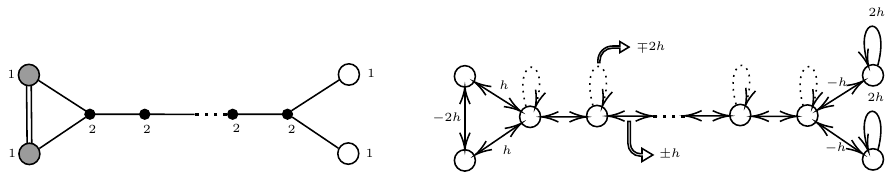},
\end{equation}
\begin{equation}
	\includegraphics[width=12cm]{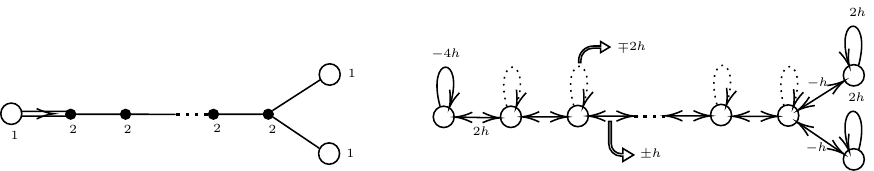}.
\end{equation}
In these cases, if there is the part
\begin{equation}
	\includegraphics[width=5cm]{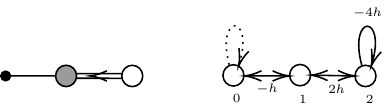},
\end{equation}
where the numbers label the nodes in the quiver. Then there is an extra superpotential term $X_{12}X_{21}X_{10}X_{01}X_{10}X_{01}$, which is consistent with the weight assignments.

\paragraph{Case 2: $\mathfrak{osp}(2m+1|2n)^{(1)}$} There are three types of configurations for $\mathcal{F}=0$:
\begin{equation}
	\includegraphics[width=12cm]{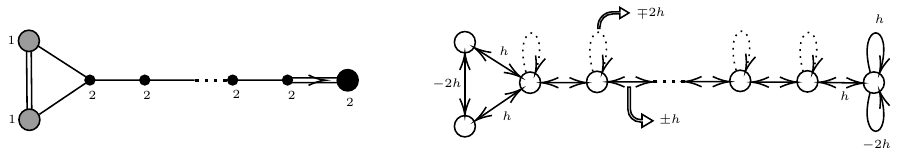},
\end{equation}
\begin{equation}
	\includegraphics[width=12cm]{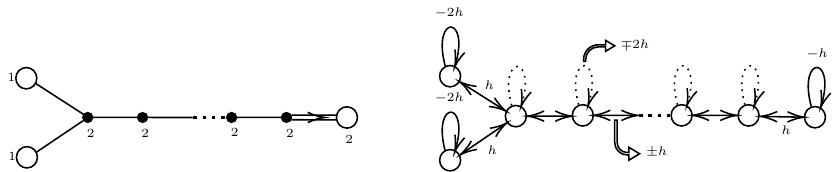},
\end{equation}
\begin{equation}
	\includegraphics[width=12cm]{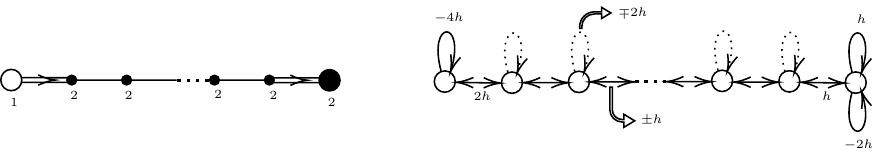}.
\end{equation}
There are three types of configurations for $\mathcal{F}=1$:
\begin{equation}
	\includegraphics[width=12cm]{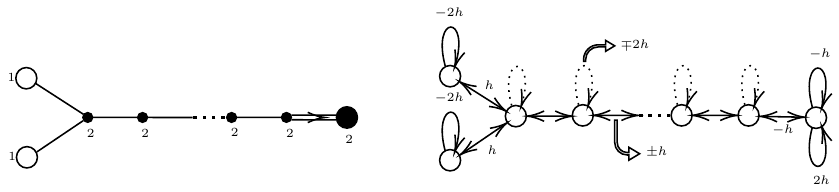},
\end{equation}
\begin{equation}
	\includegraphics[width=12cm]{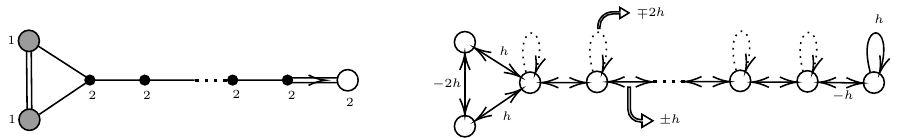},
\end{equation}
\begin{equation}
	\includegraphics[width=12cm]{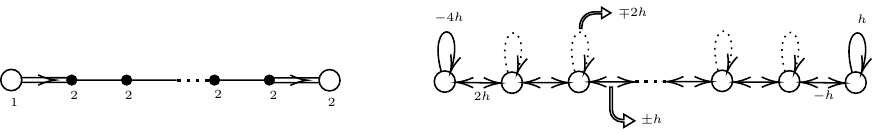}.
\end{equation}

\addcontentsline{toc}{section}{References}
\bibliographystyle{utphys}
\bibliography{references}

\providecommand{\href}[2]{#2}\begingroup\raggedright\begin{thebibliography}{100}

\bibitem{Li:2020rij}
W.~Li and M.~Yamazaki, ``{Quiver Yangian from Crystal Melting},''
  \href{http://dx.doi.org/10.1007/JHEP11(2020)035}{{\em JHEP} {\bfseries 11}
  (2020) 035}, \href{http://arxiv.org/abs/2003.08909}{{\ttfamily
  arXiv:2003.08909 [hep-th]}}.

\bibitem{Galakhov:2020vyb}
D.~Galakhov and M.~Yamazaki, ``{Quiver Yangian and Supersymmetric Quantum
  Mechanics},'' \href{http://dx.doi.org/10.1007/s00220-022-04490-y}{{\em
  Commun. Math. Phys.} {\bfseries 396} no.~2, (2022) 713--785},
  \href{http://arxiv.org/abs/2008.07006}{{\ttfamily arXiv:2008.07006
  [hep-th]}}.

\bibitem{Harvey:1995fq}
J.~A. Harvey and G.~W. Moore, ``{Algebras, BPS states, and strings},''
  \href{http://dx.doi.org/10.1016/0550-3213(95)00605-2}{{\em Nucl. Phys. B}
  {\bfseries 463} (1996) 315--368},
  \href{http://arxiv.org/abs/hep-th/9510182}{{\ttfamily arXiv:hep-th/9510182}}.

\bibitem{Harvey:1996gc}
J.~A. Harvey and G.~W. Moore, ``{On the algebras of BPS states},''
  \href{http://dx.doi.org/10.1007/s002200050461}{{\em Commun. Math. Phys.}
  {\bfseries 197} (1998) 489--519},
  \href{http://arxiv.org/abs/hep-th/9609017}{{\ttfamily arXiv:hep-th/9609017}}.

\bibitem{Okounkov:2003sp}
A.~Okounkov, N.~Reshetikhin, and C.~Vafa, ``{Quantum Calabi-Yau and classical
  crystals},'' \href{http://dx.doi.org/10.1007/0-8176-4467-9_16}{{\em Prog.
  Math.} {\bfseries 244} (2006) 597},
  \href{http://arxiv.org/abs/hep-th/0309208}{{\ttfamily arXiv:hep-th/0309208}}.

\bibitem{Iqbal:2003ds}
A.~Iqbal, N.~Nekrasov, A.~Okounkov, and C.~Vafa, ``{Quantum foam and
  topological strings},''
  \href{http://dx.doi.org/10.1088/1126-6708/2008/04/011}{{\em JHEP} {\bfseries
  04} (2008) 011}, \href{http://arxiv.org/abs/hep-th/0312022}{{\ttfamily
  arXiv:hep-th/0312022}}.

\bibitem{Ooguri:2009ijd}
H.~Ooguri and M.~Yamazaki, ``{Crystal Melting and Toric Calabi-Yau
  Manifolds},'' \href{http://dx.doi.org/10.1007/s00220-009-0836-y}{{\em Commun.
  Math. Phys.} {\bfseries 292} (2009) 179--199},
  \href{http://arxiv.org/abs/0811.2801}{{\ttfamily arXiv:0811.2801 [hep-th]}}.

\bibitem{Yamazaki:2022cdg}
M.~Yamazaki, ``{Quiver Yangians and crystal meltings: A concise summary},''
  \href{http://dx.doi.org/10.1063/5.0089785}{{\em J. Math. Phys.} {\bfseries
  64} no.~1, (2023) 011101}, \href{http://arxiv.org/abs/2203.14314}{{\ttfamily
  arXiv:2203.14314 [hep-th]}}.

\bibitem{Galakhov:2021xum}
D.~Galakhov, W.~Li, and M.~Yamazaki, ``{Shifted quiver Yangians and
  representations from BPS crystals},''
  \href{http://dx.doi.org/10.1007/JHEP08(2021)146}{{\em JHEP} {\bfseries 08}
  (2021) 146}, \href{http://arxiv.org/abs/2106.01230}{{\ttfamily
  arXiv:2106.01230 [hep-th]}}.

\bibitem{Aganagic:2010qr}
M.~Aganagic and K.~Schaeffer, ``{Wall Crossing, Quivers and Crystals},''
  \href{http://dx.doi.org/10.1007/JHEP10(2012)153}{{\em JHEP} {\bfseries 10}
  (2012) 153}, \href{http://arxiv.org/abs/1006.2113}{{\ttfamily arXiv:1006.2113
  [hep-th]}}.

\bibitem{Bao:2022oyn}
J.~Bao, Y.-H. He, and A.~Zahabi, ``{Crystal melting, BPS quivers and
  plethystics},'' \href{http://dx.doi.org/10.1007/JHEP06(2022)016}{{\em JHEP}
  {\bfseries 06} (2022) 016}, \href{http://arxiv.org/abs/2202.12850}{{\ttfamily
  arXiv:2202.12850 [hep-th]}}.

\bibitem{Noshita:2021ldl}
G.~Noshita and A.~Watanabe, ``{A note on quiver quantum toroidal algebra},''
  \href{http://dx.doi.org/10.1007/JHEP05(2022)011}{{\em JHEP} {\bfseries 05}
  (2022) 011}, \href{http://arxiv.org/abs/2108.07104}{{\ttfamily
  arXiv:2108.07104 [hep-th]}}.

\bibitem{Galakhov:2021vbo}
D.~Galakhov, W.~Li, and M.~Yamazaki, ``{Toroidal and elliptic quiver BPS
  algebras and beyond},'' \href{http://dx.doi.org/10.1007/JHEP02(2022)024}{{\em
  JHEP} {\bfseries 02} (2022) 024},
  \href{http://arxiv.org/abs/2108.10286}{{\ttfamily arXiv:2108.10286
  [hep-th]}}.

\bibitem{Galakhov:2023aev}
D.~Galakhov, ``{BPS States Meet Generalized Cohomology},''
  \href{http://arxiv.org/abs/2303.05538}{{\ttfamily arXiv:2303.05538
  [hep-th]}}.

\bibitem{Li:2023zub}
W.~Li, ``{Quiver algebras and their representations for arbitrary quivers},''
  \href{http://arxiv.org/abs/2303.05521}{{\ttfamily arXiv:2303.05521
  [hep-th]}}.

\bibitem{Joyce:2008pc}
D.~Joyce and Y.~Song, ``{A Theory of generalized Donaldson-Thomas
  invariants},'' \href{http://arxiv.org/abs/0810.5645}{{\ttfamily
  arXiv:0810.5645 [math.AG]}}.

\bibitem{Kontsevich:2008fj}
M.~Kontsevich and Y.~Soibelman, ``{Stability structures, motivic
  Donaldson-Thomas invariants and cluster transformations},''
  \href{http://arxiv.org/abs/0811.2435}{{\ttfamily arXiv:0811.2435 [math.AG]}}.

\bibitem{Kontsevich:2010px}
M.~Kontsevich and Y.~Soibelman, ``{Cohomological Hall algebra, exponential
  Hodge structures and motivic Donaldson-Thomas invariants},''
  \href{http://dx.doi.org/10.4310/CNTP.2011.v5.n2.a1}{{\em Commun. Num. Theor.
  Phys.} {\bfseries 5} (2011) 231--352},
  \href{http://arxiv.org/abs/1006.2706}{{\ttfamily arXiv:1006.2706 [math.AG]}}.

\bibitem{Rapcak:2018nsl}
M.~Rapcak, Y.~Soibelman, Y.~Yang, and G.~Zhao, ``{Cohomological Hall algebras,
  vertex algebras and instantons},''
  \href{http://dx.doi.org/10.1007/s00220-019-03575-5}{{\em Commun. Math. Phys.}
  {\bfseries 376} no.~3, (2019) 1803--1873},
  \href{http://arxiv.org/abs/1810.10402}{{\ttfamily arXiv:1810.10402
  [math.QA]}}.

\bibitem{Galakhov:2018lta}
D.~Galakhov, ``{BPS Hall Algebra of Scattering Hall States},''
  \href{http://dx.doi.org/10.1016/j.nuclphysb.2019.114693}{{\em Nucl. Phys. B}
  {\bfseries 946} (2019) 114693},
  \href{http://arxiv.org/abs/1812.05801}{{\ttfamily arXiv:1812.05801
  [hep-th]}}.

\bibitem{Denef:2002ru}
F.~Denef, ``{Quantum quivers and Hall / hole halos},''
  \href{http://dx.doi.org/10.1088/1126-6708/2002/10/023}{{\em JHEP} {\bfseries
  10} (2002) 023}, \href{http://arxiv.org/abs/hep-th/0206072}{{\ttfamily
  arXiv:hep-th/0206072}}.

\bibitem{Alday:2009aq}
L.~F. Alday, D.~Gaiotto, and Y.~Tachikawa, ``{Liouville Correlation Functions
  from Four-dimensional Gauge Theories},''
  \href{http://dx.doi.org/10.1007/s11005-010-0369-5}{{\em Lett. Math. Phys.}
  {\bfseries 91} (2010) 167--197},
  \href{http://arxiv.org/abs/0906.3219}{{\ttfamily arXiv:0906.3219 [hep-th]}}.

\bibitem{Wyllard:2009hg}
N.~Wyllard, ``{A(N-1) conformal Toda field theory correlation functions from
  conformal N = 2 SU(N) quiver gauge theories},''
  \href{http://dx.doi.org/10.1088/1126-6708/2009/11/002}{{\em JHEP} {\bfseries
  11} (2009) 002}, \href{http://arxiv.org/abs/0907.2189}{{\ttfamily
  arXiv:0907.2189 [hep-th]}}.

\bibitem{Bao:2022jhy}
J.~Bao, ``{Quiver Yangians and $\mathcal{W}$-Algebras for Generalized
  Conifolds},'' \href{http://arxiv.org/abs/2208.13395}{{\ttfamily
  arXiv:2208.13395 [hep-th]}}.

\bibitem{ragoucy2001twisted}
E.~Ragoucy, ``{Twisted Yangians and folded-algebras},'' {\em International
  Journal of Modern Physics A} {\bfseries 16} no.~13, (2001) 2411--2433,
  \href{http://arxiv.org/abs/math/0012182}{{\ttfamily arXiv:math/0012182}}.

\bibitem{brown2009twisted}
J.~Brown, ``{Twisted Yangians and finite W-algebras},'' {\em Transformation
  Groups} {\bfseries 14} (2009) 87--114,
  \href{http://arxiv.org/abs/0710.2918}{{\ttfamily arXiv:0710.2918 [math.QA]}}.

\bibitem{ueda2021twisted}
M.~Ueda, ``{Twisted Affine Yangian and Rectangular $W$-algebra of type $D$},''
  \href{http://arxiv.org/abs/2107.09999}{{\ttfamily arXiv:2107.09999
  [math.QA]}}.

\bibitem{Szendroi:2007nu}
B.~Szendroi, ``{Non-commutative Donaldson\textendash{}Thomas invariants and the
  conifold},'' \href{http://dx.doi.org/10.2140/gt.2008.12.1171}{{\em Geom.
  Topol.} {\bfseries 12} no.~2, (2008) 1171--1202},
  \href{http://arxiv.org/abs/0705.3419}{{\ttfamily arXiv:0705.3419 [math.AG]}}.

\bibitem{Bershadsky:1996nh}
M.~Bershadsky, K.~A. Intriligator, S.~Kachru, D.~R. Morrison, V.~Sadov, and
  C.~Vafa, ``{Geometric singularities and enhanced gauge symmetries},''
  \href{http://dx.doi.org/10.1016/S0550-3213(96)90131-5}{{\em Nucl. Phys. B}
  {\bfseries 481} (1996) 215--252},
  \href{http://arxiv.org/abs/hep-th/9605200}{{\ttfamily arXiv:hep-th/9605200}}.

\bibitem{Cecotti:2012gh}
S.~Cecotti and M.~Del~Zotto, ``{4d N=2 Gauge Theories and Quivers: the
  Non-Simply Laced Case},''
  \href{http://dx.doi.org/10.1007/JHEP10(2012)190}{{\em JHEP} {\bfseries 10}
  (2012) 190}, \href{http://arxiv.org/abs/1207.7205}{{\ttfamily arXiv:1207.7205
  [hep-th]}}.

\bibitem{slodowy2006simple}
P.~Slodowy, {\em {Simple Singularities and Simple Algebraic Groups}}.
\newblock Springer, 1980.

\bibitem{stekolshchik2005notes}
R.~Stekolshchik, ``Notes on coxeter transformations and the mckay
  correspondence,'' \href{http://arxiv.org/abs/math/0510216}{{\ttfamily
  arXiv:math/0510216}}.

\bibitem{McKay1980}
J.~McKay, ``{Graphs, singularities, and finite groups},'' {\em Proc. Symp. Pure
  Math.} {\bfseries 37} no.~183, (1980) .

\bibitem{sagemath}
{The Sage Developers}, {\em {S}ageMath, the {S}age {M}athematics {S}oftware
  {S}ystem ({V}ersion 9.8)}, 2023.
\newblock {\tt https://www.sagemath.org}.

\bibitem{Cabrera:2019izd}
S.~Cabrera, A.~Hanany, and M.~Sperling, ``{Magnetic quivers, Higgs branches,
  and 6d $N$=(1,0) theories},''
  \href{http://dx.doi.org/10.1007/JHEP06(2019)071}{{\em JHEP} {\bfseries 06}
  (2019) 071}, \href{http://arxiv.org/abs/1904.12293}{{\ttfamily
  arXiv:1904.12293 [hep-th]}}. [Erratum: JHEP 07, 137 (2019)].

\bibitem{Bourget:2019aer}
A.~Bourget, S.~Cabrera, J.~F. Grimminger, A.~Hanany, M.~Sperling, A.~Zajac, and
  Z.~Zhong, ``{The Higgs mechanism \textemdash{} Hasse diagrams for symplectic
  singularities},'' \href{http://dx.doi.org/10.1007/JHEP01(2020)157}{{\em JHEP}
  {\bfseries 01} (2020) 157}, \href{http://arxiv.org/abs/1908.04245}{{\ttfamily
  arXiv:1908.04245 [hep-th]}}.

\bibitem{Bourget:2021siw}
A.~Bourget, J.~F. Grimminger, A.~Hanany, M.~Sperling, and Z.~Zhong, ``{Branes,
  Quivers, and the Affine Grassmannian},''
  \href{http://arxiv.org/abs/2102.06190}{{\ttfamily arXiv:2102.06190
  [hep-th]}}.

\bibitem{Bourget:2022ehw}
A.~Bourget, J.~F. Grimminger, A.~Hanany, and Z.~Zhong, ``{The Hasse diagram of
  the moduli space of instantons},''
  \href{http://dx.doi.org/10.1007/JHEP08(2022)283}{{\em JHEP} {\bfseries 08}
  (2022) 283}, \href{http://arxiv.org/abs/2202.01218}{{\ttfamily
  arXiv:2202.01218 [hep-th]}}.

\bibitem{guay2018coproduct}
N.~Guay, H.~Nakajima, and C.~Wendlandt, ``{Coproduct for Yangians of affine
  Kac--Moody algebras},'' {\em Advances in Mathematics} {\bfseries 338} (2018)
  865--911, \href{http://arxiv.org/abs/1701.05288}{{\ttfamily arXiv:1701.05288
  [math.QA]}}.

\bibitem{Negut:2023iia}
A.~Negu\c{t}, ``{Reduced quiver quantum toroidal algebras},''
  \href{http://arxiv.org/abs/2301.00703}{{\ttfamily arXiv:2301.00703
  [hep-th]}}.

\bibitem{gholampour2009counting}
A.~Gholampour and Y.~Jiang, ``{Counting invariants for the ADE McKay
  quivers},'' \href{http://arxiv.org/abs/0910.5551}{{\ttfamily arXiv:0910.5551
  [math.AG]}}.

\bibitem{Young:2008hn}
B.~Young and J.~Bryan, ``{Generating functions for colored 3D Young diagrams
  and the Donaldson-Thomas invariants of orbifolds},''
  \href{http://dx.doi.org/10.1215/00127094-2010-009}{{\em Duke Math. J.}
  {\bfseries 152} (2010) 115--153},
  \href{http://arxiv.org/abs/0802.3948}{{\ttfamily arXiv:0802.3948 [math.CO]}}.

\bibitem{drinfeld1985hopf}
V.~G. Drinfeld, ``{Hopf algebras and the quantum Yang-Baxter equation},'' in
  {\em Dokl. Akad. Nauk SSSR}, vol.~283, pp.~1060--1064.
\newblock 1985.

\bibitem{olshanskii2006twisted}
G.~Olshanskii, ``{Twisted Yangians and infinite-dimensional classical Lie
  algebras},'' in {\em Quantum Groups: Proceedings of Workshops held in the
  Euler International Mathematical Institute, Leningrad, Fall 1990},
  pp.~104--119, Springer.
\newblock 2006.

\bibitem{Molev:1994rs}
A.~Molev, M.~Nazarov, and G.~Olshansky, ``{Yangians and classical Lie
  algebras},'' \href{http://dx.doi.org/10.1070/RM1996v051n02ABEH002772}{{\em
  Russ. Math. Surveys} {\bfseries 51} (1996) 205},
  \href{http://arxiv.org/abs/hep-th/9409025}{{\ttfamily arXiv:hep-th/9409025}}.

\bibitem{Molev:1997wp}
A.~Molev, ``{Finite dimensional irreducible representations of twisted
  Yangians},'' \href{http://dx.doi.org/10.1063/1.532551}{{\em J. Math. Phys.}
  {\bfseries 39} (1998) 5559--5600},
  \href{http://arxiv.org/abs/q-alg/9711022}{{\ttfamily arXiv:q-alg/9711022}}.

\bibitem{guay2016twisted}
N.~Guay and V.~Regelskis, ``{Twisted Yangians for symmetric pairs of types B,
  C, D},'' {\em Mathematische Zeitschrift} {\bfseries 284} (2016) 131--166,
  \href{http://arxiv.org/abs/1407.5247}{{\ttfamily arXiv:1407.5247 [math.QA]}}.

\bibitem{Gerrard:2017igy}
A.~Gerrard, N.~MacKay, and V.~Regelskis, ``{Nested algebraic Bethe ansatz for
  open spin chains with even twisted Yangian symmetry},''
  \href{http://dx.doi.org/10.1007/s00023-018-0731-1}{{\em Annales Henri
  Poincare} {\bfseries 20} no.~2, (2019) 339--392},
  \href{http://arxiv.org/abs/1710.08409}{{\ttfamily arXiv:1710.08409
  [math-ph]}}.

\bibitem{DeLeeuw:2019ohp}
M.~De~Leeuw, T.~Gombor, C.~Kristjansen, G.~Linardopoulos, and B.~Pozsgay,
  ``{Spin Chain Overlaps and the Twisted Yangian},''
  \href{http://dx.doi.org/10.1007/JHEP01(2020)176}{{\em JHEP} {\bfseries 01}
  (2020) 176}, \href{http://arxiv.org/abs/1912.09338}{{\ttfamily
  arXiv:1912.09338 [hep-th]}}.

\bibitem{Belliard:2014uja}
S.~Belliard and V.~Regelskis, ``{Drinfeld J Presentation of Twisted
  Yangians},'' \href{http://dx.doi.org/10.3842/SIGMA.2017.011}{{\em SIGMA}
  {\bfseries 13} (2017) 011}, \href{http://arxiv.org/abs/1401.2143}{{\ttfamily
  arXiv:1401.2143 [math.QA]}}.

\bibitem{helgason1979differential}
S.~Helgason, {\em {Differential geometry, Lie groups, and symmetric spaces}}.
\newblock Academic press, 1979.

\bibitem{kac1990infinite}
V.~G. Kac, {\em {Infinite-dimensional Lie algebras}}.
\newblock Cambridge university press, 1990.

\bibitem{musson2012lie}
I.~M. Musson, {\em {Lie superalgebras and enveloping algebras}}, vol.~131.
\newblock American Mathematical Soc., 2012.

\bibitem{ueda2019affine}
M.~Ueda, ``{Affine Super Yangian},''
  \href{http://arxiv.org/abs/1911.06666}{{\ttfamily arXiv:1911.06666
  [math.RT]}}.

\bibitem{ueda2022surjectivity}
M.~Ueda, ``{The surjectivity of the evaluation map of the affine super
  Yangian},'' {\em Osaka Journal of Mathematics} {\bfseries 59} no.~3, (2022)
  481--493, \href{http://arxiv.org/abs/2001.06398}{{\ttfamily arXiv:2001.06398
  [math.RT]}}.

\bibitem{ueda2022affine}
M.~Ueda, ``{Affine super Yangians and rectangular W-superalgebras},'' {\em
  Journal of Mathematical Physics} {\bfseries 63} no.~5, (2022) 051701,
  \href{http://arxiv.org/abs/2002.03479}{{\ttfamily arXiv:2002.03479
  [math.RT]}}.

\bibitem{frappat1989structure}
L.~Frappat, A.~Sciarrino, and P.~Sorba, ``{Structure of basic Lie superalgebras
  and of their affine extensions},'' {\em Communications in Mathematical
  Physics} {\bfseries 121} (1989) 457--500.

\bibitem{Frappat:1996pb}
L.~Frappat, P.~Sorba, and A.~Sciarrino, ``{Dictionary on Lie superalgebras},''
  \href{http://arxiv.org/abs/hep-th/9607161}{{\ttfamily arXiv:hep-th/9607161}}.

\bibitem{yamane1999defining}
H.~Yamane, ``{On defining relations of affine Lie superalgebras and affine
  quantized universal enveloping superalgebras},'' {\em Publications of the
  Research Institute for Mathematical Sciences} {\bfseries 35} no.~3, (1999)
  321--390, \href{http://arxiv.org/abs/q-alg/9603015}{{\ttfamily
  arXiv:q-alg/9603015}}.

\bibitem{serganova2011kac}
V.~Serganova, ``{Kac--Moody superalgebras and integrability},'' {\em
  Developments and trends in infinite-dimensional Lie theory} (2011) 169--218.

\bibitem{Seiberg:1994pq}
N.~Seiberg, ``{Electric - magnetic duality in supersymmetric nonAbelian gauge
  theories},'' \href{http://dx.doi.org/10.1016/0550-3213(94)00023-8}{{\em Nucl.
  Phys. B} {\bfseries 435} (1995) 129--146},
  \href{http://arxiv.org/abs/hep-th/9411149}{{\ttfamily arXiv:hep-th/9411149}}.

\bibitem{Franco:2003ja}
S.~Franco, A.~Hanany, Y.-H. He, and P.~Kazakopoulos, ``{Duality walls, duality
  trees and fractional branes},''
  \href{http://arxiv.org/abs/hep-th/0306092}{{\ttfamily arXiv:hep-th/0306092}}.

\bibitem{Feng:2000mi}
B.~Feng, A.~Hanany, and Y.-H. He, ``{D-brane gauge theories from toric
  singularities and toric duality},''
  \href{http://dx.doi.org/10.1016/S0550-3213(00)00699-4}{{\em Nucl. Phys. B}
  {\bfseries 595} (2001) 165--200},
  \href{http://arxiv.org/abs/hep-th/0003085}{{\ttfamily arXiv:hep-th/0003085}}.

\bibitem{Beasley:2001zp}
C.~E. Beasley and M.~R. Plesser, ``{Toric duality is Seiberg duality},''
  \href{http://dx.doi.org/10.1088/1126-6708/2001/12/001}{{\em JHEP} {\bfseries
  12} (2001) 001}, \href{http://arxiv.org/abs/hep-th/0109053}{{\ttfamily
  arXiv:hep-th/0109053}}.

\bibitem{Feng:2001bn}
B.~Feng, A.~Hanany, Y.-H. He, and A.~M. Uranga, ``{Toric duality as Seiberg
  duality and brane diamonds},''
  \href{http://dx.doi.org/10.1088/1126-6708/2001/12/035}{{\em JHEP} {\bfseries
  12} (2001) 035}, \href{http://arxiv.org/abs/hep-th/0109063}{{\ttfamily
  arXiv:hep-th/0109063}}.

\bibitem{Feng:2002zw}
B.~Feng, S.~Franco, A.~Hanany, and Y.-H. He, ``{Symmetries of toric duality},''
  \href{http://dx.doi.org/10.1088/1126-6708/2002/12/076}{{\em JHEP} {\bfseries
  12} (2002) 076}, \href{http://arxiv.org/abs/hep-th/0205144}{{\ttfamily
  arXiv:hep-th/0205144}}.

\bibitem{frenkel1992vertex}
I.~B. Frenkel and Y.~Zhu, ``{Vertex operator algebras associated to
  representations of affine and Virasoro algebras},''.

\bibitem{matsuo2010quasi}
A.~Matsuo, K.~Nagatomo, and A.~Tsuchiya, ``{Quasi-finite Algebras Graded by
  Hamiltonian and Vertex Operator Algebras},'' in {\em Moonshine-The First
  Quarter Century and Beyond: Proceedings of a Workshop on the Moonshine
  Conjectures and Vertex Algebras}, no.~372, p.~282, Cambridge University
  Press.
\newblock 2010.
\newblock \href{http://arxiv.org/abs/math/0505071}{{\ttfamily
  arXiv:math/0505071}}.

\bibitem{schiffmann2013cherednik}
O.~Schiffmann and E.~Vasserot, ``{Cherednik algebras, W-algebras and the
  equivariant cohomology of the moduli space of instantons on $A^2$},'' {\em
  Publications math{\'e}matiques de l'IH{\'E}S} {\bfseries 118} no.~1, (2013)
  213--342, \href{http://arxiv.org/abs/1202.2756}{{\ttfamily arXiv:1202.2756
  [math.QA]}}.

\bibitem{Gaberdiel:2017dbk}
M.~R. Gaberdiel, R.~Gopakumar, W.~Li, and C.~Peng, ``{Higher Spins and Yangian
  Symmetries},'' \href{http://dx.doi.org/10.1007/JHEP04(2017)152}{{\em JHEP}
  {\bfseries 04} (2017) 152}, \href{http://arxiv.org/abs/1702.05100}{{\ttfamily
  arXiv:1702.05100 [hep-th]}}.

\bibitem{Prochazka:2019dvu}
T.~Proch\'azka, ``{Instanton R-matrix and $ \mathcal{W} $-symmetry},''
  \href{http://dx.doi.org/10.1007/JHEP12(2019)099}{{\em JHEP} {\bfseries 12}
  (2019) 099}, \href{http://arxiv.org/abs/1903.10372}{{\ttfamily
  arXiv:1903.10372 [hep-th]}}.

\bibitem{Bouwknegt:1992wg}
P.~Bouwknegt and K.~Schoutens, ``{W symmetry in conformal field theory},''
  \href{http://dx.doi.org/10.1016/0370-1573(93)90111-P}{{\em Phys. Rept.}
  {\bfseries 223} (1993) 183--276},
  \href{http://arxiv.org/abs/hep-th/9210010}{{\ttfamily arXiv:hep-th/9210010}}.

\bibitem{Keller:2011ek}
C.~A. Keller, N.~Mekareeya, J.~Song, and Y.~Tachikawa, ``{The ABCDEFG of
  Instantons and W-algebras},''
  \href{http://dx.doi.org/10.1007/JHEP03(2012)045}{{\em JHEP} {\bfseries 03}
  (2012) 045}, \href{http://arxiv.org/abs/1111.5624}{{\ttfamily arXiv:1111.5624
  [hep-th]}}.

\bibitem{guay2007affine}
N.~Guay, ``{Affine Yangians and deformed double current algebras in type A},''
  {\em Advances in Mathematics} {\bfseries 211} no.~2, (2007) 436--484.

\bibitem{kodera2021guay}
R.~Kodera, ``{On Guay’s evaluation map for affine Yangians},'' {\em Algebras
  and Representation Theory} {\bfseries 24} (2021) 253--267,
  \href{http://arxiv.org/abs/1806.09884}{{\ttfamily arXiv:1806.09884
  [math.RT]}}.

\bibitem{arakawa2017explicit}
T.~Arakawa and A.~Molev, ``{Explicit generators in rectangular affine
  W-algebras of type A},'' {\em Letters in Mathematical Physics} {\bfseries
  107} no.~1, (2017) 47--59, \href{http://arxiv.org/abs/1403.1017}{{\ttfamily
  arXiv:1403.1017 [math.RT]}}.

\bibitem{arakawa2017introduction}
T.~Arakawa, ``{Introduction to W-algebras and their representation theory},''
  {\em Perspectives in Lie theory} (2017) 179--250,
  \href{http://arxiv.org/abs/1605.00138}{{\ttfamily arXiv:1605.00138
  [math.RT]}}.

\bibitem{Litvinov:2016mgi}
A.~Litvinov and L.~Spodyneiko, ``{On W algebras commuting with a set of
  screenings},'' \href{http://dx.doi.org/10.1007/JHEP11(2016)138}{{\em JHEP}
  {\bfseries 11} (2016) 138}, \href{http://arxiv.org/abs/1609.06271}{{\ttfamily
  arXiv:1609.06271 [hep-th]}}.

\bibitem{Prochazka:2018tlo}
T.~Proch\'azka and M.~Rap\v{c}\'ak, ``{$ \mathcal{W} $ -algebra modules, free
  fields, and Gukov-Witten defects},''
  \href{http://dx.doi.org/10.1007/JHEP05(2019)159}{{\em JHEP} {\bfseries 05}
  (2019) 159}, \href{http://arxiv.org/abs/1808.08837}{{\ttfamily
  arXiv:1808.08837 [hep-th]}}.

\bibitem{Rapcak:2019wzw}
M.~Rap\v{c}\'ak, ``{On extensions of $
  \mathfrak{gl}\widehat{\left(\left.m\right|n\right)} $ Kac-Moody algebras and
  Calabi-Yau singularities},''
  \href{http://dx.doi.org/10.1007/JHEP01(2020)042}{{\em JHEP} {\bfseries 01}
  (2020) 042}, \href{http://arxiv.org/abs/1910.00031}{{\ttfamily
  arXiv:1910.00031 [hep-th]}}.

\bibitem{Eberhardt:2019xmf}
L.~Eberhardt and T.~Proch\'azka, ``{The matrix-extended $W_{1+\infty}$
  algebra},'' \href{http://dx.doi.org/10.1007/JHEP12(2019)175}{{\em JHEP}
  {\bfseries 12} (2019) 175}, \href{http://arxiv.org/abs/1910.00041}{{\ttfamily
  arXiv:1910.00041 [hep-th]}}.

\bibitem{Creutzig:2019qos}
T.~Creutzig and Y.~Hikida, ``{Rectangular W algebras and superalgebras and
  their representations},''
  \href{http://dx.doi.org/10.1103/PhysRevD.100.086008}{{\em Phys. Rev. D}
  {\bfseries 100} no.~8, (2019) 086008},
  \href{http://arxiv.org/abs/1906.05868}{{\ttfamily arXiv:1906.05868
  [hep-th]}}.

\bibitem{Prochazka:2017qum}
T.~Proch\'azka and M.~Rap\v{c}\'ak, ``{Webs of W-algebras},''
  \href{http://dx.doi.org/10.1007/JHEP11(2018)109}{{\em JHEP} {\bfseries 11}
  (2018) 109}, \href{http://arxiv.org/abs/1711.06888}{{\ttfamily
  arXiv:1711.06888 [hep-th]}}.

\bibitem{Gaiotto:2017euk}
D.~Gaiotto and M.~Rap\v{c}\'ak, ``{Vertex Algebras at the Corner},''
  \href{http://dx.doi.org/10.1007/JHEP01(2019)160}{{\em JHEP} {\bfseries 01}
  (2019) 160}, \href{http://arxiv.org/abs/1703.00982}{{\ttfamily
  arXiv:1703.00982 [hep-th]}}.

\bibitem{Yang:2007zzb}
W.-L. Yang, Y.-Z. Zhang, and X.~Liu, ``{Free field realization of current
  superalgebra gl(m|n)(k)},'' \href{http://dx.doi.org/10.1063/1.2739306}{{\em
  J. Math. Phys.} {\bfseries 48} (2007) 053514},
  \href{http://arxiv.org/abs/0806.0190}{{\ttfamily arXiv:0806.0190 [hep-th]}}.

\bibitem{Fujitsu:1994np}
A.~Fujitsu, ``{ope.math: Operator product expansions in free field realizations
  of conformal field theory},''
  \href{http://dx.doi.org/10.1016/0010-4655(94)90231-3}{{\em Comput. Phys.
  Commun.} {\bfseries 79} (1994) 78--99}.

\bibitem{Creutzig:2019wfe}
T.~Creutzig, Y.~Hikida, and T.~Uetoko, ``{Rectangular W-algebras of types
  $so(M)$ and $sp(2M)$ and dual coset CFTs},''
  \href{http://dx.doi.org/10.1007/JHEP10(2019)023}{{\em JHEP} {\bfseries 10}
  (2019) 023}, \href{http://arxiv.org/abs/1906.05872}{{\ttfamily
  arXiv:1906.05872 [hep-th]}}.

\bibitem{Yang:2008hd}
W.-L. Yang and Y.-Z. Zhang, ``{Free field realization of the osp(2n|2n) current
  algebra},'' \href{http://dx.doi.org/10.1103/PhysRevD.78.106004}{{\em Phys.
  Rev. D} {\bfseries 78} (2008) 106004},
  \href{http://arxiv.org/abs/0806.2477}{{\ttfamily arXiv:0806.2477 [hep-th]}}.

\bibitem{Yang:2008ghr}
W.-L. Yang and Y.-Z. Zhang, ``{On explicit free field realizations of current
  algebras},'' \href{http://dx.doi.org/10.1016/j.nuclphysb.2008.02.011}{{\em
  Nucl. Phys. B} {\bfseries 800} (2008) 527--546},
  \href{http://arxiv.org/abs/0806.1996}{{\ttfamily arXiv:0806.1996 [hep-th]}}.

\bibitem{Yang:2008vb}
W.-L. Yang, Y.-Z. Zhang, and S.~Kault, ``{Differential operator realizations of
  superalgebras and free field representations of corresponding current
  algebras},'' \href{http://dx.doi.org/10.1016/j.nuclphysb.2009.06.029}{{\em
  Nucl. Phys. B} {\bfseries 823} (2009) 372--402},
  \href{http://arxiv.org/abs/0810.3719}{{\ttfamily arXiv:0810.3719 [hep-th]}}.

\bibitem{Chen:2011us}
X.~Chen, W.-L. Yang, X.-M. Ding, J.~Feng, S.-M. Ke, K.~Wu, and Y.-Z. Zhang,
  ``{Free field realization of the exceptional current superalgebra
  $\hat{D(2,1:\alpha)}_k$},''
  \href{http://dx.doi.org/10.1088/1751-8113/45/40/405204}{{\em J. Phys. A}
  {\bfseries 45} (2012) 405204},
  \href{http://arxiv.org/abs/1108.2093}{{\ttfamily arXiv:1108.2093 [hep-th]}}.

\bibitem{bezerra2021braid}
L.~Bezerra and E.~Mukhin, ``{Braid actions on quantum toroidal
  superalgebras},'' {\em Journal of Algebra} {\bfseries 585} (2021) 338--369,
  \href{http://arxiv.org/abs/1912.08729}{{\ttfamily arXiv:1912.08729
  [math.QA]}}.

\bibitem{Bao:2023kkh}
J.~Bao, ``{A Survey of Toric Quivers and BPS Algebras},''
  \href{http://arxiv.org/abs/2301.00663}{{\ttfamily arXiv:2301.00663
  [hep-th]}}.

\bibitem{bezerra2021quantum}
L.~Bezerra and E.~Mukhin, ``{Quantum Toroidal Algebra Associated with
  $\mathfrak{gl}_{m|n}$},'' {\em Algebras and Representation Theory} {\bfseries
  24} (2021) 541--564, \href{http://arxiv.org/abs/1904.07297}{{\ttfamily
  arXiv:1904.07297 [math.QA]}}.

\bibitem{Nekrasov:2009uh}
N.~A. Nekrasov and S.~L. Shatashvili, ``{Supersymmetric vacua and Bethe
  ansatz},'' \href{http://dx.doi.org/10.1016/j.nuclphysbps.2009.07.047}{{\em
  Nucl. Phys. B Proc. Suppl.} {\bfseries 192-193} (2009) 91--112},
  \href{http://arxiv.org/abs/0901.4744}{{\ttfamily arXiv:0901.4744 [hep-th]}}.

\bibitem{Nekrasov:2009ui}
N.~A. Nekrasov and S.~L. Shatashvili, ``{Quantum integrability and
  supersymmetric vacua},'' \href{http://dx.doi.org/10.1143/PTPS.177.105}{{\em
  Prog. Theor. Phys. Suppl.} {\bfseries 177} (2009) 105--119},
  \href{http://arxiv.org/abs/0901.4748}{{\ttfamily arXiv:0901.4748 [hep-th]}}.

\bibitem{Galakhov:2022uyu}
D.~Galakhov, W.~Li, and M.~Yamazaki, ``{Gauge/Bethe correspondence from quiver
  BPS algebras},'' \href{http://dx.doi.org/10.1007/JHEP11(2022)119}{{\em JHEP}
  {\bfseries 11} (2022) 119}, \href{http://arxiv.org/abs/2206.13340}{{\ttfamily
  arXiv:2206.13340 [hep-th]}}.

\bibitem{Bao:2022fpk}
J.~Bao, ``{A note on quiver Yangians and \ensuremath{\mathscr{R}}-matrices},''
  \href{http://dx.doi.org/10.1007/JHEP08(2022)219}{{\em JHEP} {\bfseries 08}
  (2022) 219}, \href{http://arxiv.org/abs/2206.06186}{{\ttfamily
  arXiv:2206.06186 [hep-th]}}.

\bibitem{Litvinov:2020zeq}
A.~Litvinov and I.~Vilkoviskiy, ``{Liouville reflection operator, affine
  Yangian and Bethe ansatz},''
  \href{http://dx.doi.org/10.1007/JHEP12(2020)100}{{\em JHEP} {\bfseries 12}
  (2020) 100}, \href{http://arxiv.org/abs/2007.00535}{{\ttfamily
  arXiv:2007.00535 [hep-th]}}.

\bibitem{Chistyakova:2021yyd}
E.~Chistyakova, A.~Litvinov, and P.~Orlov, ``{Affine Yangian of $ \mathfrak{gl}
  $(2) and integrable structures of superconformal field theory},''
  \href{http://dx.doi.org/10.1007/JHEP03(2022)102}{{\em JHEP} {\bfseries 03}
  (2022) 102}, \href{http://arxiv.org/abs/2110.05870}{{\ttfamily
  arXiv:2110.05870 [hep-th]}}.

\bibitem{Kolyaskin:2022tqi}
D.~Kolyaskin, A.~Litvinov, and A.~Zhukov, ``{R-matrix formulation of affine
  Yangian of gl\textasciicircum{}(1|1)},''
  \href{http://dx.doi.org/10.1016/j.nuclphysb.2022.116023}{{\em Nucl. Phys. B}
  {\bfseries 985} (2022) 116023},
  \href{http://arxiv.org/abs/2206.01636}{{\ttfamily arXiv:2206.01636
  [hep-th]}}.

\bibitem{Frappat:1992xz}
L.~Frappat, E.~Ragoucy, and P.~Sorba, ``{Folding the W algebras},''
  \href{http://dx.doi.org/10.1016/0550-3213(93)90598-J}{{\em Nucl. Phys. B}
  {\bfseries 404} (1993) 805--838},
  \href{http://arxiv.org/abs/hep-th/9301040}{{\ttfamily arXiv:hep-th/9301040}}.

\bibitem{Harada:2021xnm}
K.~Harada, Y.~Matsuo, G.~Noshita, and A.~Watanabe, ``{$q$-deformation of corner
  vertex operator algebras by Miura transformation},''
  \href{http://dx.doi.org/10.1007/JHEP04(2021)202}{{\em JHEP} {\bfseries 04}
  (2021) 202}, \href{http://arxiv.org/abs/2101.03953}{{\ttfamily
  arXiv:2101.03953 [hep-th]}}.

\bibitem{Noshita:2022dxv}
G.~Noshita, ``{5d AGT correspondence of supergroup gauge theories from quantum
  toroidal $ \mathfrak{gl} _{1}$},''
  \href{http://dx.doi.org/10.1007/JHEP12(2022)157}{{\em JHEP} {\bfseries 12}
  (2022) 157}, \href{http://arxiv.org/abs/2209.08313}{{\ttfamily
  arXiv:2209.08313 [hep-th]}}.

\bibitem{Avan:2018pyf}
J.~Avan, L.~Frappat, and E.~Ragoucy, ``{Elliptic deformation of
  $\mathcal{W}_N$-algebras},''
  \href{http://dx.doi.org/10.21468/SciPostPhys.6.5.054}{{\em SciPost Phys.}
  {\bfseries 6} no.~5, (2019) 054},
  \href{http://arxiv.org/abs/1810.11410}{{\ttfamily arXiv:1810.11410
  [math.QA]}}.

\bibitem{Negut:2019agq}
A.~Negu\c{t}, ``{Toward AGT for Parabolic Sheaves},''
  \href{http://dx.doi.org/10.1093/imrn/rnaa308}{{\em Int. Math. Res. Not.}
  {\bfseries 2022} no.~9, (2022) 6512--6539},
  \href{http://arxiv.org/abs/1911.02963}{{\ttfamily arXiv:1911.02963
  [math.AG]}}.

\bibitem{negut2022deformed}
A.~Negu\c{t}, ``{Deformed W-algebras in type A for rectangular nilpotent},''
  {\em Communications in Mathematical Physics} {\bfseries 389} no.~1, (2022)
  153--195, \href{http://arxiv.org/abs/2004.02737}{{\ttfamily arXiv:2004.02737
  [math.RT]}}.

\end{thebibliography}\endgroup

\end{document}